\documentclass[preprint,amsmath,amssymb,aps,prab]{revtex4-2}

\usepackage{graphicx}
\usepackage{graphics}

\begin{document}
\title{A Formalism for the Transport and Matching of Coupled Beams in Accelerators}
\author{Onur~Gilanliogullari$^{*,1,2}$, Brahim~Mustapha$^1$ and Pavel~Snopok$^2$}
\affiliation{$^1$Physics Division, Argonne National Laboratory 
$^2$Physics Department, Illinois Institute of Technology}
\date{\today}
\begin{abstract}
	Understanding transverse coupling dynamics is crucial for beam physics, accelerator design, and operations. Currently, most accelerators are designed for uncoupled beams, and coupling is treated as an error or perturbation. Many transverse ($x$-$y$) coupling parametrizations exist: Edward-Teng, Mais-Ripken, Levedev-Bogacz, and others. Here, we present an explicit and complete formalism for transporting coupled beam optics functions based on Mais-Ripken and Lebedev-Bogacz parametrizations. The formalism allows for matching generally coupled beam optics functions but applies to uncoupled optics as well. A complete transformation method for coupled optics provides easy matching routines that can be added to known beam optics codes that lack this feature.
    For fully coupled lattices, we present methods for extracting eigenmode emittances and other beam parameters from observables that can be measured, which is essential to diagnose and characterize the beam in a real machine. We express the linear difference resonance in terms of coupled optics functions and relate it to the coupling strength parameter with explicit emittance exchange formalism realized from the generating functions discussed here.
\end{abstract} 

\maketitle

\section{Introduction}
Conventional accelerators are typically based on uncoupled beam dynamics. Courant and Snyder developed the uncoupled beam parametrization and defined the well-known beam optics functions, also called Twiss parameters~\cite{courant1958theory}. Uncoupled beam dynamics have been studied extensively, and important beam properties have been determined and characterized, such as beam stability and perturbations~\cite{lee2018accelerator,wiedemann2015particle}. Single-particle dynamics shed light on beam resonances and the importance of machine operating points. Uncoupled single-particle dynamics are relatively easy to deal with due to the separation of transverse planes, where the primary source of coupling is magnet alignment errors in a linear machine. Usually, the weak coupling created in the machine is corrected using skew quadrupoles. However, coupling never disappears in the lattice since higher-order magnets are used for various corrections, and fringe fields always introduce a source of coupling to the system, whether linear or nonlinear.  

Single-particle coupled parametrizations have already been introduced. First, Mais-Ripken introduced their coupled parametrization as an extension of the Courant-Snyder  parametrization~\cite{willeke1989methods}. Edwards-Teng introduced a general decoupling parametrization through a transformation that removes the coupling and considers the optics as in the uncoupled case of Courant-Snyder parametrization~\cite{edwards1973parametrization}. Lebedev-Bogacz introduced an extension of Mais-Ripken parametrization by parameterizing the transfer map and writing down the eigenvectors~\cite{lebedev2010betatron}. Most of these parametrizations are presented and summarized in this review on coupling~\cite{vanwelde2023parametrizations}. New coupled parametrization approaches have been published recently, shedding more light on beam coupling~\cite{tzenov2024linear}. In~\cite{tzenov2024linear}, they also show the tune map, which suggests that the resonance diagram will be surfaces rather than resonance lines, which is an interesting phenomenon. Recent advancements in 4D phase space reconstruction make strongly coupled lattices possible~\cite{kim2024four,hoover2024four}. Turn-by-turn data can also give clues about the coupled parameters, as shown in~\cite{morozov2024coupled}. There is also interesting work suggesting that specific strong correlations between $x$ and $y$ planes can lead to better space-charge performance including self-consistent beam distributions~\cite{burov2013circular,gilanliogullaricircular,gilanliogullarirotational,danilov2003self}. A fully coupled beam with non-zero angular momentum creates a delta function correlation between the $x$--$y'$ and $y$--$x'$ planes, which leads to intrinsic flatness with a round beam cross-section, called circular modes. Circular mode beams inherit strong $x$--$y$ plane coupling and intrinsic flatness through Derbenev's Adapter, as presented in the Applications section. Danilov distribution is a class of coupled beam distributions that are self-consistent under space charge forces, similar to Kapchinskij-Vladimirskij (KV) distribution~\cite{kapchinskij1959limitations}. Furthermore, it was shown that synchrotron light sources would benefit from round beams rather than flat beams, especially for low-emittance machines~\cite{li2022designing}. In these machines, the beam shape can be manipulated by managing coupling scales and strengths in a lattice to achieve the desired beam projections. 

The introduction of linear coupling elements to the lattice adds new resonance conditions in addition to the integer resonances, namely sum and difference resonances originating from skew quadrupoles. The difference resonance is responsible for the emittance exchange mechanism in flat beams, where the emittance planes can be exchanged, resulting in a swap between $x$ and $y$ planes. Typically, a skew element is treated as a perturbation to the system, and the effect of the exchange rate is studied with respect to the resonance proximity parameter, $\Delta= Q_{x}-Q_{y}-\delta$, discussed in~\cite{franchi2007emittance,franchi2011vertical}. The emittance exchange mechanism is described in terms of the resonance proximity parameter and the integrated skew strengths, where the invariant of the system is replaced by the sum of plane invariants. It is interesting to note that the stable difference resonance is studied from the perspective of uncoupled elements using perturbation theory. In this manuscript, we will introduce the emittance exchange mechanism for difference resonance from the perspective of coupled functions and parameters.
	
The most important challenge in designing a linear coupled lattice is the behavior of the optics functions. This paper follows Mais-Ripken and Lebedev-Bogacz parametrizations of the optics functions. Their periodic behavior is essential in designing coupled lattices. First, we review the Courant-Snyder parametrization and introduce the generating vector formalism, explaining how the generating vectors transform to reproduce the known transfer matrix formalism. Secondly, we extend the generating vector formalism to the coupled case using a method similar to Mais-Ripken parametrization. We also make a connection between the Mais-Ripken and Lebedev-Bogacz parameters. We then write the explicit transfer matrix method for the coupled optics functions of eigenmodes $1$ and $2$, similar to the Courant-Snyder transfer matrix. The transfer of optics functions is derived from the generating vectors. Using the Lebedev-Bogacz parametrization, we show that the generating vectors also decouple using the decoupling matrix, similar to the Edwards-Teng parametrization. Following the description of the formalism, we present some application cases of the transformation and compare them with existing beam optics codes such as MAD-X and OptiMX. An important application is the periodic matching of coupled optics functions. Popular optics codes such as MAD-X and OptiMX come with the calculation of coupled optics functions for a given lattice and initial uncoupled functions. Both MAD-X and OptiMX provide matching routines that can compute periodic solutions for uncoupled optics functions using Courant-Snyder parametrization or Edward-Teng parametrization. However, there is no matching routine for coupled optics functions. The formulation in this paper was used to construct a matching routine for generally coupled beams, which could be adapted for popular optics and design codes. Using the formalism described here, we also apply it to the linear difference resonance and express the exchange mechanism from the perspective of coupled functions. Finally, using this formalism, we developed a fully coupled and periodic ring design.

\section{Brief Review of Uncoupled Beam Optics}
\label{sec:background}

The Courant-Snyder parametrization employs Floquet theory to solve Hill's equation~\cite{courant1958theory,lee2018accelerator}, where the beam particle coordinates are parameterized as follows:
\begin{equation*}
	\begin{split}
		x& = \sqrt{\epsilon_{x}\beta_{x}}\cos\psi_{x},\quad y= \sqrt{\epsilon_{y}\beta_{y}}\cos\psi_{y}, \\
		x'&= -\sqrt{\frac{\epsilon_{x}}{\beta_{x}}}(\alpha_{x}\cos\psi_{x} + \sin\psi_{x}), \\
		y'&= - \sqrt{\frac{\epsilon_{y}}{\beta_{y}}}(\alpha_{y}\cos\psi_{y} + \sin\psi_{y}).
	\end{split}
\end{equation*}
Here, $\beta_{x,y}$ are the beta functions, $\alpha_{x,y}=-\frac{1}{2}\beta_{x,y}'$ the alpha functions, and $\psi_{x,y}$ the betatron phases. In order to obtain the expressions for $x'$ and $y'$, we used $\frac{d\psi_{x,y}}{ds}= \frac{1}{\beta_{x,y}}$. Combining the position and momentum coordinates in each plane leads to the beam emittances as invariants of motion:
\begin{equation}
	\begin{split}
		\left(\frac{1+\alpha^{2}}{\beta}\right)x^{2} + 2\alpha xx' + \beta x^{'2} = \epsilon.
	\end{split}
    \label{eq:cs_invariant}
\end{equation}
Equation~\eqref{eq:cs_invariant} indicates that individual particles follow elliptical trajectories in phase space. The parametrization of the coordinates can be expressed as a function of the eigenvectors of the one-turn map, $\mathcal{M}\vec{v}=\lambda \vec{v}$. The phase space coordinate vector, $\vec{z}=[x,x',y,y']^{T}$, can be represented as:\begin{equation}
	\begin{split}
		\vec{z} &= \frac{1}{2}\sqrt{\epsilon_{x}}\vec{v}_{x}e^{i\psi_{x0}} + \frac{1}{2}\sqrt{\epsilon_{y}}\vec{v}_{y}e^{i\psi_{y0}} + \frac{1}{2}\sqrt{\epsilon_{x}}\vec{v}_{x}^{*}e^{-i\psi_{x0}} + \frac{1}{2}\sqrt{\epsilon_{y}}\vec{v}_{y}^{*}e^{-i\psi_{y0}},
	\end{split}
	\label{Eq:eigenvecunc}
\end{equation}
where $\vec{v}_{x,y}$ are the eigenvectors, $\psi_{x,y0}$ the initial betatron phases. The eigenvectors can be written as:
\begin{equation}
	\vec{v}_{x} = \begin{pmatrix}
		\sqrt{\beta_{x}} \\ 
		-\frac{i + \alpha_{x}}{\sqrt{\beta_{x}}} \\
        0 \\
        0
	\end{pmatrix}, \quad 
	\vec{v}_{y} = \begin{pmatrix}
            0 \\
            0 \\
		\sqrt{\beta_{y}} \\
		-\frac{i + \alpha_{y}}{\sqrt{\beta_{y}}}
	\end{pmatrix}.
\end{equation}
They are normalized using $\vec{v}^{\dagger}S\vec{v}=-2i$, where $S$ is the $4\times 4$ symplectic unit matrix. We can extract the geometric invariant from the parametrization using the normalization condition by applying $\vec{v}^{\dagger}S$ to the left side of Eq.~\eqref{Eq:eigenvecunc}, which will result in
\begin{equation} 
	\begin{split}
		2J_{x} &= \vec{v}_{x}^{\dagger}S\vec{z}\vec{z}^{T}S\vec{v}_{x},\quad S=\begin{pmatrix}
		    0 & 1 & 0 & 0\\
            -1 & 0 & 0 & 0 \\
            0 & 0 & 0 & 1 \\
            0 & 0 & -1 & 0
		\end{pmatrix}, \\
				&= \left(\frac{1+\alpha_{x}^{2}}{\beta_{x}}\right)x^{2} + 2\alpha_{x}xx' + \beta_{x}x^{'2}.
	\end{split}
\end{equation}
There are various methods to introduce transfer maps for beam optics functions. One particular case is to use the system's invariant, $\epsilon$. Using the invariant and the transfer matrix method, we can write
\begin{equation}
	\begin{split}
		\Sigma &= \begin{pmatrix}
			\beta & -\alpha \\
			-\alpha & \gamma
		\end{pmatrix}, \quad \Sigma_{f} = \mathcal{M} \Sigma_{i}\mathcal{M}^{T},
	\end{split}
\end{equation}
where $\mathcal{M}$ is the $2\times 2$ transfer matrix, $\Sigma$ is $2\times 2$ second moment beam matrix for each plane and $\gamma =(1+\alpha^{2})/\beta$. Transformation of the beam matrix leads to the transfer of optics functions that can be written in the following form:
\begin{equation}
\textstyle
\begin{pmatrix}
    \beta \\
    \alpha \\
    \gamma
\end{pmatrix}_{f} = \begin{pmatrix}
    M_{11}^{2} & -2M_{11}M_{12} & M_{12}^{2} \\
    -M_{11}M_{21} & (M_{11}M_{22} + M_{12}M_{21}) & -M_{12}M_{22} \\
    M_{21}^{2} & -2M_{21}M_{22} & M_{22}^{2}
\end{pmatrix}\begin{pmatrix}
    \beta \\
    \alpha \\
    \gamma
\end{pmatrix}_{i}.
\label{Eq:unctransfermatrix}
\end{equation}

Equation \eqref{Eq:unctransfermatrix} illustrates the transformation of the Twiss parameters. A key aspect of this transformation is its independence from phase advance. This property makes it particularly useful for beam matching for various beam optics conditions. Additionally, it can be easily integrated into optimization routines to identify matched solutions.

It is also worth noting that an alternative method exists to achieve the same outcome presented in Eq.~\eqref{Eq:unctransfermatrix}. By employing the eigenvectors of the transfer map, one can derive the transformation of the phase space vector as follows:
\begin{equation}
	\vec{z}(s) = \frac{1}{2}\sqrt{\epsilon_{x}}\vec{v}_{x}e^{-i(\mu_{x}+\psi_{x0})} + \frac{1}{2}\sqrt{\epsilon_{y}}\vec{v}_{y}e^{-i(\mu_{y}+\psi_{y0})} + \frac{1}{2}\sqrt{\epsilon_{x}}\vec{v}_{x}^{*}e^{i(\mu_{x}+\psi_{x0})} + \frac{1}{2}\sqrt{\epsilon_{y}}\vec{v}_{y}^{*}e^{i(\mu_{y}+\psi_{y0})}, 
	\label{Eq:phaseadvancephasevecunc}
\end{equation}
where $\mu_{x,y}$ are the phase advances derived from the eigenvector equation. Focusing on the $x$ component with $\vec{x}=[x,x']^{T}$, we can introduce generating vectors as
\begin{equation}
	\begin{split}
		\vec{x}_{1} &= \frac{1}{2}(\vec{v}_{x} + \vec{v}_{x}^{*})\cos\mu_{x} - \frac{i}{2}(\vec{v}_{x}-\vec{v}_{x}^{*})\sin\mu_{x}, \\ 
		\vec{x}_{2} &= \frac{1}{2}(\vec{v}_{x} + \vec{v}_{x}^{*})\sin\mu_{x} + \frac{i}{2}(\vec{v}_{x}-\vec{v}_{x}^{*})\cos\mu_{x}. 
	\end{split}
	\label{Eq:genvecunc}
\end{equation}
This leads to the generating vectors for the $x$ plane:
\begin{equation}
	\vec{x}_{1}= \begin{pmatrix}
		\sqrt{\beta_{x}}\cos\mu_{x} \\
		\frac{-\alpha_{x}\cos\mu_{x}-\sin\mu_{x}}{\sqrt{\beta_{x}}}
	\end{pmatrix}, \qquad \vec{x}_{2}= \begin{pmatrix}
		\sqrt{\beta_{x}}\sin\mu_{x} \\
		\frac{-\alpha_{x}\sin\mu_{x}+\cos\mu_{x}}{\sqrt{\beta_{x}}}.
	\end{pmatrix}
\end{equation}

To express the generating vectors in a more familiar form, we can define variables $\tilde{\mu}_{x} = \mu_{x} - \arctan{(1/\alpha_{x})}$ and $\gamma_{x}=(1+\alpha_{x}^{2})/\beta_{x}$. This transformation allows us to rewrite the generating vectors in the following form:
\begin{equation}
	\vec{x}_{1}= \begin{pmatrix}
		\sqrt{\beta_{x}}\cos\mu_{x} \\
		-\sqrt{\gamma_{x}}\cos\overline{\mu}_{x}
	\end{pmatrix}, \qquad \vec{x}_{2}=\begin{pmatrix}
		\sqrt{\beta_{x}}\sin\mu_{x} \\
		-\sqrt{\gamma_{x}}\sin\overline{\mu}_{x}
	\end{pmatrix},
	\label{Eq:genvecsfamform}
\end{equation}
where the phase space vector can be written as $\vec{x} = \sqrt{\epsilon_{x}}(\vec{x}_{1}\cos\psi_{x} - \vec{x}_{2}\sin\psi_{x})$. Since the transformation of the phase space vector is done via $\vec{x}(s) = \mathcal{M}\vec{x}(0)$, where $\mathcal{M}$ is a $2\times 2$ transfer matrix, each generating vector is transferred using the same method. Applying equations~\eqref{Eq:genvecsfamform} yields
\begin{equation}
	\begin{split}
		\beta_{x} &= x_{1}^{2} + x_{2}^{2}, \\
		\alpha_{x} &= -(x_{1}x_{1}' + x_{2}x_{2}'), \\
		\gamma_{x} &= x_{1}'^{2} + x_{2}'^{2}.
	\end{split}
\end{equation}
Since the right-hand side of the equations above depends on the components of the generating vectors, transforming those vectors will yield the optics functions at the $s$ location. This transformation is identical to the transformation of Eq.~\eqref{Eq:unctransfermatrix}. We will not discuss phase stability, one-turn map behavior, and other aspects of beam dynamics here since they are well-established from~\cite{courant1958theory,lee2018accelerator,wiedemann2015particle}.

\section{Coupled Beam Optics parametrizations}
\label{sec:coupledparams}
As mentioned, Mais-Ripken parametrization extends Courant-Snyder parametrization, this is done by generalizing and defining the generating vectors for the coupled optics case. Mais-Ripken parametrization introduces two eigenmodes named $1$ and $2$ and their projections onto the real phase planes $x$ and $y$. It expands the generating vectors, adapts them to 4D motion variables and uses the generating vectors to formulate the rest of the linear coupled motion from the four generating vectors. The extended optics functions are: $\beta_{1x},\beta_{2x},\beta_{1y},\beta_{2y}$ as the betatron functions, $\alpha_{1x},\alpha_{2x}.\alpha_{1y},\alpha_{2y}$ as the alpha functions and $\Phi_{1x},\Phi_{1y},\Phi_{2x},\Phi_{2y}$ as separate betatron phases. In this section, we start with Lebedev-Bogacz parametrization, an extension of Mais-Ripken parametrization that formulates eigenvectors based on the one-turn transfer map. The phase space vector, $\vec{z}=[x,x',y,y']^{T}$ is given as~\cite{lebedev2010betatron}
\begin{equation}
	\vec{z}= \frac{1}{2}\sqrt{2J_{1}}e^{-i\psi_{1}}\vec{v}_{1} + \frac{1}{2}\sqrt{2J_{2}}e^{-i\psi_{2}}\vec{v}_{2} + \frac{1}{2}\sqrt{2J_{1}}e^{i\psi_{1}}\vec{v}_{1}^{*} + \frac{1}{2}\sqrt{2J_{2}}e^{i\psi_{2}}\vec{v}_{2}^{*}.
\end{equation}
Here, $J_{1,2}$ are the invariants of the system, $\psi_{1,2}$ are the initial betatron phases, $\vec{v}_{1,2}$ are the eigenvectors. The eigenvectors are given by:
\begin{equation}
	\vec{v}_{1} = \begin{pmatrix}
		\sqrt{\beta_{1x}} \\
		-\frac{i(1-u)+\alpha_{1x}}{\sqrt{\beta_{1x}}} \\
		\sqrt{\beta_{1y}}e^{i\nu_{1}} \\
		-\frac{iu + \alpha_{1y}}{\sqrt{\beta_{1y}}}e^{i\nu_{1}}
	\end{pmatrix}, \qquad \vec{v}_{2}= \begin{pmatrix}
		\sqrt{\beta_{2x}}e^{i\nu_{2}} \\
		-\frac{iu + \alpha_{2x}}{\sqrt{\beta_{2x}}}e^{i\nu_{2}} \\
		\sqrt{\beta_{2y}} \\
		-\frac{i(1-u) + \alpha_{2y}}{\sqrt{\beta_{2y}}}
	\end{pmatrix}, 
    \label{eq:eigenvecscoupled}
\end{equation}
where $\nu_{1,2}$ are defined as the phases of coupling and $u$ the coupling strength parameter as defined in Appendix~\ref{appendix:B}. The coupling strength parameter indicates the weight of the mode contributions to the phase planes, while phases of coupling indicate how different modes project onto phase planes. The eigenvectors are normalized with respect to the symplectic unit matrix, $\vec{v}_{1,2}^{\dagger}S\vec{v}_{1,2}=-2i$. Application of the transfer matrix to the eigenvectors yields the phase advances as eigenvalues. Therefore, this leads to an equation similar to Eq.~\eqref{Eq:phaseadvancephasevecunc}:
\begin{equation}
	\vec{z}(s) = \frac{1}{2}\sqrt{\epsilon_{1}}\vec{v}_{1}e^{-i(\mu_{1}+\psi_{1})} + \frac{1}{2}\vec{v}_{2}e^{-i(\mu_{2}+\psi_{2})} + \frac{1}{2}\sqrt{\epsilon_{1}}e^{i(\mu_{1}+\psi_{1})}\vec{v}_{1}^{*} + \frac{1}{2}\sqrt{\epsilon_{2}}e^{i(\mu_{2}+\psi_{2})}\vec{v}_{2}^{*}. 
\end{equation}
Performing a similar operation as in Eq.~\eqref{Eq:genvecunc} on the phase space vector results in the new coupled generating vectors below:
\begin{equation}
	\begin{split}
		\vec{z}_{1} &= \frac{1}{2}(\vec{v}_{1} + \vec{v}_{1}^{*})\cos\mu_{1} - \frac{i}{2}(\vec{v}_{1}-\vec{v}_{1}^{*})\sin\mu_{1}, \\
		\vec{z}_{2} &= \frac{1}{2}(\vec{v}_{1} + \vec{v}_{1}^{*})\sin\mu_{1} + \frac{i}{2}(\vec{v}_{1}-\vec{v}_{1}^{*})\cos\mu_{1}, \\
		\vec{z}_{3} &= \frac{1}{2}(\vec{v}_{2} + \vec{v}_{2}^{*})\cos\mu_{2} - \frac{i}{2}(\vec{v}_{2}-\vec{v}_{2}^{*})\sin\mu_{2}, \\
		\vec{z}_{4} &= \frac{1}{2}(\vec{v}_{2} + \vec{v}_{2}^{*})\sin\mu_{2} + \frac{i}{2}(\vec{v}_{2}-\vec{v}_{2}^{*})\cos\mu_{2}. \\
	\end{split}
\end{equation}
Generating vectors of mode $1$ can be written as:
\begin{equation}
	\vec{z}_{1} = \begin{pmatrix}
		\sqrt{\beta_{1x}}\cos\mu_{1} \\
		\frac{-\alpha_{1x}\cos\mu_{1}-(1-u)\sin\mu_{1}}{\sqrt{\beta_{1x}}} \\
		\sqrt{\beta_{1y}}\cos(\mu_{1}-\nu_{1}) \\
		\frac{-\alpha_{1y}\cos(\mu_{1}-\nu_{1})-u\sin(\mu_{1}-\nu_{1})}{\sqrt{\beta_{1y}}}
	\end{pmatrix}, \qquad \vec{z}_{2}= \begin{pmatrix}
		\sqrt{\beta_{1x}}\sin\mu_{1} \\
		\frac{(1-u)\cos\mu_{1} - \alpha_{1x}\sin\mu_{1}}{\sqrt{\beta_{1x}}} \\
		\sqrt{\beta_{1y}}\sin(\mu_{1}-\nu_{1})\\
		\frac{u\cos(\mu_{1}-\nu_{1})-\alpha_{1y}\sin(\mu_{1}-\nu_{1})}{\sqrt{\beta_{1y}}}
	\end{pmatrix}.
\end{equation}
Similarly for mode $2$, the generating vectors are:
\begin{equation}
	\vec{z}_{3}= \begin{pmatrix}
		\sqrt{\beta_{2x}}\cos(\mu_{2}-\nu_{2}) \\
		\frac{-\alpha_{2x}\cos(\mu_{2}-\nu_{2})-u\sin(\mu_{2}-\nu_{2})}{\sqrt{\beta_{2x}}} \\
		\sqrt{\beta_{2y}}\cos\mu_{2} \\
		\frac{-\alpha_{2y}\cos\mu_{2}-(1-u)\sin\mu_{2}}{\sqrt{\beta_{2y}}}
	\end{pmatrix}, \qquad \vec{z}_{4}= \begin{pmatrix}
		\sqrt{\beta_{2x}}\sin(\mu_{2}-\nu_{2}) \\
		\frac{u\cos(\mu_{2}-\nu_{2})-\alpha_{2x}\sin(\mu_{2}-\nu_{2})}{\sqrt{\beta_{2x}}} \\
		\sqrt{\beta_{2y}}\sin\mu_{2} \\
		\frac{(1-u)\cos\mu_{2}-\alpha_{2y}\sin\mu_{2}}{\sqrt{\beta_{2y}}}
	\end{pmatrix}.
\end{equation}
These generating vectors bring the phase space vector to:
\begin{equation}
	\vec{z} = \sqrt{\epsilon_{1}}(\vec{z}_{1}\cos\psi_{1}-\vec{z}_{2}\sin\psi_{1}) + \sqrt{\epsilon_{2}}(\vec{z}_{3}\cos\psi_{2}-\vec{z}_{4}\sin\psi_{2}).
\end{equation}
By introducing the following phase transformations,
\begin{equation}
	\begin{split}
		\overline{\mu}_{1} &= \mu_{1} - \arctan\left(\frac{(1-u)}{\alpha_{1x}}\right), \quad \overline{\mu}_{2} = \mu_{2} - \arctan\left(\frac{1-u}{\alpha_{2y}}\right), \\
		\overline{(\mu_{1}-\nu_{1})} &= (\mu_{1}-\nu_{1}) - \arctan\left(\frac{u}{\alpha_{1y}}\right), \quad \overline{(\mu_{2}-\nu_{2})} = (\mu_{2}-\nu_{2}) - \arctan \left(\frac{u}{\alpha_{2x}}\right). \\
	\end{split}
\end{equation}
The generating vectors can be simplified to the familiar forms:
\begin{equation}
	\begin{split}
		\vec{z}_{1}&= \begin{pmatrix}
			\sqrt{\beta_{1x}}\cos\mu_{1} \\
			-\sqrt{\gamma_{1x}}\cos\overline{\mu}_{1} \\
			\sqrt{\beta_{1y}}\cos(\mu_{1}-\nu_{1}) \\
			-\sqrt{\gamma_{1y}}\cos(\overline{\mu_{1}-\nu_{1}})
		\end{pmatrix}, \quad \vec{z}_{2}= \begin{pmatrix}
			\sqrt{\beta_{1x}}\sin\mu_{1} \\
			-\sqrt{\gamma_{1x}}\sin\overline{\mu}_{1} \\
			\sqrt{\beta_{1y}}\sin(\mu_{1}-\nu_{1}) \\
			-\sqrt{\gamma_{1y}}\sin\overline{(\mu_{1}-\nu_{1})}
		\end{pmatrix} , \\
		\vec{z}_{3}&= \begin{pmatrix}
			\sqrt{\beta_{2x}}\cos(\mu_{2}-\nu_{2})\\
			-\sqrt{\gamma_{2x}}\cos\overline{(\mu_{2}-\nu_{2})} \\
			\sqrt{\beta_{2y}}\cos\mu_{2} \\
			-\sqrt{\gamma_{2y}}\cos\overline{\mu}_{2}
		\end{pmatrix}, \quad \vec{z}_{4} = \begin{pmatrix}
			\sqrt{\beta_{2x}}\sin(\mu_{2}-\nu_{2}) \\
			-\sqrt{\gamma_{2x}}\sin\overline{(\mu_{2}-\nu_{2})} \\
			\sqrt{\beta_{2y}}\sin\mu_{2} \\
			-\sqrt{\gamma_{2y}}\sin\overline{\mu}_{2}
		\end{pmatrix}.
	\end{split}
	\label{Eq:genvecscoupledLB}
\end{equation}
Eq.~\eqref{Eq:genvecscoupledLB} has the exact same form as in Mais-Ripken paper~\cite{willeke1989methods}. The generating vectors are normalized with respect to the symplectic unit matrix using
\begin{equation}
	\begin{split}
		\vec{z}_{1}^{T}S\vec{z}_{2} &= 1, \quad \vec{z}_{3}^{T}S\vec{z}_{4} = 1, \\
		\vec{z}_{1}^{T}S\vec{z}_{4} &= 0, \quad \vec{z}_{2}^{T}S\vec{z}_{3} =0 . 
	\end{split}
	\label{Eq:normalizationgenvecs}
\end{equation}
 This establishes a direct connection between the two parameterizations, Lebedev-Bogacz and Mais-Ripken making them essentially equivalent. The phases introduced in Mais-Ripken parametrization are now identified as $\Phi_{1x}=\mu_{1}$, $\Phi_{2y}=\mu_{2}$, $\Phi_{2x}=\mu_{2}-\nu_{2}$ and $\Phi_{1y}=\mu_{1}-\nu_{1}$ as functions of the phases in the Lebedev-Bogacz parameterization; $\mu_{1}$, $\mu_{2}$, $\nu_{1}$ and $\nu_{2}$. In addition, the gamma functions are defined as
\begin{equation}
	\begin{split}
		\gamma_{1x} &= \frac{(1-u)^{2} + \alpha_{1x}^{2}}{\beta_{1x}}, \qquad \gamma_{2x}=\frac{u^{2}+\alpha_{2x}^{2}}{\beta_{2x}}, \\
		\gamma_{2y}&= \frac{(1-u)^{2}+\alpha_{2y}^{2}}{\beta_{2y}}, \qquad \gamma_{1y}= \frac{u^{2}+\alpha_{1y}^{2}}{\beta_{1y}}. 
	\end{split}
\end{equation}
The optics functions are represented in terms of generating functions as
\begin{equation}
	\begin{split}
		\beta_{1x}&= x_{1}^{2} + x_{2}^{2}, \qquad \beta_{2x}= x_{3}^{2} + x_{4}^{2}, \\
		\beta_{1y}&= y_{1}^{2} + y_{2}^{2}, \qquad \beta_{2y}= y_{3}^{2} + y_{4}^{2}, \\
		\gamma_{1x}&= x_{1}'^{2} + x_{2}'^{2}, \qquad \gamma_{2x}= x_{3}'^{2} + x_{4}'^{2}, \\
		\gamma_{1y}&= y_{1}'^{2} + y_{2}'^{2}, \qquad \gamma_{2y}= y_{3}'^{2} + y_{4}'^{2}, \\
		\alpha_{1x}&= -(x_{1}x_{1}' + x_{2}x_{2}'), \qquad \alpha_{2x}=-(x_{3}x_{3}' + x_{4}x_{4}'),\\
		\alpha_{1y}&= -(y_{1}y_{1}' + y_{2}y_{2}'), \qquad \alpha_{2y}=-(y_{3}y_{3}' + y_{4}y_{4}').
	\end{split}
	\label{Eq:genvectorscoupledfunctions}
\end{equation}
Using the transformation of generating functions, we can establish a transfer matrix for the coupled optics functions. The coupled transfer matrix is given by
\begin{equation}
	\mathcal{M} = \begin{pmatrix}
		M & m \\
		n & N
	\end{pmatrix},
\end{equation}
where $M,m,N,n$ are $2\times 2$ sub-matrices. The overall transfer matrix for mode $1$ is expressed as 
\begin{equation}
	\begin{split}
		\begin{pmatrix}
			\beta_{1x}\\
			\alpha_{1x}\\
			\gamma_{1x}\\
			\beta_{1y}\\
			\alpha_{1y}\\
			\gamma_{1y}
		\end{pmatrix}_{f} = \begin{pmatrix}
			C_{M} & C_{m} \\
			C_{n} & C_{N}
		\end{pmatrix} \begin{pmatrix}
			\beta_{1x}\\
			\alpha_{1x}\\
			\gamma_{1x}\\
			\beta_{1y}\\
			\alpha_{1y}\\
			\gamma_{1y}
		\end{pmatrix}_{i} + \begin{pmatrix}
			C_{Mm} \\
			C_{nN} 
		\end{pmatrix}\begin{pmatrix}
			\xi_{1xy} \\
			\xi_{1xy'} \\
			\xi_{1yx'} \\
			\xi_{1x'y'}
		\end{pmatrix}_{i} .
	\end{split}
	\label{eq:coupledcoptictransfer}
\end{equation}
Here, the matrices $C_{M},C_{m},C_{n},C_{N}$ are $3\times 3 $ matrices and $C_{Mm},C_{nN}$ are $4\times 3$ matrices, they are given below as functions of the sub-matrices $M, m, N$ and $n$:
\begin{widetext}
\begin{equation}
	\begin{split}
		C_{M} &= \begin{pmatrix}
			M_{11}^{2} & -2M_{11}M_{12} & M_{12}^{2} \\
			-2M_{11}M_{21} & (M_{11}M_{22} + M_{12}M_{21}) & -2M_{22}M_{12} \\
			M_{21}^{2} & -2M_{21}M_{22} & M_{22}^{2}
		\end{pmatrix}, \\
		C_{m} &= \begin{pmatrix}
			m_{11}^{2} & -2m_{11}m_{12} & m_{12}^{2} \\
			-2m_{11}m_{21} & (m_{11}m_{22} + m_{12}m_{21}) & -2m_{22}m_{12} \\
			m_{21}^{2} & -2m_{21}m_{22} & m_{22}^{2}
		\end{pmatrix}, \\
		C_{N} &= \begin{pmatrix}
			N_{11}^{2} & -2N_{11}N_{12} & N_{12}^{2} \\
			-2N_{11}N_{21} & (N_{11}N_{22} + N_{12}N_{21}) & -2N_{22}N_{12} \\
			N_{21}^{2} & -2N_{21}N_{22} & N_{22}^{2}
		\end{pmatrix}, \\
		C_{n} &= \begin{pmatrix}
			n_{11}^{2} & -2n_{11}n_{12} & n_{12}^{2} \\
			-2n_{11}n_{21} & (n_{11}n_{22} + n_{12}n_{21}) & -2n_{22}n_{12} \\
			n_{21}^{2} & -2n_{21}n_{22} & n_{22}^{2}
		\end{pmatrix}, \\          
             C_{Mm} &= \left(\begin{matrix}
		2M_{11}m_{11} & 2M_{11}m_{12}  \\
		-(M_{11}m_{21} + m_{11}M_{21}) & -(m_{12}M_{21} + M_{11}m_{22}) \\
			2m_{21}M_{21} & 2M_{21}m_{22} 
		\end{matrix}\right. \\
        & \left.\begin{matrix}
        \qquad\qquad\qquad 2M_{12}m_{11} & 2M_{12}m_{12} \\
        \qquad\qquad\qquad-(M_{12}m_{21} + m_{11}M_{22}) & -(M_{12}m_{22} + m_{12}M_{22}) \\
        \qquad\qquad\qquad 2M_{22}m_{21} & 2M_{22}m_{22}
        \end{matrix}\right) \\
		C_{Nn} &= \left(\begin{matrix}
			2n_{11}N_{11} & 2n_{11}N_{12} \\
			-(n_{11}N_{21} + N_{11}n_{21}) & -(N_{12}n_{21} + n_{11}N_{22}) \\
			2N_{21}n_{21} & 2n_{21}N_{22}
		\end{matrix}\right.\\
        & \left.\begin{matrix}
        \qquad\qquad\qquad 2n_{12}N_{11} & 2n_{12}N_{12} \\
        \qquad\qquad\qquad -(n_{12}N_{21} + N_{11}n_{22}) & -(n_{12}N_{22} + N_{12}n_{22}) \\
        \qquad\qquad\qquad 2n_{22}N_{21} & 2n_{22}N_{22}
        \end{matrix}\right).
	\end{split}
	\label{Eq:allmatrices}
\end{equation}
\end{widetext}
The main matrices in Eq.~\eqref{Eq:allmatrices} are in the same format as the Courant-Synder transfer matrix shown earlier in Eq.~\eqref{Eq:unctransfermatrix}. The coupling vector $\vec{\xi}_{1}$ for mode $1$ in Eq.~(25) is given by:
\begin{equation}
	\begin{split}
		\xi_{1xy} &= \sqrt{\beta_{1x}\beta_{1y}}\cos\nu_{1}, \\
		\xi_{1xy'} &= \sqrt{\frac{\beta_{1x}}{\beta_{1y}}}\left( -\alpha_{1y}\cos\nu_{1} + u\sin\nu_{1} \right ),\\
		\xi_{1yx'} &= -\sqrt{\frac{\beta_{1y}}{\beta_{1x}}}\left( \alpha_{1x}\cos\nu_{1} + (1-u)\sin\nu_{1}\right ), \\
		\xi_{1x'y'}&= \frac{1}{\sqrt{\beta_{1x}\beta_{1y}}}\left ( (u(1-u) + \alpha_{1x}\alpha_{1y})\cos\nu_{1} + (\alpha_{1y} -u(\alpha_{1x} + \alpha_{1y}))\sin\nu_{1} \right ).
	\end{split}
\end{equation}
For mode $2$, it will be the same transformation as Eq.~(25), by changing the subscript from $1\rightarrow 2$, except for the coupling vector $\vec{\xi}_{2}$, which is given by:
\begin{equation}
	\begin{split}
		\xi_{2xy} &= \sqrt{\beta_{2x}\beta_{2y}}\cos\nu_{2}, \\
		\xi_{2xy'} &= -\sqrt{\frac{\beta_{2x}}{\beta_{2y}}}\left( \alpha_{2y}\cos\nu_{2} + (1-u)\sin\nu_{2} \right ),\\
		\xi_{2yx'} &= \sqrt{\frac{\beta_{2y}}{\beta_{2x}}}\left( -\alpha_{2x}\cos\nu_{2} + u\sin\nu_{2}\right ), \\
		\xi_{2x'y'}&= \frac{1}{\sqrt{\beta_{2x}\beta_{2y}}}\left ( (u(1-u) + \alpha_{2x}\alpha_{2y})\cos\nu_{2} + ((1-u)\alpha_{2y} + u\alpha_{2x})\sin\nu_{2} \right ).
	\end{split}
\end{equation} 
The above formulation is for a generally coupled beam, however, simplifications can be made for particular cases of interest. First case; when the initial optics functions (beam) are uncoupled but the transfer matrix (lattice) is coupled, the second term in Eq.~\eqref{eq:coupledcoptictransfer} disappears, resulting in:
\begin{equation}
	\begin{split}
		\begin{pmatrix}
			\beta_{1x} \\
			\alpha_{1x} \\
			\gamma_{1x} \\
			\beta_{1y} \\
			\alpha_{1y} \\
			\gamma_{1y}
		\end{pmatrix}_{f} &= \begin{pmatrix}
			C_{M} & C_{m} \\
			C_{n} & C_{N}
		\end{pmatrix} \begin{pmatrix}
			\beta_{x} \\
			\alpha_{x} \\
			\gamma_{x} \\
			0 \\
			0\\
			0
		\end{pmatrix}_{i}, \\
		\begin{pmatrix}
			\beta_{2x}\\
			\alpha_{2x} \\
			\gamma_{2x} \\
			\beta_{2y} \\
			\alpha_{2y} \\
			\gamma_{2y}
		\end{pmatrix}_{f} &= \begin{pmatrix}
			C_{M} & C_{m} \\
			C_{n} & C_{N}
		\end{pmatrix}\begin{pmatrix}
			0 \\
			0\\
			0\\
			\beta_{y} \\
			\alpha_{y}\\
			\gamma_{y}
		\end{pmatrix}_{i}.
	\end{split}
	\label{Eq:fromunctocoupled}
\end{equation}
Second case; when the optics functions (beam) are coupled but the transfer matrix (lattice) is uncoupled:
\begin{equation}
	\begin{split}
		\begin{pmatrix}
			\beta_{1x}\\
			\alpha_{1x}\\
			\gamma_{1x}\\
			\beta_{1y}\\
			\alpha_{1y}\\
			\gamma_{1y}
		\end{pmatrix}_{f}&= \begin{pmatrix}
			C_{M} & 0 \\
			0 & C_{N}
		\end{pmatrix}\begin{pmatrix}
			\beta_{1x}\\
			\alpha_{1x}\\
			\gamma_{1x}\\
			\beta_{1y}\\
			\alpha_{1y}\\
			\gamma_{1y}
		\end{pmatrix}_{i}, \\ 
		\begin{pmatrix}
			\beta_{2x}\\
			\alpha_{2x}\\
			\gamma_{2x}\\
			\beta_{2y}\\
			\alpha_{2y}\\
			\gamma_{2y}
		\end{pmatrix} &= \begin{pmatrix}
			C_{M} & 0 \\
			0 & C_{N}
		\end{pmatrix}\begin{pmatrix}
			\beta_{2x}\\
			\alpha_{2x}\\
			\gamma_{2x}\\
			\beta_{2y}\\
			\alpha_{2y}\\
			\gamma_{2y}
		\end{pmatrix}_{i}.
		\label{Eq:Fromcoupledtocoupleduncmatrix}
	\end{split}
\end{equation}
Equation~\eqref{Eq:fromunctocoupled} shows how the coupled optics functions are created starting from uncoupled functions, whereas Eq.~\eqref{Eq:Fromcoupledtocoupleduncmatrix} shows the propagation of coupled parameters in the absence of coupling elements. One can also deduce the phase advances from the transfer matrix:
\begin{widetext}
\begin{equation}
	\begin{split}
		\tan(\Delta\mu_{1}) &= \frac{M_{12}(1-u) - m_{11}\sqrt{\beta_{1x}\beta_{1y}}\sin\nu_{1} +m_{12}\sqrt{\frac{\beta_{1x}}{\beta_{1y}}}\left (u\cos\nu_{1} + \alpha_{1y}\sin\nu_{1} \right )}{M_{11}\beta_{1x}-M_{12}\alpha_{1x} + m_{11}\xi_{1xy} + m_{12}\xi_{1xy'}}, \\
		\tan(\Delta\mu_{2}) &= \frac{N_{12}(1-u) -n_{11}\sqrt{\beta_{2x}\beta_{2y}}\sin\nu_{2} + n_{12}\sqrt{\frac{\beta_{2y}}{\beta_{2x}}}(u\cos\nu_{2} + \alpha_{2y}\sin\nu_{2})}{N_{11}\beta_{2y}-N_{12}\alpha_{2y}+n_{11}\xi_{2xy} + n_{12}\xi_{2yx'}}, \\
		\tan(\Delta(\mu_{1}-\nu_{1}))&= \frac{N_{12}u + n_{11}\sqrt{\beta_{1x}\beta_{1y}}\sin\nu_{1} -n_{12}\sqrt{\frac{\beta_{1y}}{\beta_{1x}}}(-(1-u)\cos\nu_{1} + \alpha_{1x}\sin\nu_{1})}{N_{11}\beta_{1y}-N_{12}\alpha_{1y} + n_{11}\xi_{1xy} + n_{12}\xi_{1yx'}}, \\
		\tan(\Delta(\mu_{2}-\nu_{2}))&= \frac{M_{12}u + m_{11}\sqrt{\beta_{2x}\beta_{2y}}\sin\nu_{2} - m_{12}\sqrt{\frac{\beta_{2x}}{\beta_{2y}}}(-(1-u)\cos\nu_{2} + \alpha_{2y}\sin\nu_{2})}{M_{11}\beta_{2x}-M_{12}\alpha_{2x} +m_{11}\xi_{2xy} + m_{12}\xi_{2xy'}}. 
	\end{split}
	\label{Eq:phaseadvances}
\end{equation}
\end{widetext}
All the parameters on the right-hand side of Eq.~\eqref{Eq:phaseadvances} are the initial values used to calculate the changes in the phases. 

\subsection{Relation to Edwards-Teng parametrization}\label{subsec:edwardstenglBrel}
In the Lebedev-Bogacz parameterization, the decoupling matrix is parametrized in terms of coupled optics functions~\cite{lebedev2010betatron}. The decoupling matrix, $\tilde{R}$, applied to the eigenvectors results in decoupled eigenvectors. Since the generating vectors are defined from eigenvectors, applying the same matrix will result in decoupled generating vectors. The decoupling transformation~\cite{lebedev2010betatron} is shown below:
\begin{equation}
	\begin{split}
		\begin{pmatrix}
			\sqrt{\beta_{1}} \\
			-\frac{i + \alpha_{1}}{\sqrt{\beta_{1}}} \\
			0 \\
			0
		\end{pmatrix} = \begin{pmatrix}
			\cos\phi & 0 & -d\sin\phi & b\sin\phi \\
			0 & \cos\phi & c\sin\phi & -a\sin\phi \\
			a\sin\phi & b\sin\phi & \cos\phi & 0 \\
			c\sin\phi & d\sin\phi & 0 & \cos\phi
		\end{pmatrix}\cdot \begin{pmatrix}
			\sqrt{\beta_{1x}} \\
			-\frac{i(1-u) + \alpha_{1x}}{\sqrt{\beta_{1x}}} \\
			\sqrt{\beta_{1y}}e^{i\nu_{1}} \\
			-\frac{iu + \alpha_{1y}}{\sqrt{\beta_{1y}}}e^{i\nu_{1}}
		\end{pmatrix},
	\end{split}
\end{equation}
where the parameters $a, b, c,$ and $d$ are given by:
\begin{equation} 
	\begin{split}
		a\tan\phi &= \sqrt{\frac{\beta_{2y}}{\beta_{2x}}}\frac{\alpha_{2x}\sin\nu_{2} + u\cos\nu_{2}}{(1-u)}, \\
		b\tan\phi &= \sqrt{\beta_{1x}\beta_{1y}}\frac{\sin\nu_{1}}{1-u}, \\
		c\tan\phi &= \frac{1}{(1-u)\sqrt{\beta_{2x}\beta_{2y}}}\left( \cos\nu_{2}(\alpha_{2x}(1-u) - \alpha_{2y}u)-\sin\nu_{2}(u(1-u) + \alpha_{2x}\alpha_{2y})\right), \\
		d\tan\phi &=-\sqrt{\frac{\beta_{1x}}{\beta_{1y}}}\frac{u\cos\nu_{1}+\alpha_{1y}\sin\nu_{1}}{1-u},
	\end{split}
    \label{Eq:decouplingmat}
\end{equation}
 and the decoupling angle $\phi$ is given by $\tan^{2}\phi=u/(1-u)$. Since this transformation decouples the eigenvectors, it also decouples the generating vectors. For instance,
\begin{equation}
	\begin{split}
		\vec{\tilde{z}}_{1}&= \tilde{R}\vec{z}_{1} = \frac{1}{2}(\tilde{R}\vec{v}_{1} + \tilde{R}\vec{v}_{1}^{*})\cos\mu_{1} - \frac{i}{2}(\tilde{R}\vec{v}_{1} - \tilde{R}\vec{v}_{1}^{*}), \\
		&= \frac{1}{2}(\tilde{v}_{1} + \tilde{v}_{1}^{*})\cos\mu_{1} - \frac{i}{2}(\tilde{v}_{1}-\tilde{v}_{1}^{*})\sin\mu_{1}, \\
	\end{split}
\end{equation}
which yields for mode $1$:
\begin{equation}
	\begin{split}
	\tilde{z}_{1} &= \begin{pmatrix}
		\sqrt{\beta_{1}}\cos\mu_{1} \\
		\frac{-\alpha_{1}\cos\mu_{1}-\sin\mu_{1}}{\sqrt{\beta_{1}}} \\
		0 \\
		0
		\end{pmatrix}, \quad \tilde{z}_{2}= \begin{pmatrix}
			\sqrt{\beta_{1}}\sin\mu_{1} \\
			\frac{-\alpha_{1}\sin\mu_{1}+\cos\mu_{1}}{\sqrt{\beta_{1}}} \\
			0 \\
			0
		\end{pmatrix}. 
	\end{split}
\end{equation}
This results in the same transformation as for the uncoupled Courant-Snyder transformation presented earlier in Section~\ref{sec:background}. The  parameters shown in Eq.~\eqref{Eq:decouplingmat} depend on the coupled optics functions, which makes the decoupling matrix unique. However, in the Edwards-Teng parametrization~\cite{edwards1973parametrization}, the decoupling matrix is not unique. Edwards-Teng parametrization decouples the transfer matrix in 4D plane. Additional rotations in individual 2D planes preserve the decoupling. These ``gauge"-like rotations introduce infinitely many solutions, making the transformation non-unique. Lebedev-Bogacz parametrization fixes this ``gauge" to the physical solution by relating the rotation angle parameters to the eigenvectors.

The relationship between the beam matrix elements, $\Sigma_{4\times 4}$, and the optics functions are given in~\cite{lebedev2010betatron}. One may realize that this method of transformation also leads to transformation of the beam matrix. The beam matrix can be written as a sum of contributions of modes 1 and 2; $\Sigma = \Sigma_{1} + \Sigma_{2}$, which leads to its transport as follows:
\begin{equation}
	\begin{split}
	\Sigma_{f} &= M\cdot\Sigma_{i}\cdot M^{T}, \\ 
	\Sigma_{f} &= M\cdot\Sigma_{1i}\cdot M^{T} + M\cdot\Sigma_{2i}\cdot M^{T}, \\
	\Sigma_{f} &= \Sigma_{1f} + \Sigma_{2f},
	\end{split}
\end{equation}
where subscripts $i$ and $f$ stand for initial and final conditions. Therefore, the transformation of the second-order moments matrix is the sum of individual mode contributions after transformation. 

\subsection{Special Solutions}\label{subsec:specialsoln}

\begin{figure}[tbp]
	\includegraphics[width=0.49\linewidth]{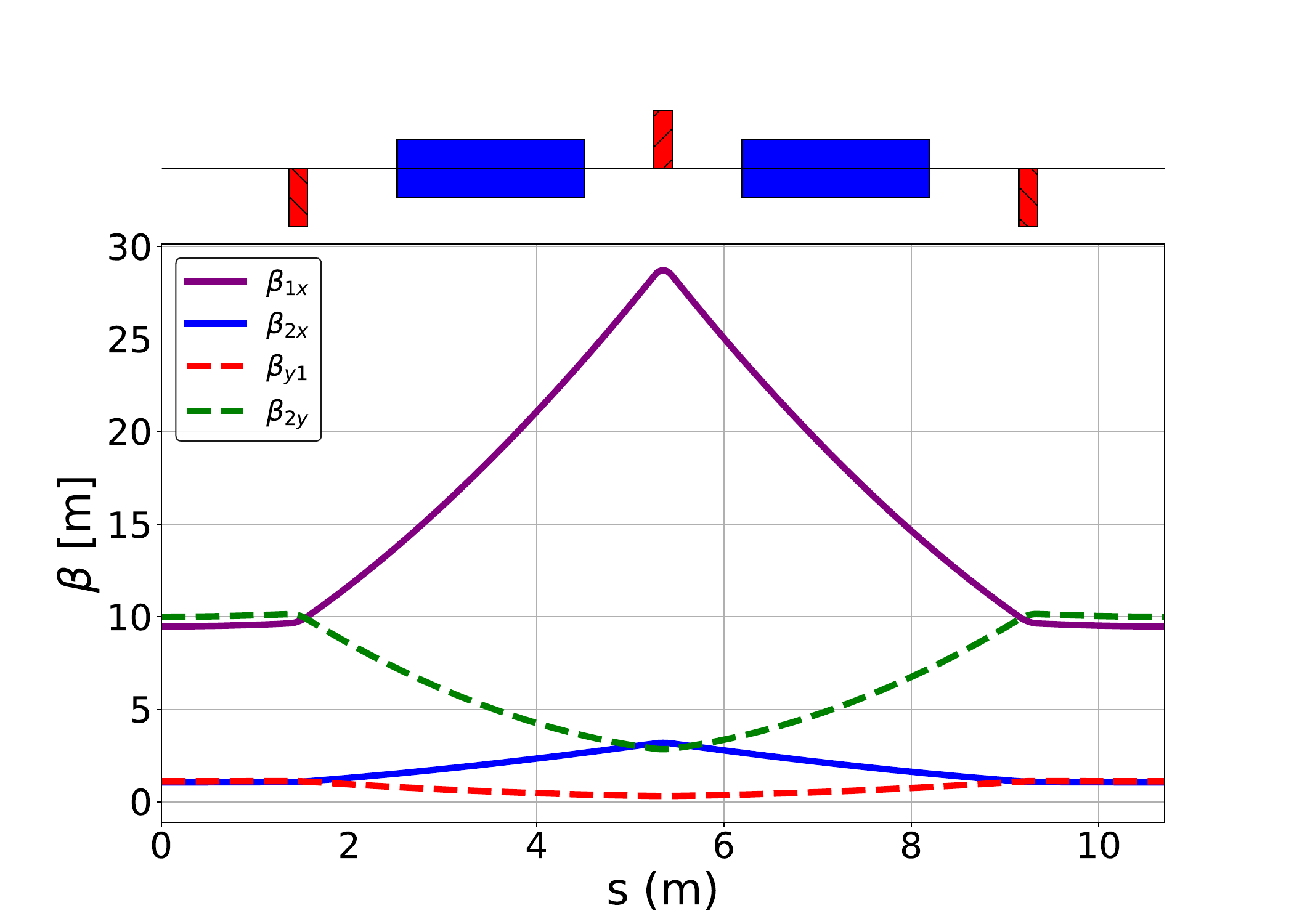}
	\includegraphics[width=0.49\linewidth]{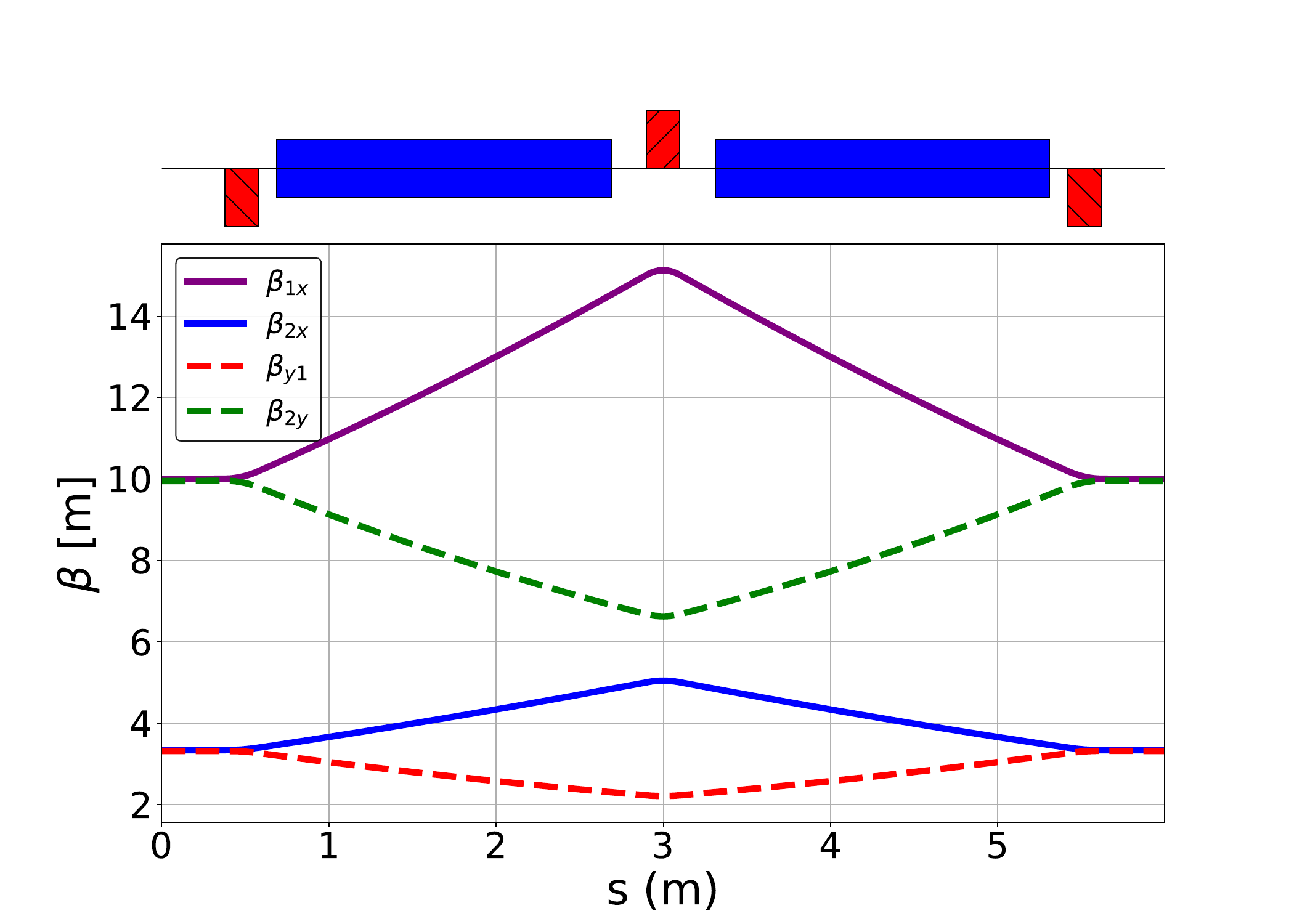}
	\includegraphics[width=0.49\linewidth]{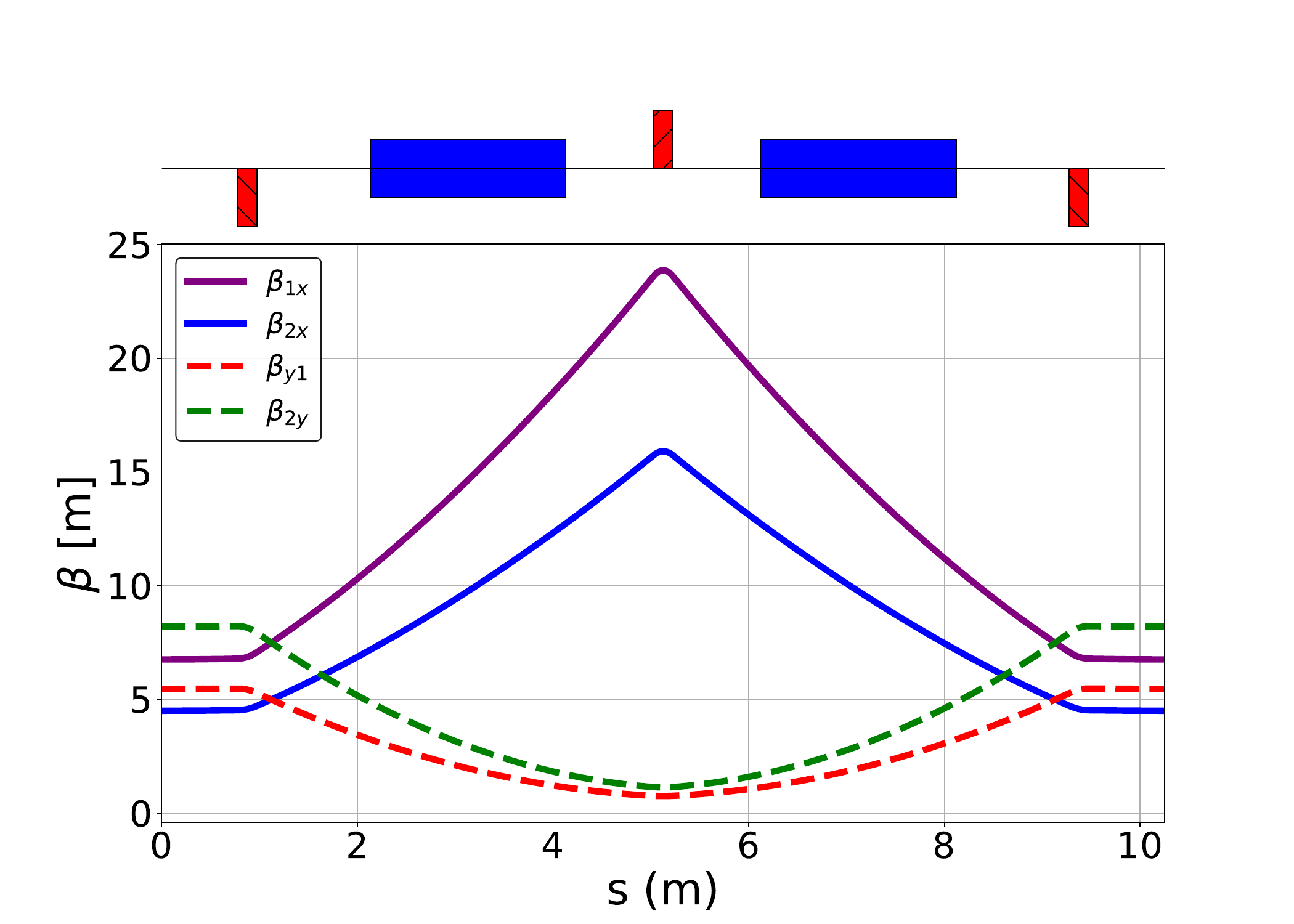}
	\includegraphics[width=0.49\linewidth]{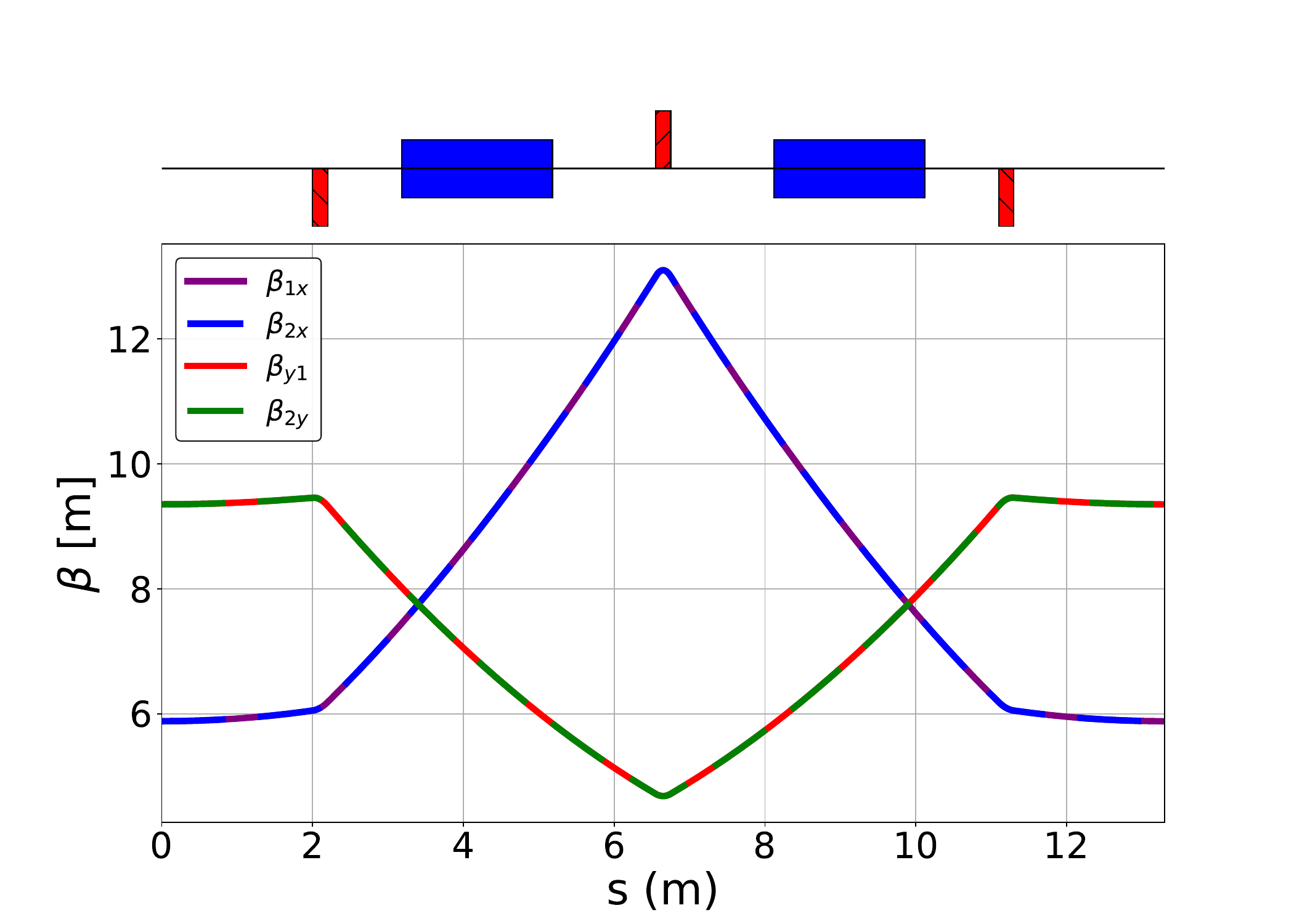}
	\caption{Coupled beta functions in normal quadrupole triplets with: $u=0.1$ (top-left), $u=0.25$ (top-right), $u=0.4$ (bottom-left), $u=0.5$ (bottom-right).}
	\label{Fig:difuparamsnormalquad}
\end{figure}

The general method of transporting the optics functions helps us understand how the coupled functions propagate. An important part of lattice design is achieving periodicity with the optics functions. Coupling can be generated due to errors within the lattice, and understanding these errors leads to correcting them. Interestingly, coupling can be generated for various purposes, such as having a larger beam cross-section to mitigate collective effects, studying the effects of coupling on beam dynamics, etc. Coupling can be created using a special section where the rest of the periodic cells are uncoupled to study coupling effects. In this case, periodicity will depend on the uncoupled cells downstream, so the created coupled functions must be matched to the uncoupled cells, or the uncoupled cells must be matched for the coupled functions created.
\begin{figure}[tbp]
    \centering
    \includegraphics[width=\linewidth]{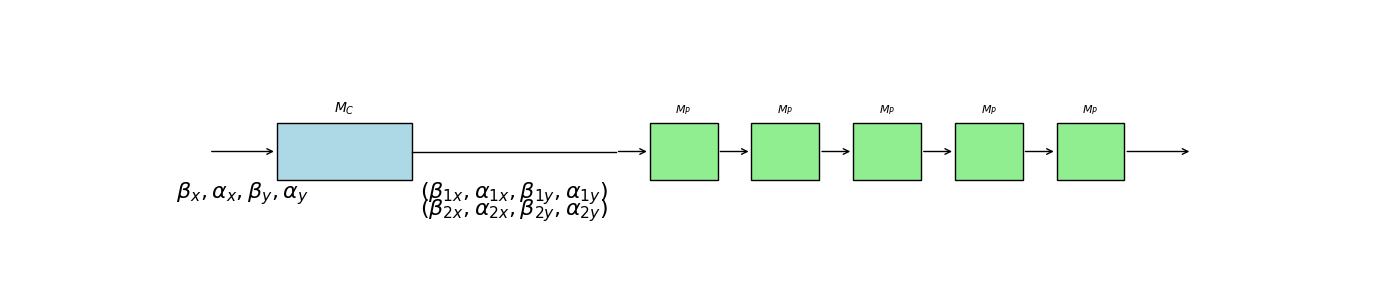}
    \caption{Creation of coupled optics functions for uncoupled periodic cells.}
    \label{fig:specialc}
\end{figure}
Based on the illustrated cell structure, one logical reason is to match the coupled functions to the uncoupled cells for periodicity while maintaining symplectic normalization within the phase-space vector parametrization. From Courant-Snyder's theory, the uncoupled downstream elements are easy to find solutions for to maintain periodic optics functions. The matched Courant-Snyder optics will serve as decoupled functions, and from the decoupled Edwards-Teng parametrization, we know the on-mode functions (mode $1$ projection on $x$ and mode $2$ projection on $y$) relations to coupled functions as:
\begin{equation}
    \begin{split}
        \beta_{1} &= \frac{\beta_{1x}}{1-u}, \qquad \beta_{2}=\frac{\beta_{2y}}{1-u}, \\
        \alpha_{1}&= \frac{\alpha_{1x}}{1-u}, \qquad \alpha_{2}=\frac{\alpha_{2y}}{1-u}.
    \end{split}
\end{equation}
$\beta_{1},\beta_{2}$ are decoupled functions where here are associated with uncoupled Twiss functions as downstream periodic cells are uncoupled. The transformation of coupled parameters for uncoupled lattices is given by Eq.~\eqref{Eq:Fromcoupledtocoupleduncmatrix}, which suggests the same phase plane projections of optics functions for each mode transform similarly. Using vector notation as, $\vec{\beta}_{1x}=[\beta_{1x},\alpha_{1x},\gamma_{1x}]^{T}$, the transformation for uncoupled matrix simplifies to:
\begin{equation}
    \begin{split}
        \vec{\beta}_{1,2,x}(s_{f}) = C_{M}\vec{\beta}_{1,2x}(s_{i}), \quad \vec{\beta}_{1,2,y}(s_{f})= C_{N}\vec{\beta}_{1,2,y}(s_{i}).
    \end{split}
\end{equation}
We can write the coupled function in terms of matched decoupled functions and use normalization constraints to propose a relation of off-mode functions, $\beta_{2x},\beta_{1y}$, as:
\begin{equation}
    \begin{split}
        \beta_{1x}&=(1-u)\beta_{1}, \quad \beta_{2y}=(1-u)\beta_{2}, \\
        \beta_{2x}&=u\beta_{1}, \quad \beta_{1y}=u\beta_{2},
    \end{split}
    \label{Eq:coupledrel}
\end{equation}
and similar relationships for $\alpha$ functions. The same phase coordinates optics functions transform similarly, such as $\beta_{1x}$ and $\beta_{2x}$; therefore, the matched coupled functions should be related to the matched uncoupled functions for uncoupled periodic cells. Interestingly, this relation introduces a relation between the same mode functions as $\beta_{2x}=u/(1-u)\beta_{1x}$, and similarly for other off-mode functions with on-mode functions. Computing emittance projections on $x$ and $y$ planes with this relationship yields:
\begin{equation}
    \begin{split}
        \epsilon_{x}&= (1-u)\epsilon_{1} + u\epsilon_{2}, \\
        \epsilon_{y}&= u\epsilon_{1} + (1-u)\epsilon_{2}.
    \end{split}
    \label{eq:projectedemittanceswithu}
\end{equation}
For small coupling to no coupling, this suggests $u=0$ leads to correct assignment of emittances with $\epsilon_{1}=\epsilon_{x}$ and $\epsilon_{2}=\epsilon_{y}$. The set of relationships shown in Eq.~\eqref{Eq:coupledrel} gives matched solutions for coupled functions while maintaining the normalization conditions between the generating vectors. When $u\rightarrow 0$, Eq.~\eqref{eq:eigenvecscoupled} do not fully reduce to the uncoupled eigenvectors, and further requirement of setting off-mode functions to zero is needed, which leads to singularity (division by zero). The special set of solutions for off-mode functions in Eq.~\eqref{Eq:coupledrel} fixes this singularity for uncoupled lattice. This set of solutions is illustrated in Fig.~\ref{Fig:difuparamsnormalquad}.

The second case where this approach can be used is for a special periodic matrix made of skew quadrupoles. A cell made of skew quadrupole type focusing elements transfer matrix is represented as:
\begin{equation}
    \mathcal{M}=R(\pi/4)\begin{pmatrix}
        M & 0 \\
        0 & N
    \end{pmatrix}R^{-1}(\pi/4) = \frac{1}{2}\begin{pmatrix}
        M+N & N-M \\
        N-M & M+N
    \end{pmatrix}.
\end{equation}
Suppose that the total transfer matrix has equal overall focusing in both planes, $M=N$. In that case, the overall transfer matrix reduces to $\mathcal{M}=diag(M,M)$ where the periodic cell made up of skew quadrupoles leads to an overall uncoupled transfer matrix. 

\section{Applications: Lattices \& Lattice Building Blocks}
 In this section, we apply this method to various examples of lattices and lattice building blocks. In order to write a compact matching and transport script, the generating functions are used, where the relations of optics functions to the generating functions are given in Eq.~\eqref{Eq:genvectorscoupledfunctions}. It is important to check the normalization condition for the transport of the generating vectors which is done through the normalization condition. The python scripts that are used for matching and propagating of the coupled lattices is available on~\cite{matchingcode}.

\subsection{Derbenev's Adapter}
\begin{figure}[tbp]
	\centering
	\includegraphics[width=0.49\linewidth]{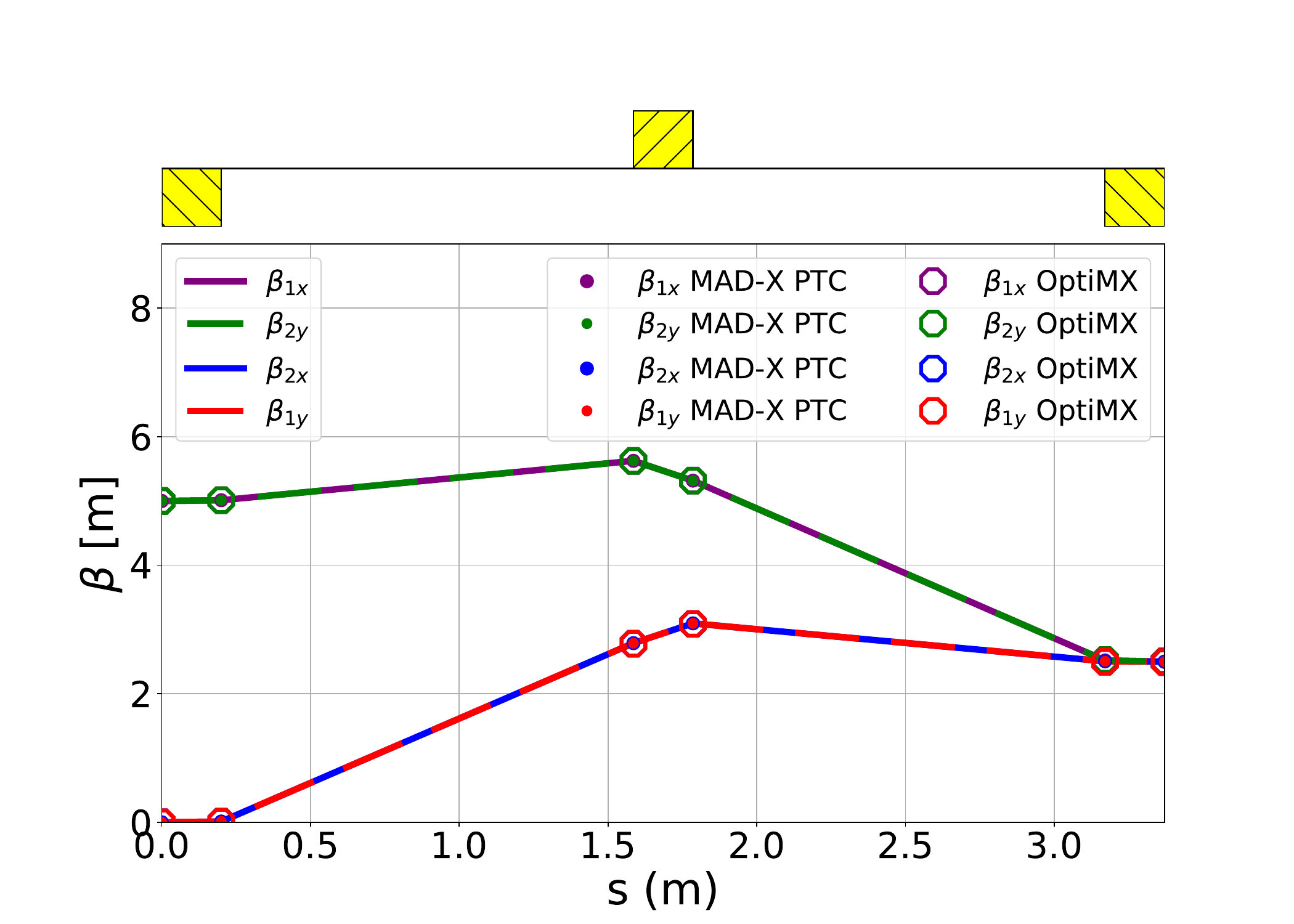}
	\includegraphics[width=0.49\linewidth]{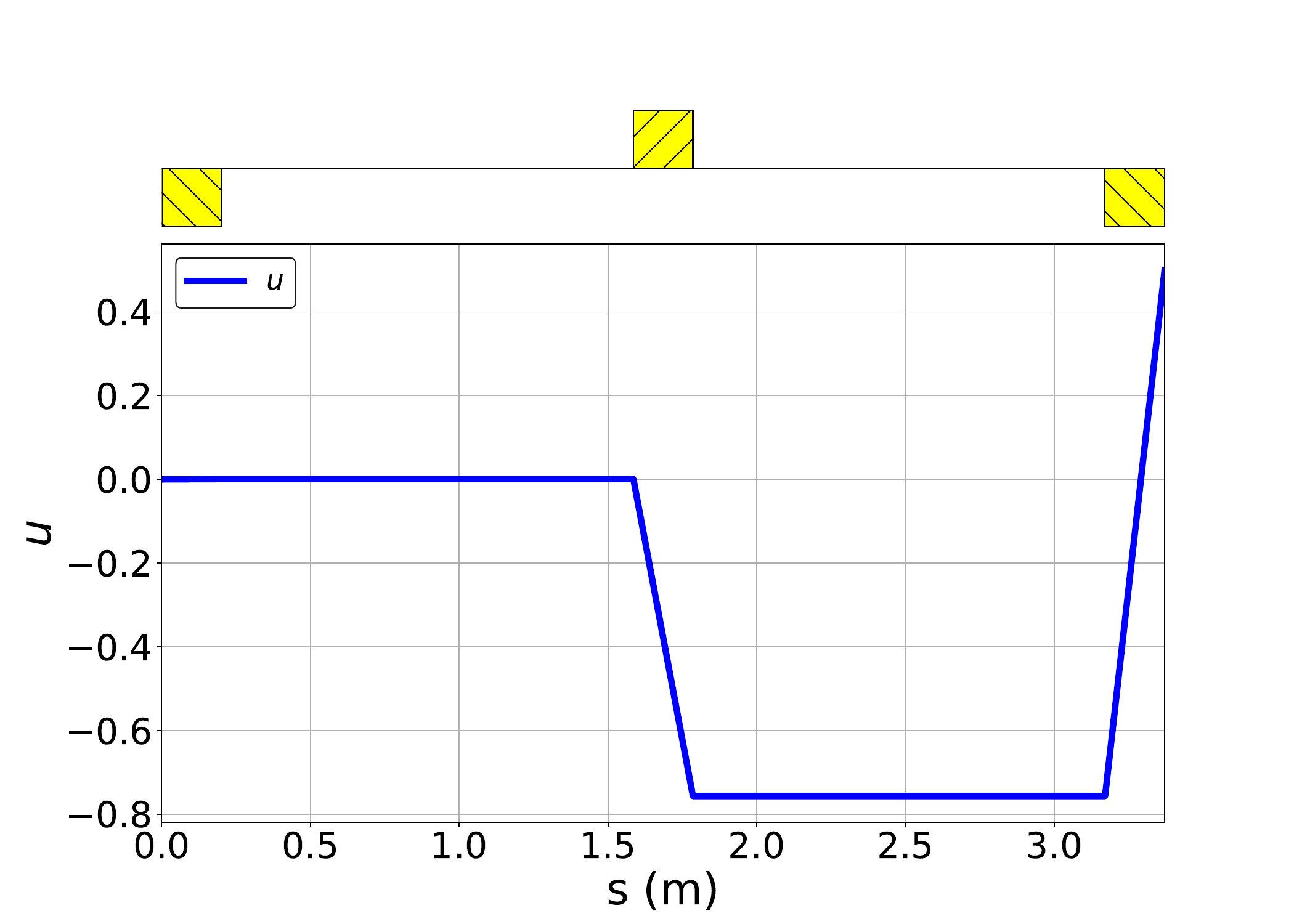}
	\caption{Derbenev Adapter Matching: Initial uncoupled beta functions, $\beta_{x,y}=5.0\mathrm{m}$ and $\alpha_{x,y}=0.0$ converts to coupled optics functions $\beta_{1x},\beta_{2x},\beta_{1y},\beta_{2y}=2.5\mathrm{m}$ and all $\alpha$ functions zero. Yellow elements represent skew quadrupoles. Coupled beta functions plot (left). Coupling strength parameter, $u$ (right).}
	\label{Fig:Derbenevadaptertransform}
\end{figure}

Derbenev's Adapter, known as a flat-to-round beam converter, is a triplet of skew quadrupoles that converts flat beams into coupled round beams with non-zero angular momentum known as circular modes~\cite{burov2002circular,burov2013circular,derbenev1998adapting}. This transformation creates a strong coupling between the $x$ and $y$ planes, which produces fully coupled optics. Fully coupled optics correspond to coupling strength parameter, $u=1/2$~\cite{lebedev2010betatron}. Using the method developed here, we can match the three skew quadrupoles to perform a flat-to-round transformation. The matched results for an initial uncoupled beam with $\beta_{x,y}=5$\,m are shown in Fig.~\ref{Fig:Derbenevadaptertransform}. Derbenev adapter reduces the uncoupled functions by half, creating a strong $x-y$ coupling. Strong coupling is seen at $u=1/2$. After the transformation, the phases of coupling are $\pi/2$ and the $x-y$ cross section is round. Transformation of the optics functions from uncoupled to coupled optics can be done using Eq.~\eqref{Eq:fromunctocoupled}:
\begin{equation}
	\begin{split}
		\begin{pmatrix}
			\beta_{1x} \\
			\alpha_{1x} \\
			\gamma_{1x}
		\end{pmatrix}_{f} & = C_{M} \begin{pmatrix}
			\beta_{x} \\
			\alpha_{x} \\
			\gamma_{x}
		\end{pmatrix}_{i}, \quad \begin{pmatrix}
			\beta_{1y} \\
			\alpha_{1y} \\
			\gamma_{1y}
		\end{pmatrix}_{f} = C_{n}\begin{pmatrix}
			\beta_{x} \\
			\alpha_{x} \\
			\gamma_{x}
			\end{pmatrix}_{i}, \\
			\begin{pmatrix}
				\beta_{2x} \\
				\alpha_{2x}\\
				\gamma_{2x}
			\end{pmatrix}_{f} &= C_{m}\begin{pmatrix}
				\beta_{y} \\
				\alpha_{y}\\
				\gamma_{y}
			\end{pmatrix}_{i}, \quad \begin{pmatrix}
				\beta_{2y} \\
				\alpha_{2y} \\
				\gamma_{2y}
			\end{pmatrix}_{f} = C_{N}\begin{pmatrix}
				\beta_{y} \\
				\alpha_{y}\\
				\gamma_{y}
			\end{pmatrix}_{i}.
	\end{split}
\end{equation}
Establishing the necessary conditions on the transfer matrix of horizontal and vertical planes is straightforward for creating Derbenev's Adapter. The condition called the vorticity condition relates the vertical and horizontal planes at the end of the transformation. In this formalism, the constraints at the end of the beam transport are $u=1/2$ and $\nu_{1,2}=\pi/2$. 

Figure~\ref{Fig:Derbenevadaptertransform} shows the creation of coupled beta functions with Derbenev's adapter. It also compares MAD-X's PTC and OptiMX optics with the calculated coupled functions. It shows the propagation of the optics functions, which are identical between the two optics codes, and the formalism here.

\subsection{Mode Flipping with Skew Quadrupoles}
\begin{figure}[tbp]
    \centering
    \includegraphics[width=0.49\linewidth]{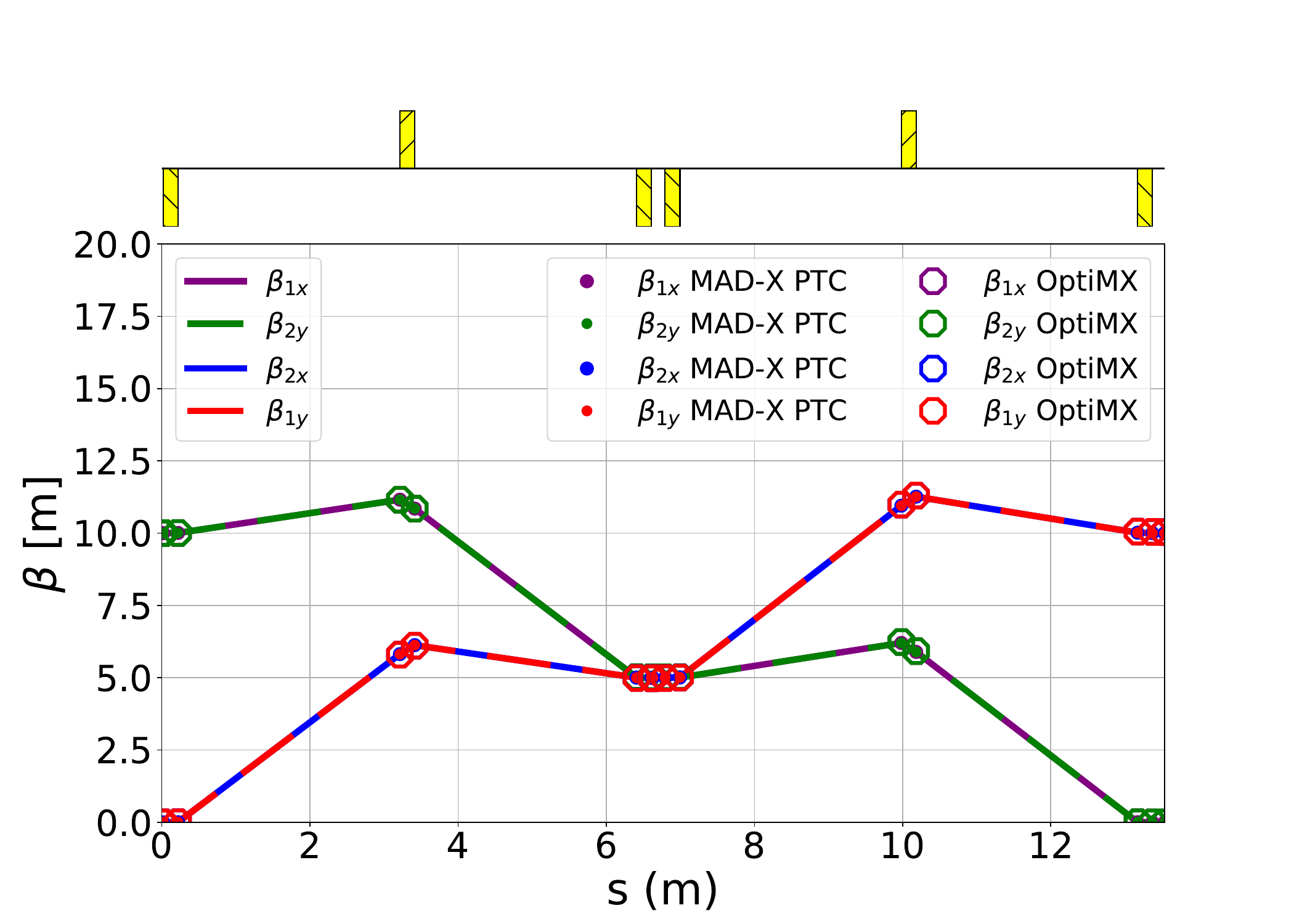}
    \includegraphics[width=0.49\linewidth]{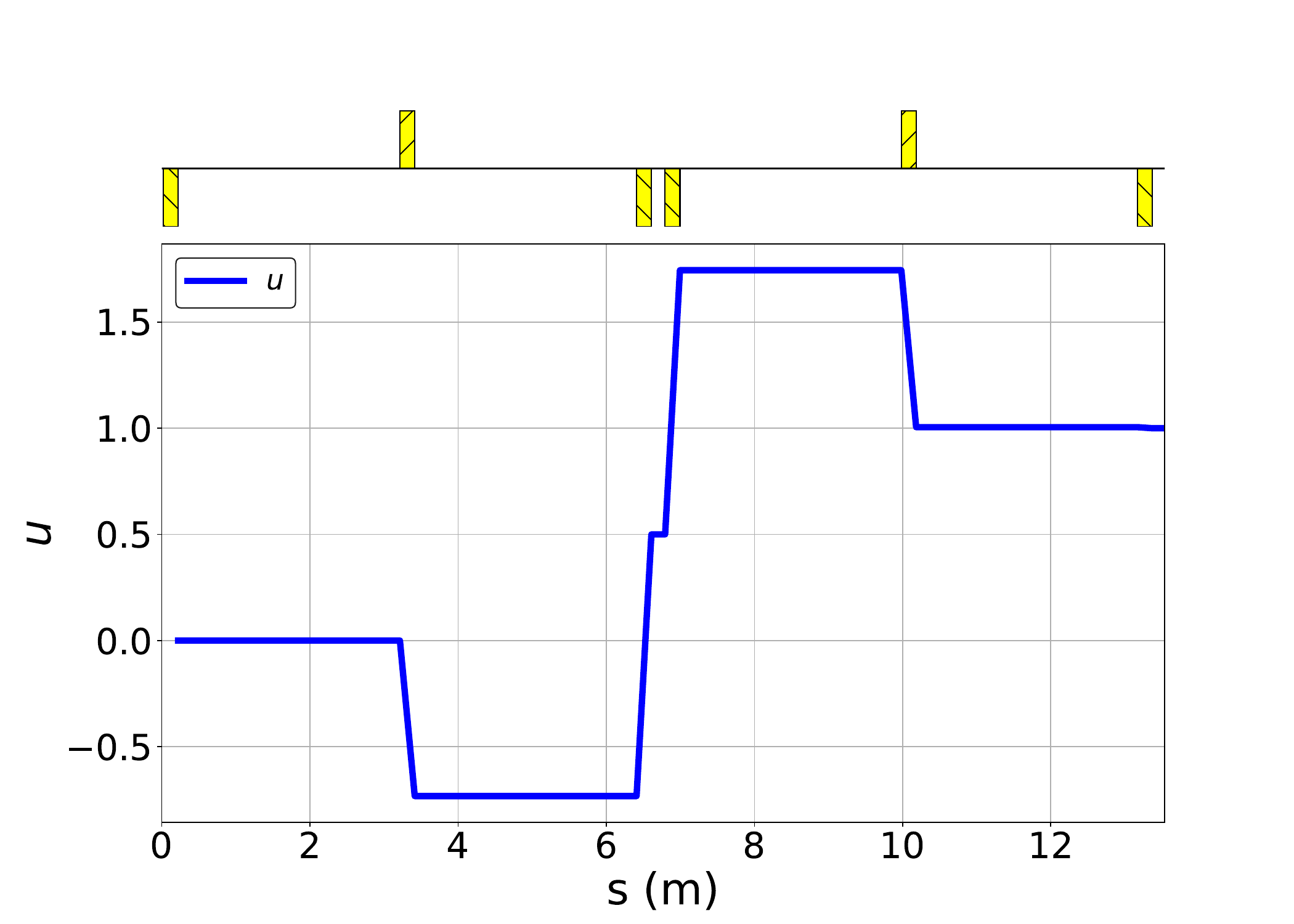}
    \includegraphics[width=0.49\linewidth]{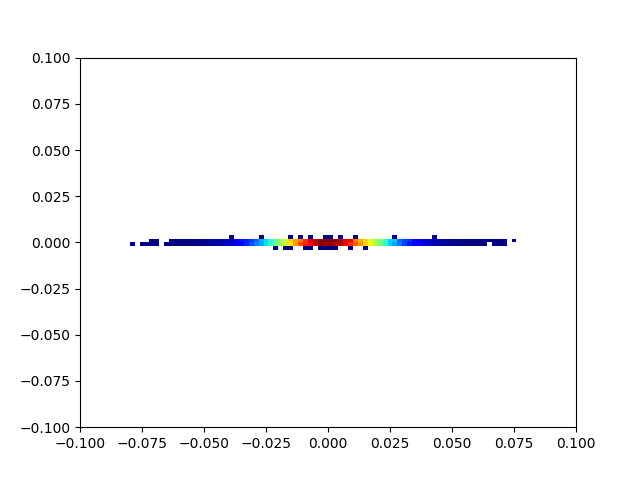}
    \includegraphics[width=0.49\linewidth]{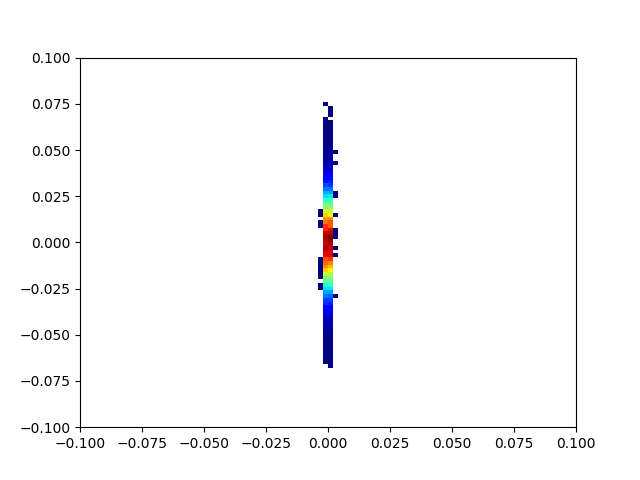}
    \caption{Mode flipping process: creation and flipping of modal beta functions (top-left). Coupling strength parameter $u$ (top-right). Initial flat beam with $\epsilon_{y}\ll\epsilon_{x}$ (bottom-left) flipped into flat beam with $\epsilon_{x}\ll\epsilon_{y}$ (bottom-right) from particle tracking simulation using the TRACK code.}
    \label{fig:modefliptscheme}
\end{figure}

Uncoupled flat beams are beams in which the emittance is much smaller in one plane than in the other, for example $\epsilon_{y}\ll\epsilon_{x}$, when the beam is flat in the $y$ direction. The beam sizes are usually larger in the $x$ dimension than in the $y$ dimension. Flatness is usually achieved for electrons in storage rings as a result of damping from synchotron radiation. The flatness can be exchanged between the $x$ and $y$ planes using a set of skew quadrupoles. As we have discussed $u=0$ corresponds to uncoupled motion. This suggests that mode labeling can be $\epsilon_{x}\rightarrow\epsilon_{1}$ and $\epsilon_{y}\rightarrow\epsilon_{2}$, for example. We also notice that $u=1$ leads to generating vectors with similar form as the case with $u=0$, but with mode flipping. Using the transformation of optics functions above, we can design a set of skew quadrupoles that achieve mode flipping by taking $u=0$ to $u=1$ as shown in Fig.~\ref{fig:modefliptscheme}. The process of mode flipping is simply to exchange the modes' contributions to x and y projections: $\epsilon_{x}\rightarrow\epsilon_{1}\rightarrow\epsilon_{y}$ and $\epsilon_{y}\rightarrow\epsilon_{2}\rightarrow\epsilon_{x}$. This flips an initial uncoupled flat beam in $y$ with $\epsilon_{y}\ll\epsilon_{x}$ into an uncoupled flat beam in $x$ with $\epsilon_{x}\ll\epsilon_{y}$.

The betatron functions are also compared with the MAD-X PTC and OptiMX design codes. Figure~\ref{fig:modefliptscheme} shows that all design codes align with the formalism discussed in this paper. The difference between the design codes and the transport formalism described here is that the latter was implemented in Python, including a matching routine for coupled beam optics. However, both MAD-X and OptiMX compute the coupled functions but do not include matching for coupled optics. Once the quadrupole strengths and drift lengths are found using our transport code, they are passed to the design codes for verification. 

\subsection{FODO With a Skew Quadrupole}
\begin{figure}[tbp]
	\centering
	\includegraphics[width=0.49\linewidth]{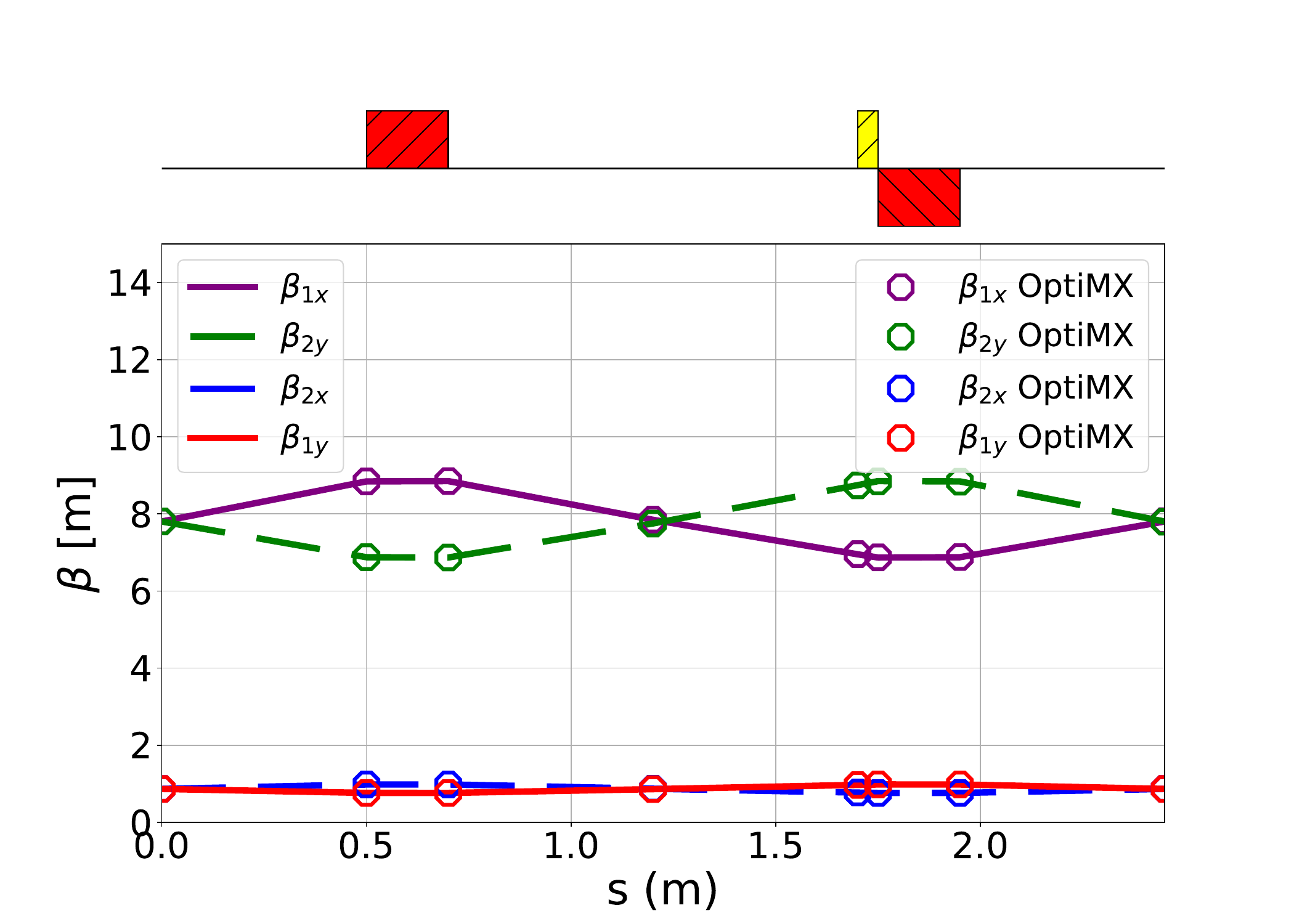}
	\includegraphics[width=0.49\linewidth]{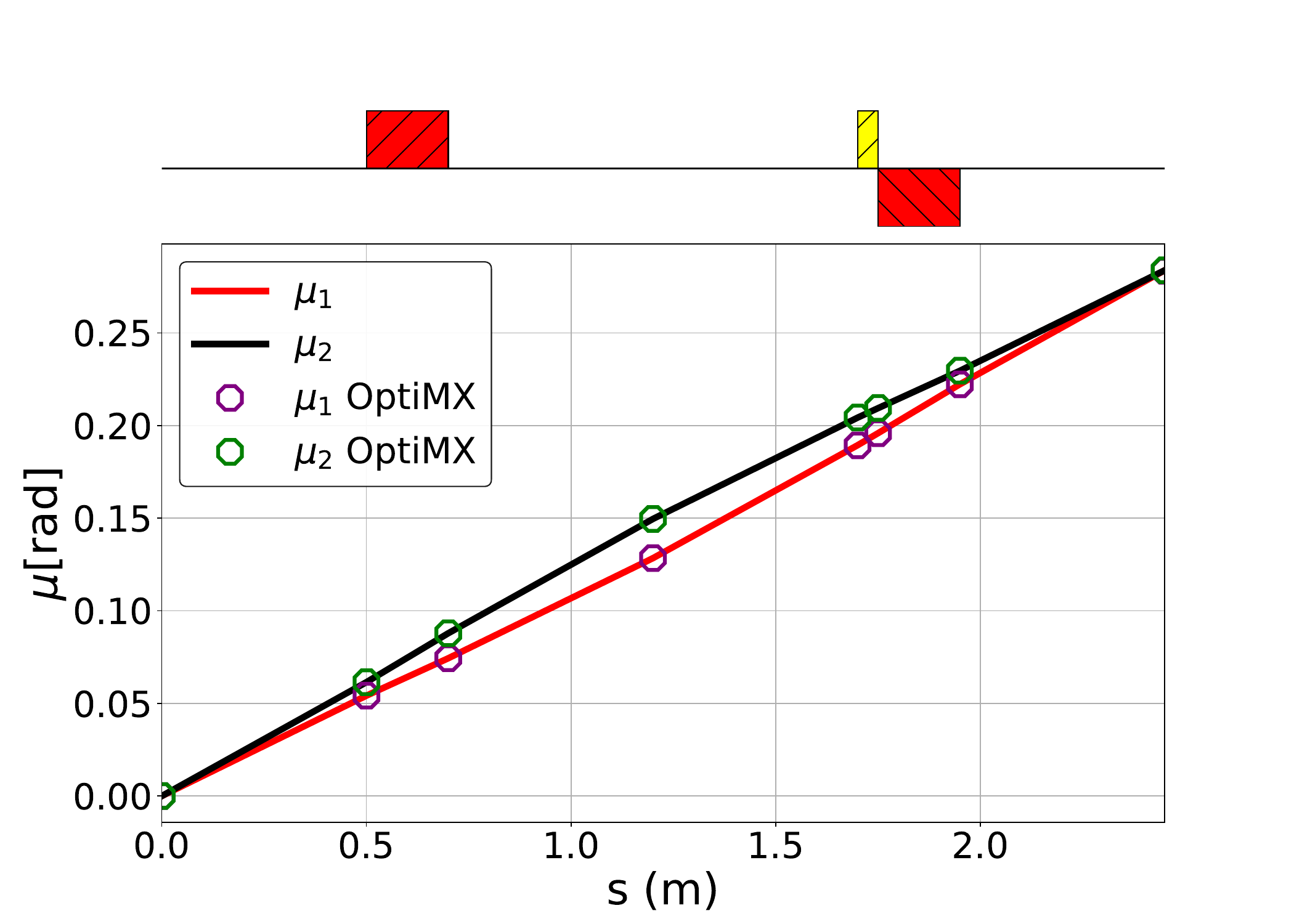}
	\includegraphics[width=0.49\linewidth]{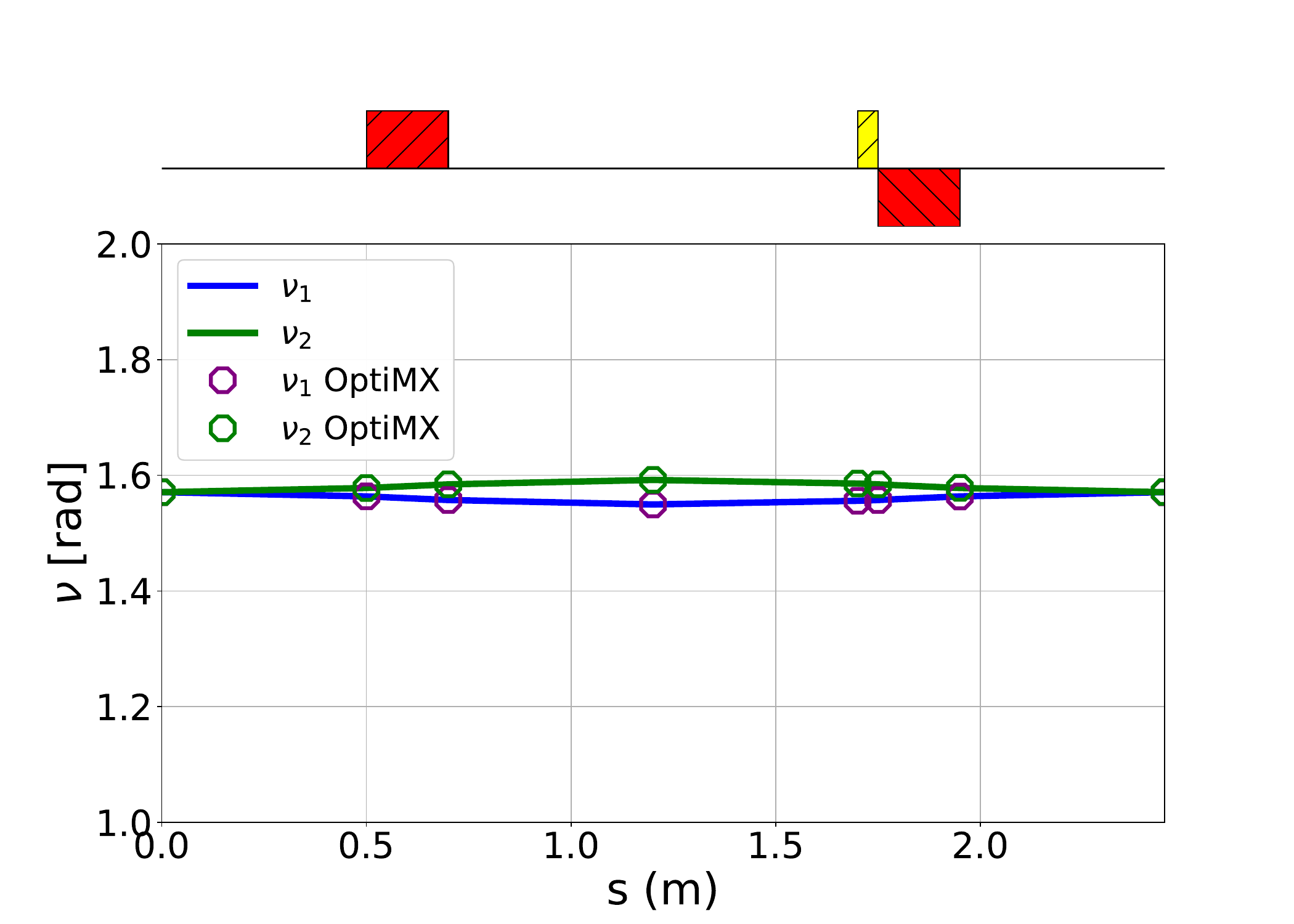}
	\includegraphics[width=0.49\linewidth]{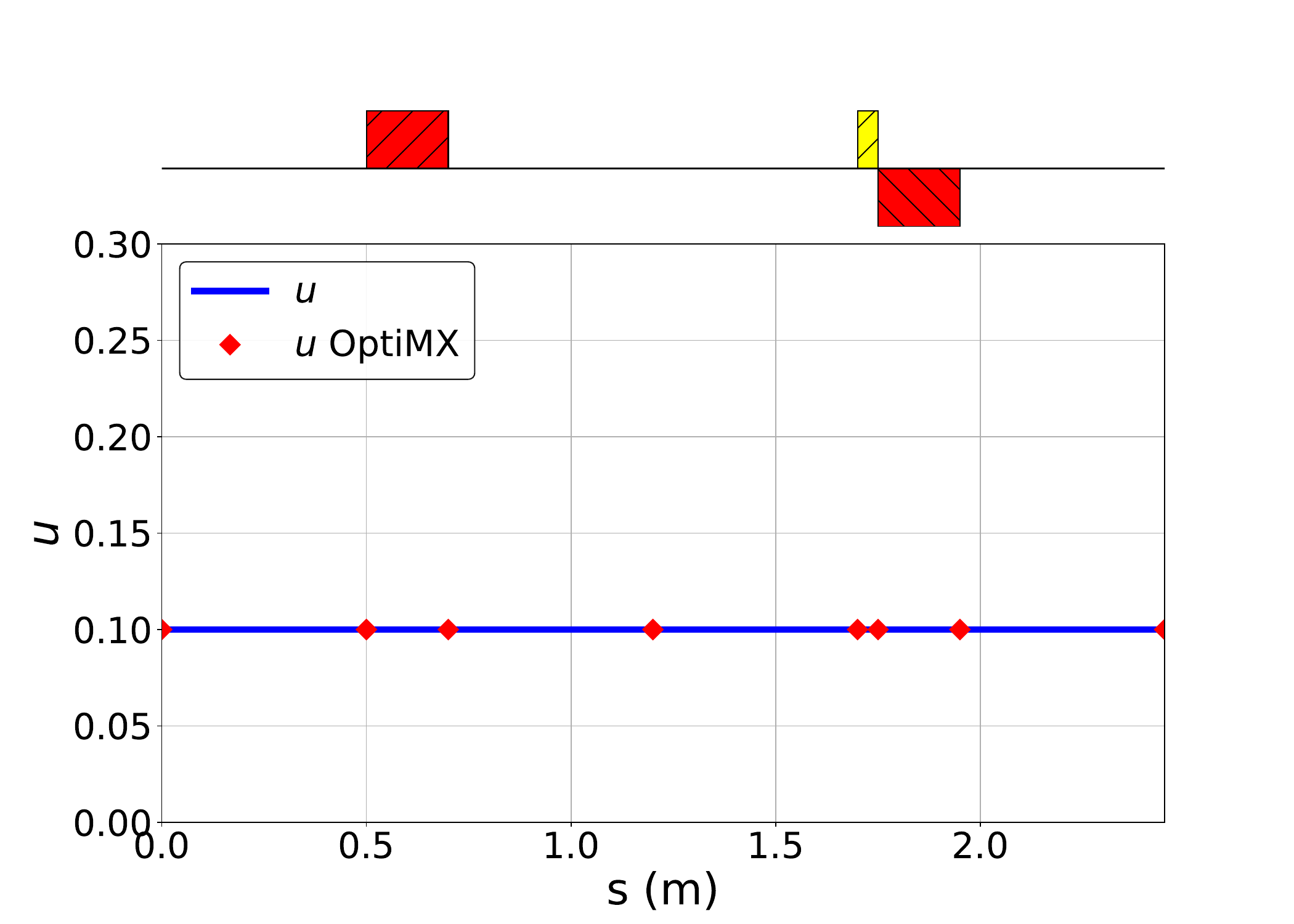}
	\caption{Optics of a FODO cell with a skew quadrupole component. Normal quadrupoles are in red, and the skew quadrupole is in yellow. Focusing quadrupole strength $k_{f}=3.043$, defocusing quadrupole $k_{d}=-3.043$, and skew quadrupole $k_{s}=0.01$. }
	\label{Fig:OFODOsingleskew}
\end{figure}

In addition to creating coupling, we can use this formalism to match and find periodic solutions for a given cell involving coupling elements, such as skew quadrupoles. This cell type provides an example of weak coupling, $u\approx 0$, which could also arise from a normal quadrupole rotation error. In this case, the main oscillation modes are mostly $1\rightarrow x$ and $2\rightarrow y$, making it quite similar to the uncoupled optics case. The result of the matching procedure is shown in Fig.~\ref{Fig:OFODOsingleskew}.  
Using the symmetry of the FODO lattice, a solution can be found by setting $\beta_{1x}=\beta_{2y}$, $\beta_{2x}=\beta_{1y}$, $\alpha_{1x}=-\alpha_{2y}$ and $\alpha_{2x}=-\alpha_{1y}$. 

\subsection{Solenoid-Based Cells}
\begin{figure}[htbp]
	\centering
    \includegraphics[width=0.49\linewidth]{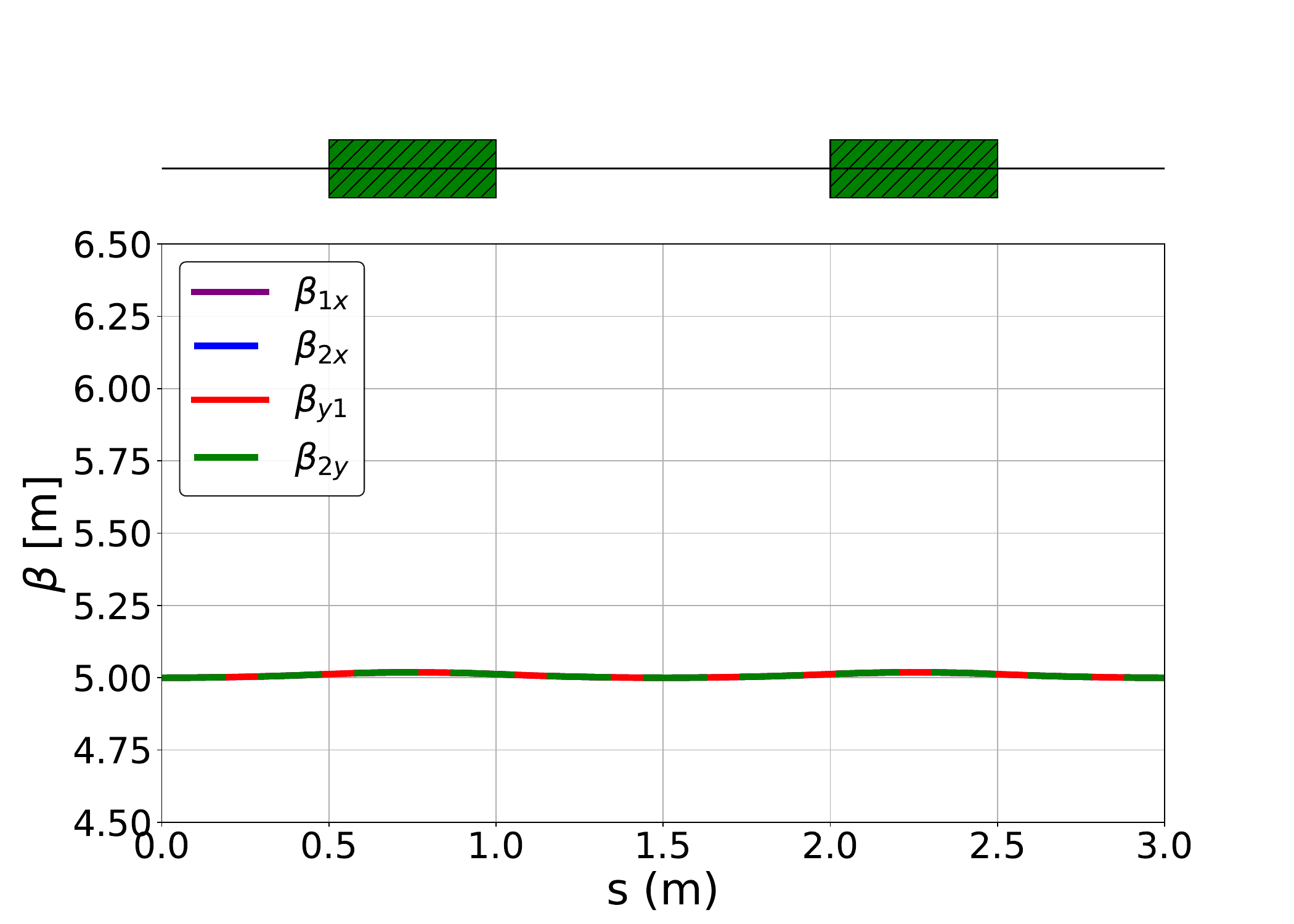}
    \includegraphics[width=0.49\linewidth]{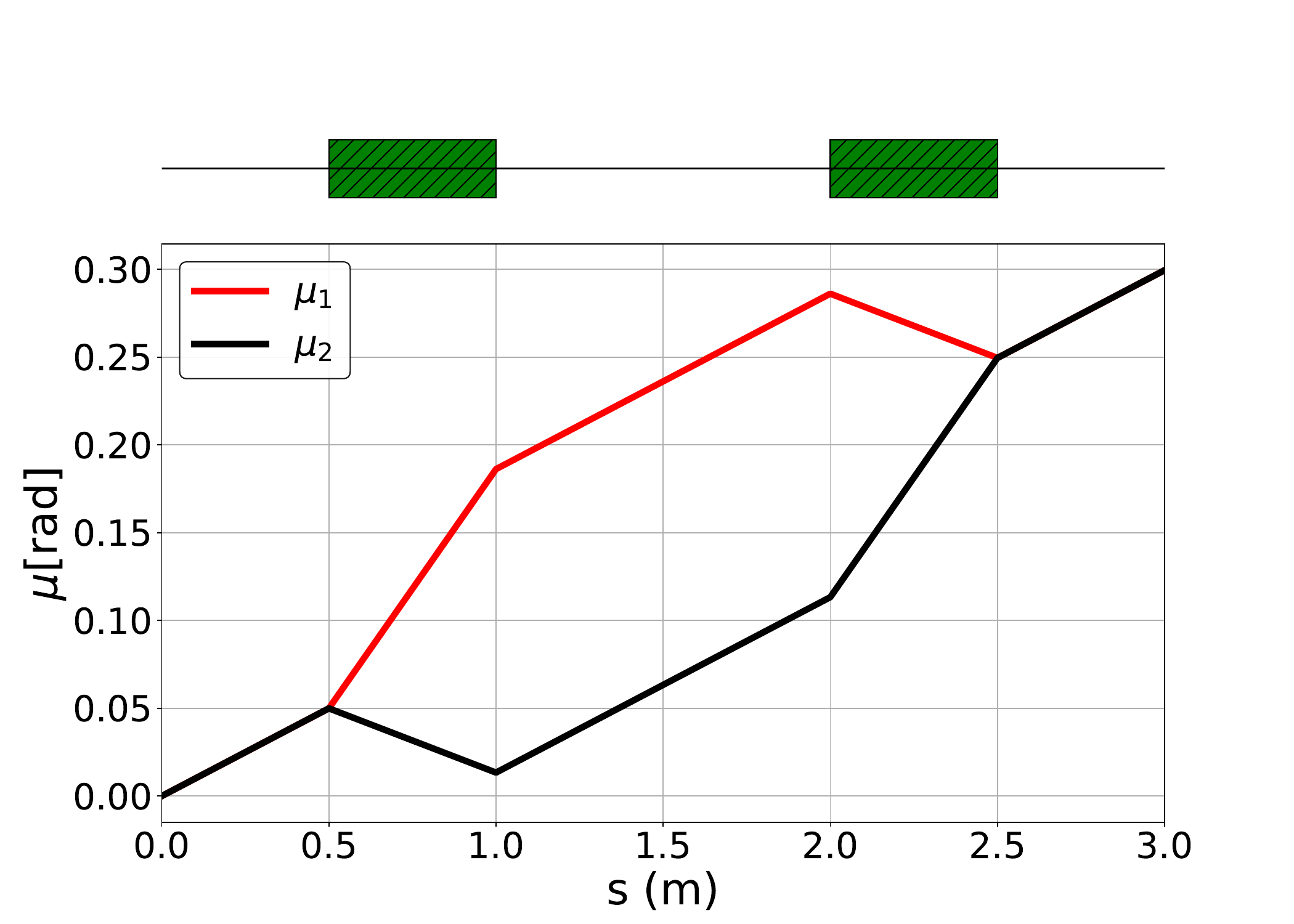}
    \includegraphics[width=0.49\linewidth]{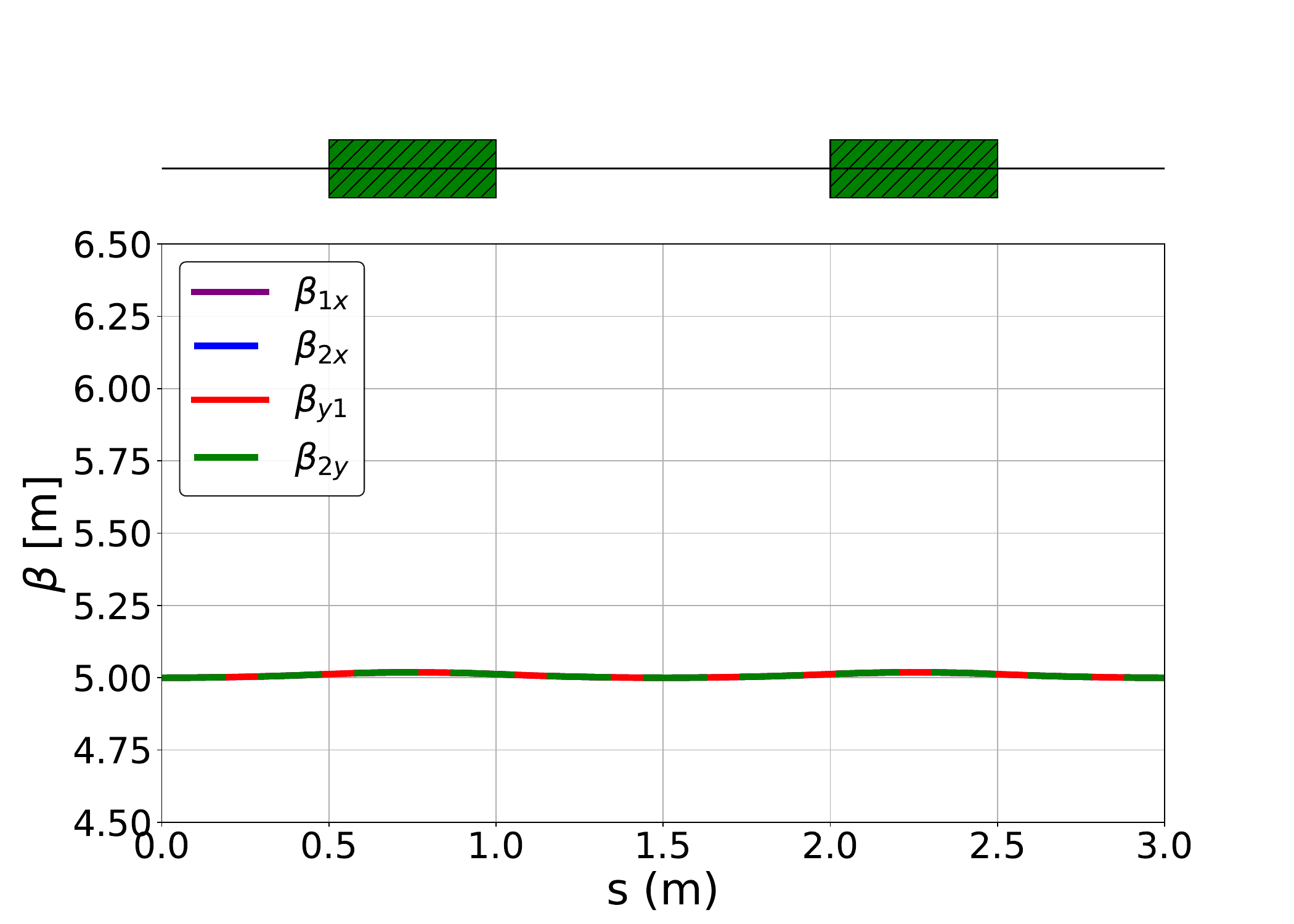}
    \includegraphics[width=0.49\linewidth]{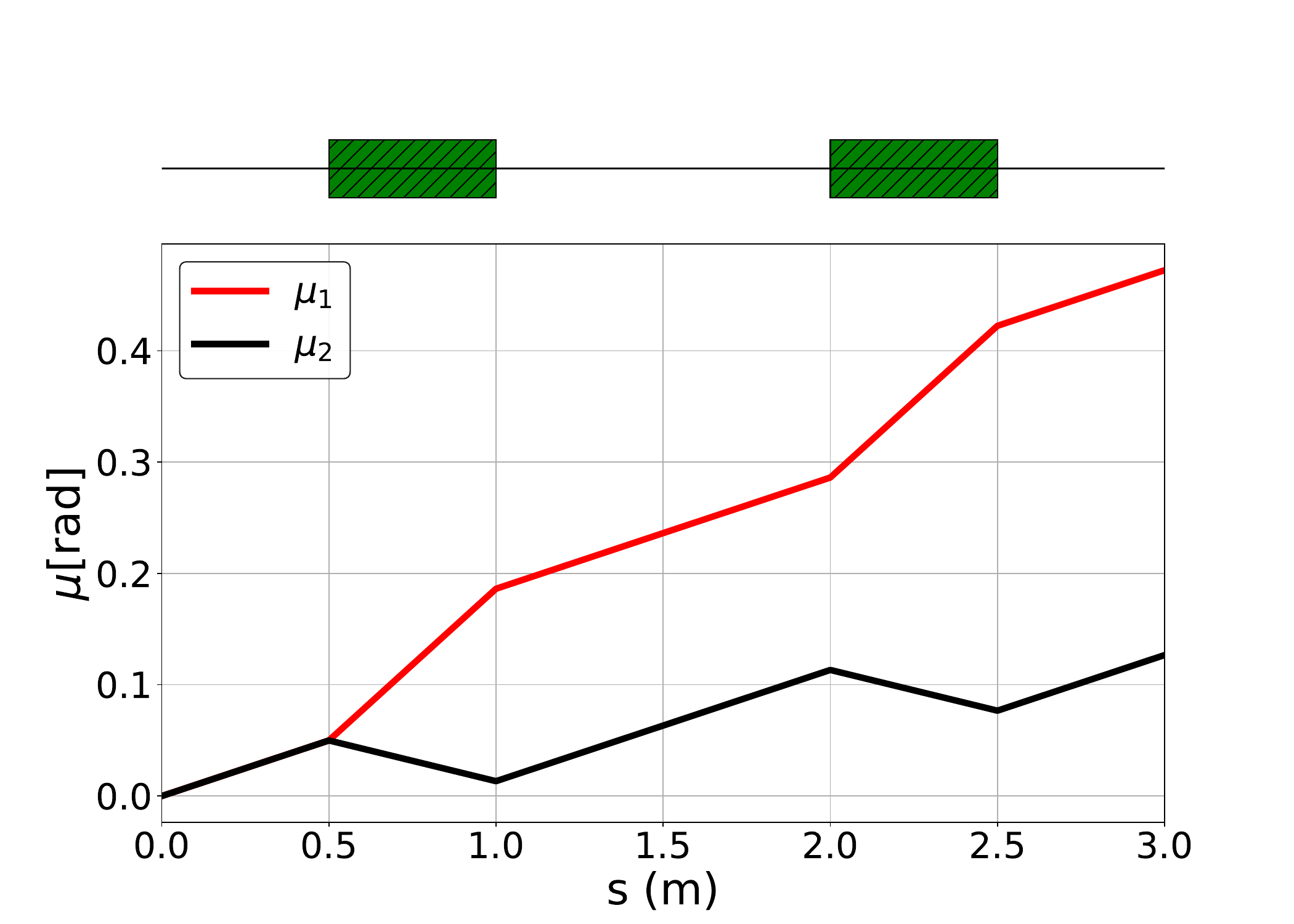}
	\caption{Solenoid cells: Beta functions of opposite polarity of solenoid fields (top-left), same polarity (bottom-left). Phase advances of modes $1$ and $2$ of opposite polarity (top-right), same polarity (bottom-right), for $u=1/2$.} 
	\label{Fig:solenoidcells}
\end{figure}

Solenoid focusing is mainly used at low energies; it is less effective and requires very high fields at high energies. Solenoids are used to propagate axisymmetric beams, to produce magnetized beams, and for beam cooling. Interestingly, solenoids can split the tunes of the coupled beams into mode $1$ and mode $2$ tunes. The splitting of tunes is illustrated in Fig.~\ref{Fig:solenoidcells} for the same cell of two solenoids, one case with opposite polarity $K=-K$ and the other with the same polarity. In both cases, the propagation of beta functions is the same, whereas the effect on the phase advance is different; a splitting is observed. The solenoid cells were matched for $u=1/2$ and equal beta functions for a round beam. The propagation of phase advance in solenoid-type cells is given by
\begin{equation}
    \begin{split}
        \frac{d\mu_{1}}{ds} &= \frac{1-u}{\beta_{1x}} - \frac{R}{2}\sqrt{\frac{\beta_{1y}}{\beta_{1x}}}\sin\nu_{1}, \\
        \frac{d\mu_{2}}{ds} &= \frac{1-u}{\beta_{2y}} + \frac{R}{2}\sqrt{\frac{\beta_{2x}}{\beta_{2y}}}\sin\nu_{2}.
    \end{split}
    \label{eq:phaseadvancesolenoids}
\end{equation}
Solenoid-type transfer matrices have equal focusing in $x$ and $y$ planes, which makes the diagonal sub-matrices equal and presents itself as equal phase advances in $x$ and $y$ projections. However, we see that the modes phase advances split based on the phases of couplings and solenoid strength. Tune splitting is very useful for round beam axisymmetric focusing in coupled lattices, it is essential to avoid resonances.

\subsection{Periodic Strongly Coupled Lattices}
\begin{figure}[tbp]
	\centering
	\includegraphics[width=0.49\linewidth]{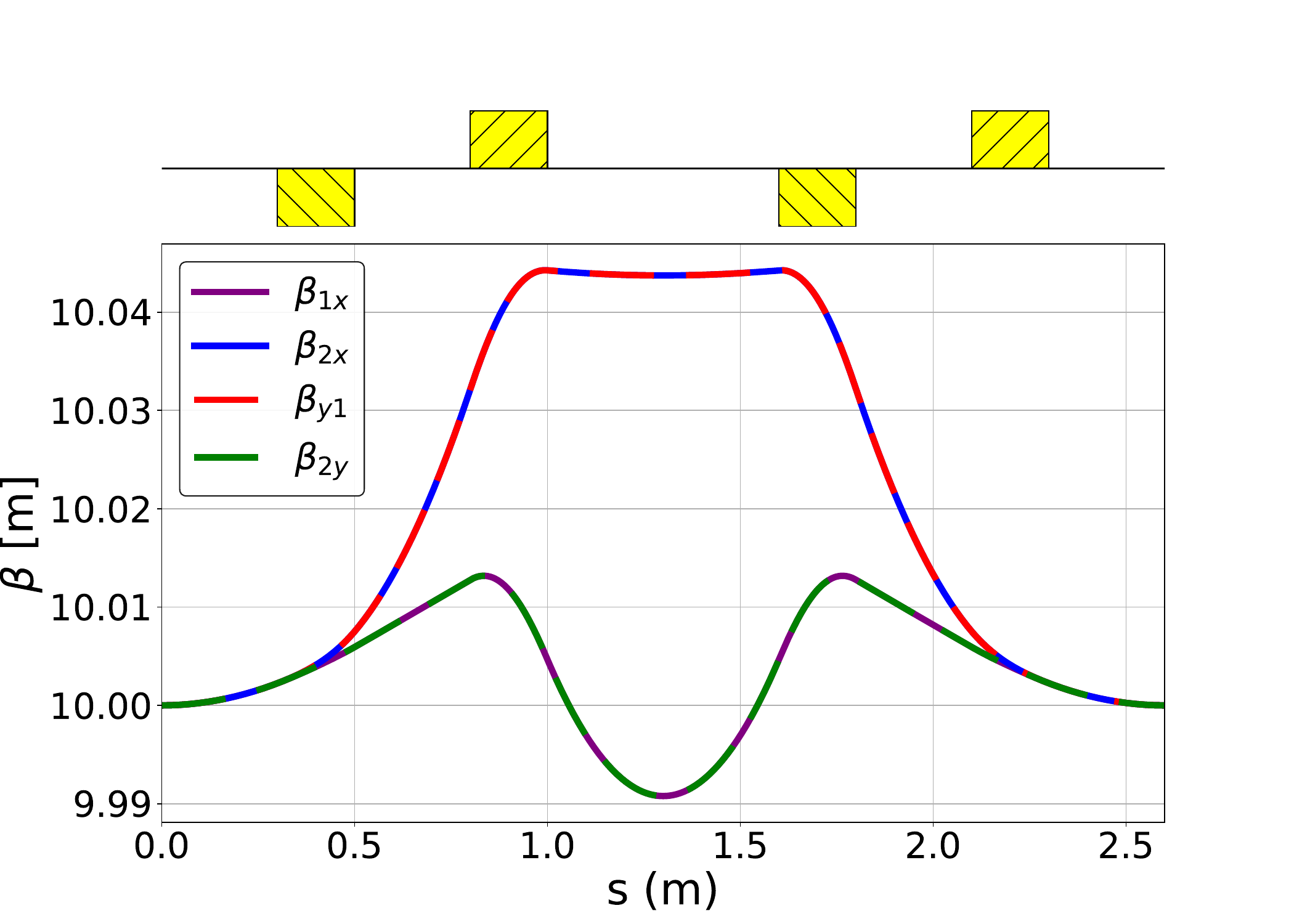}
	\includegraphics[width=0.49\linewidth]{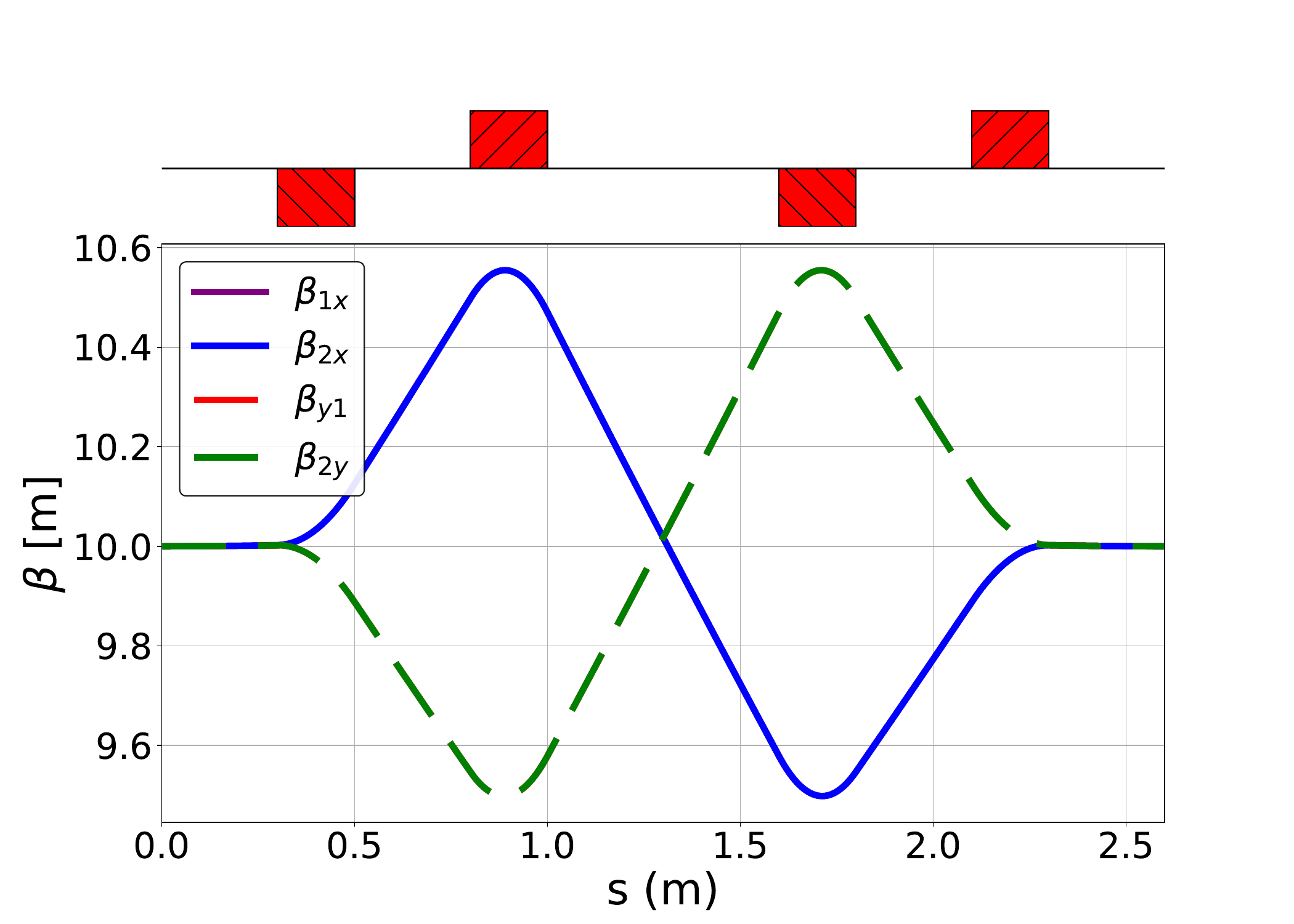}
	\includegraphics[width=0.49\linewidth]{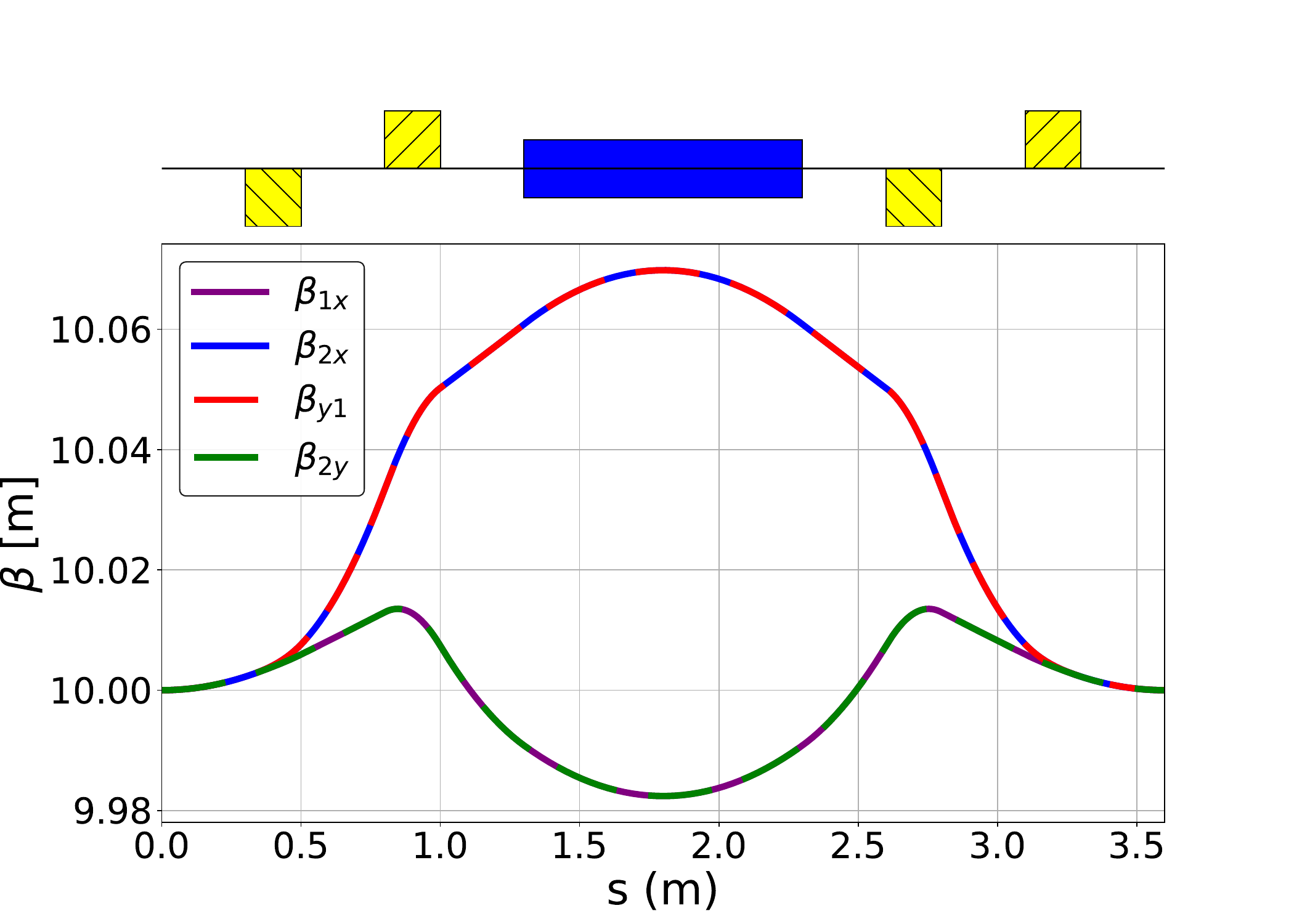}
	\includegraphics[width=0.49\linewidth]{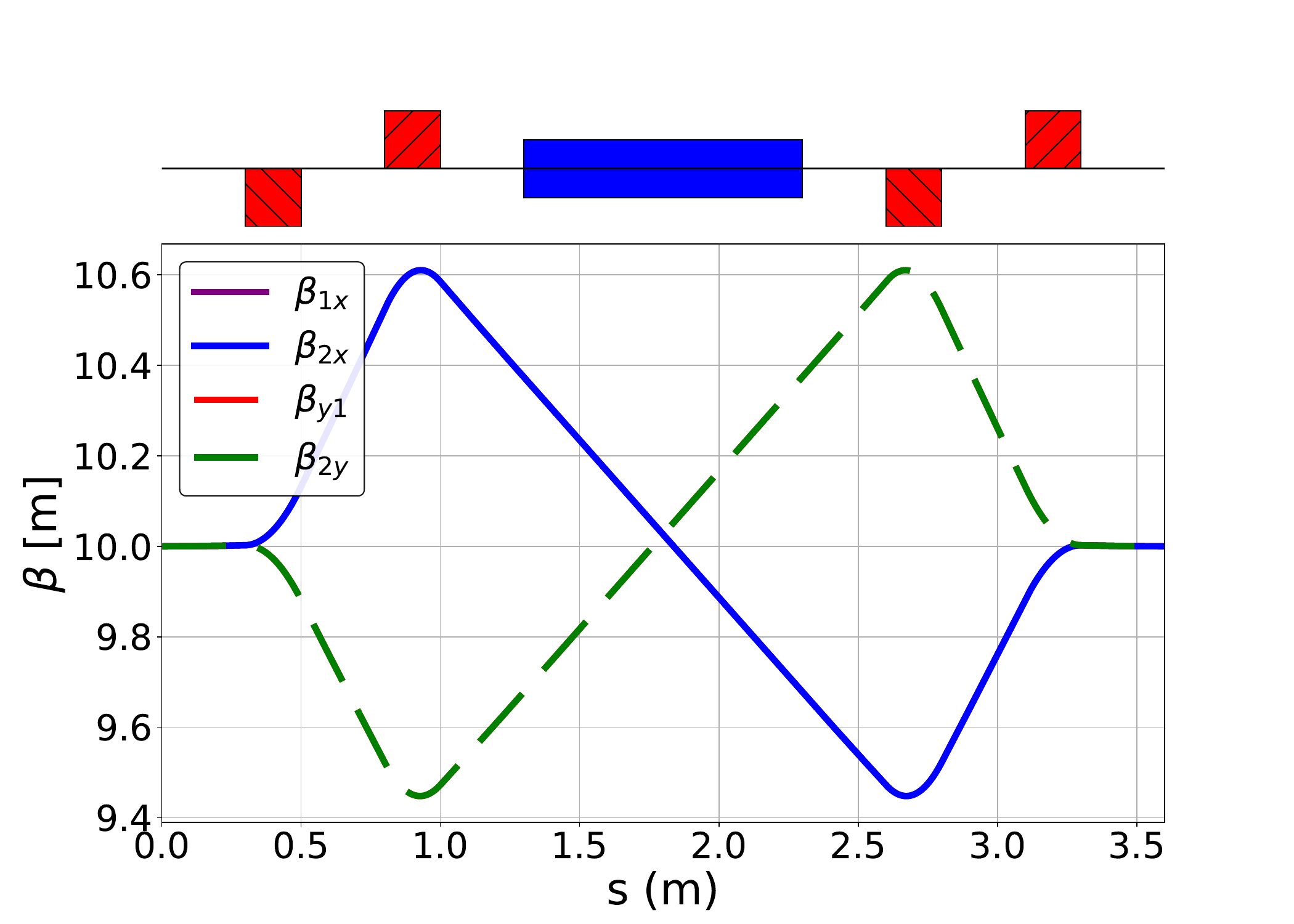}
	\caption{Quadrupole lattices with skew quadrupoles (left) and normal quadrupoles (right). Straight cells (top) and bending cells (bottom).}
	\label{fig:quadrupoledoublets}
\end{figure}

\begin{figure}[tbp]
	\centering
	\includegraphics[width=0.49\linewidth]{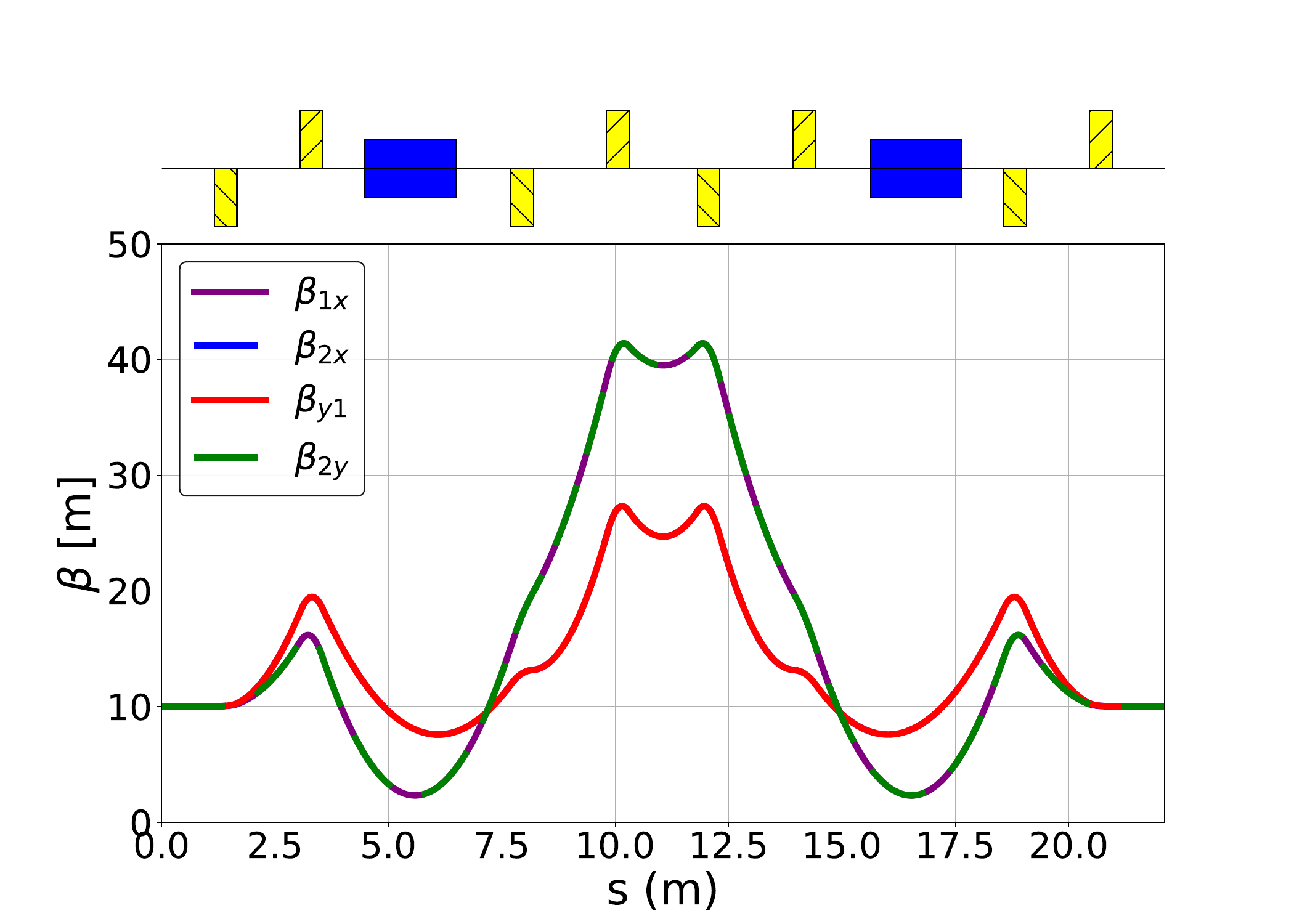}
	\includegraphics[width=0.49\linewidth]{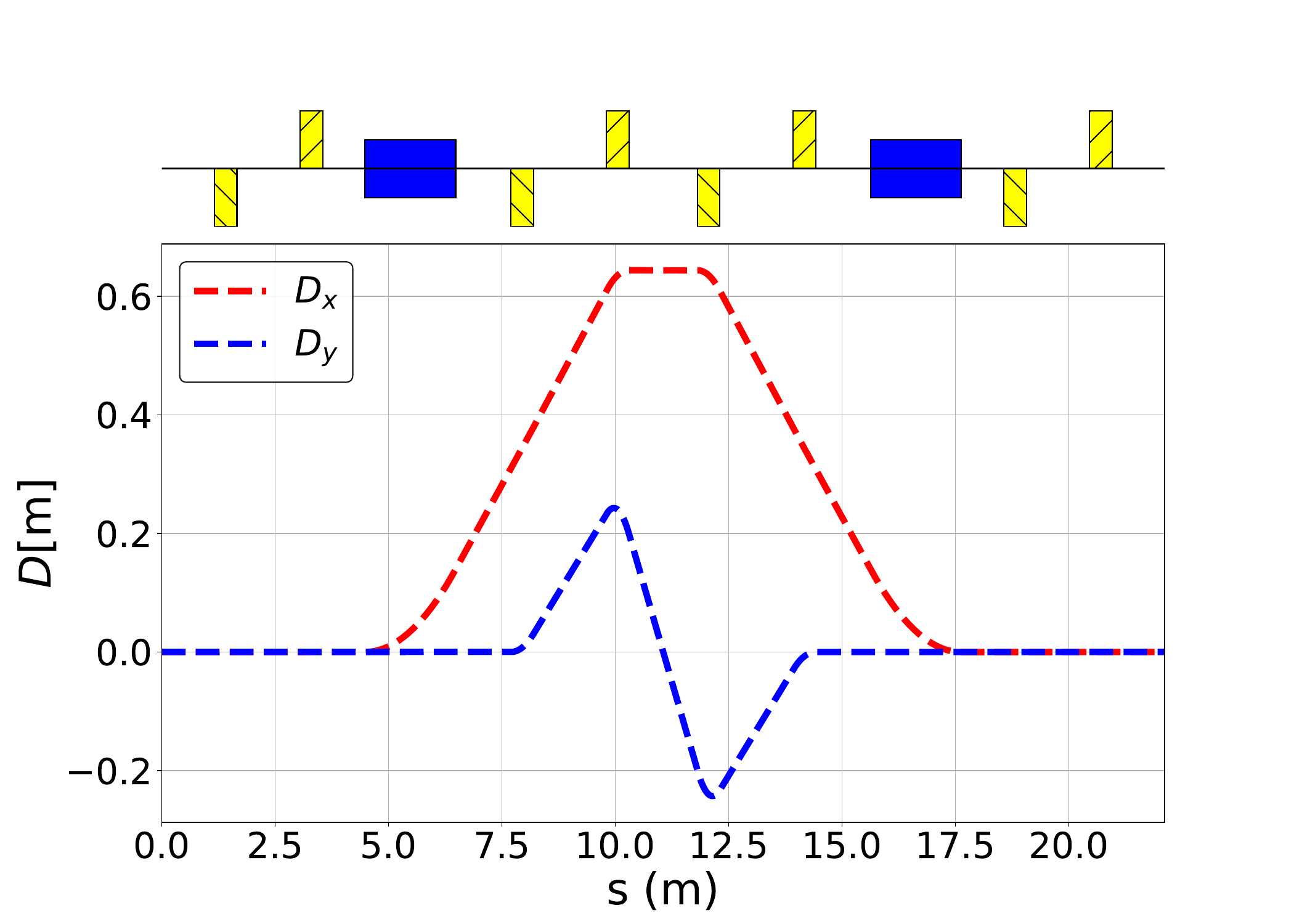}
	\includegraphics[width=0.49\linewidth]{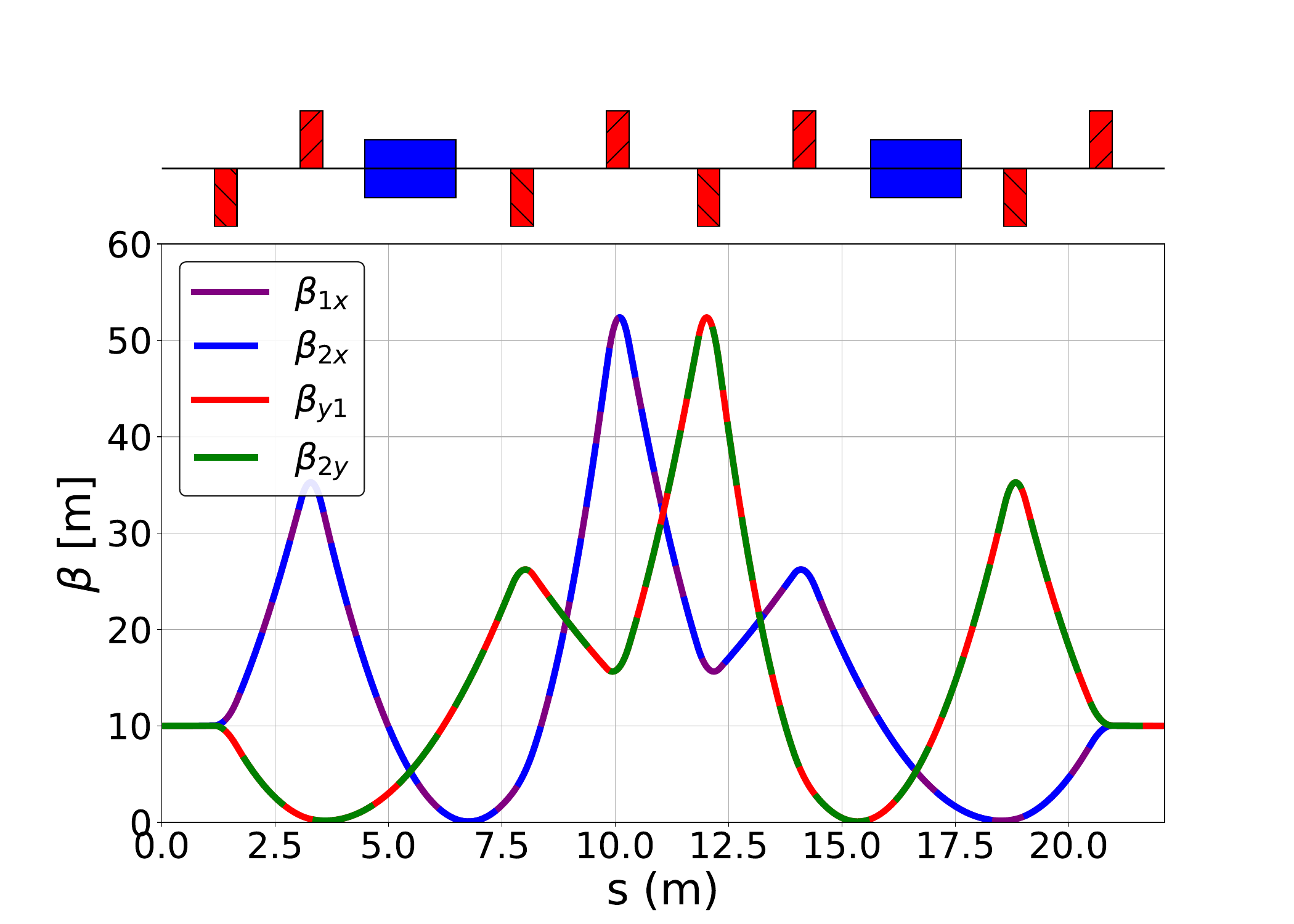}
	\includegraphics[width=0.49\linewidth]{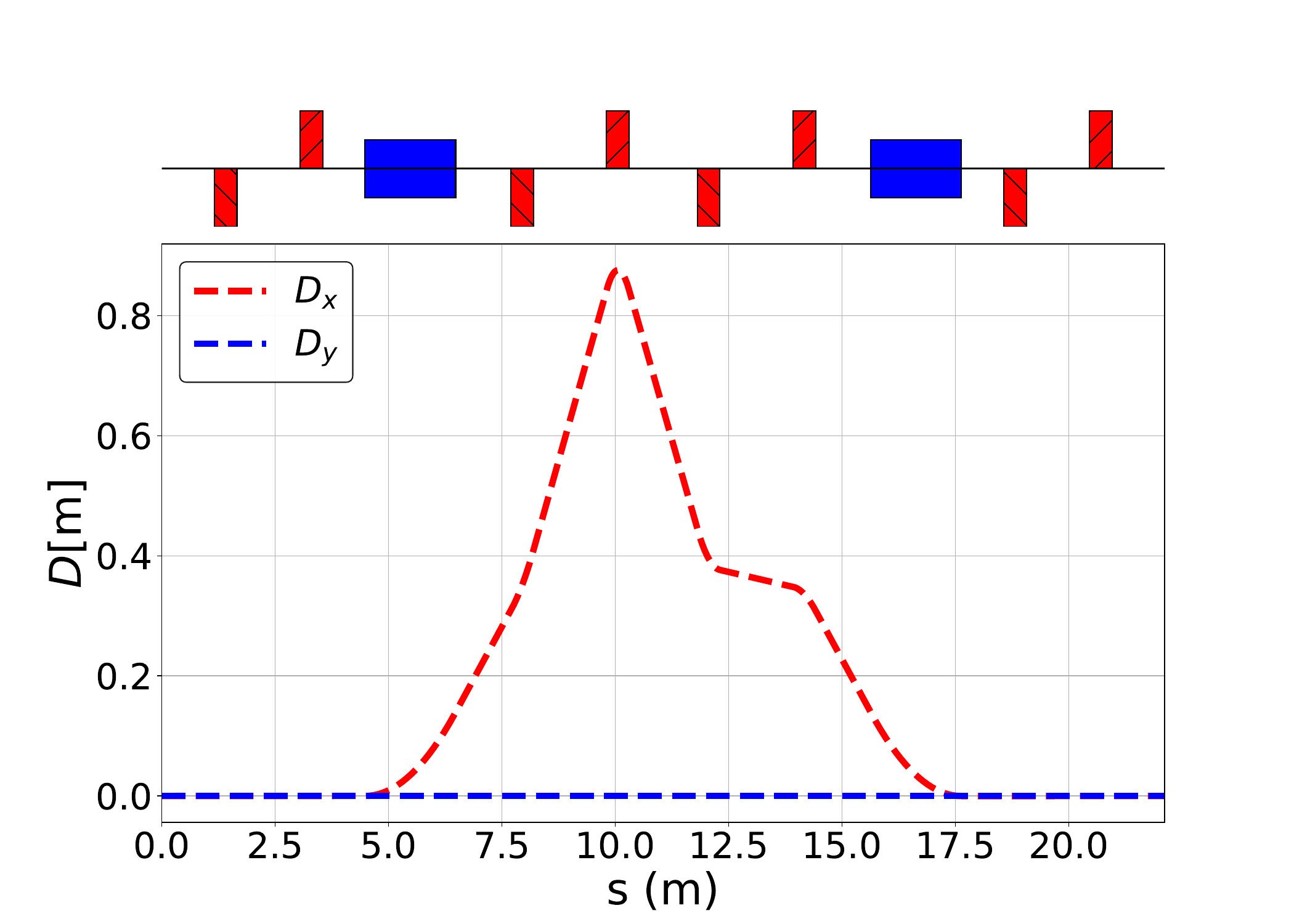}
	\caption{Optics functions for a Double Bend Achromat (DBA). Top row representing skew quadrupoles, bottom row representing normal quadrupoles. Left plots showing coupled beta functions, right plots showing dispersions.}
	\label{fig:dbas}
\end{figure}

In this section, we address the case of strongly coupled lattices and give examples that can be built based on $u=1/2$. The condition of $u=1/2$ can be realized using Derbenev's Adapter or magnetized beams, for example. To form full periodicity in coupled optics functions, a symmetric alternating gradient (SAG) design is required (see Appendix~\ref{Appendix:C}). The optics functions in Fig.~\ref{fig:quadrupoledoublets} and Fig.~\ref{fig:dbas} show that the beam maintains its roundness better in skew quadrupole settings than with normal quadrupoles, which is more favorable in terms of beam dynamics and collective effects.

Creating fully coupled rings with strong $x$-$y$ correlation can lead to better controlled collective effects and stability~\cite{burov2013circular}. The matching method mentioned above is used to create fully coupled ring lattices with $u=1/2$ and $\nu_{1,2}=\pi /2$. Achieving complete periodicity requires combined function magnets with both dipole and quadrupole components, known as indexed dipoles, with a field index, $n=1/2$. The choice of the coupling phases, $\nu_{1,2}=\pi/2$, leads to a round beam distribution in real space aligned with the $x$-$y$ axis.

\begin{figure}[tbp]
	\centering
	\includegraphics[width=0.49\linewidth]{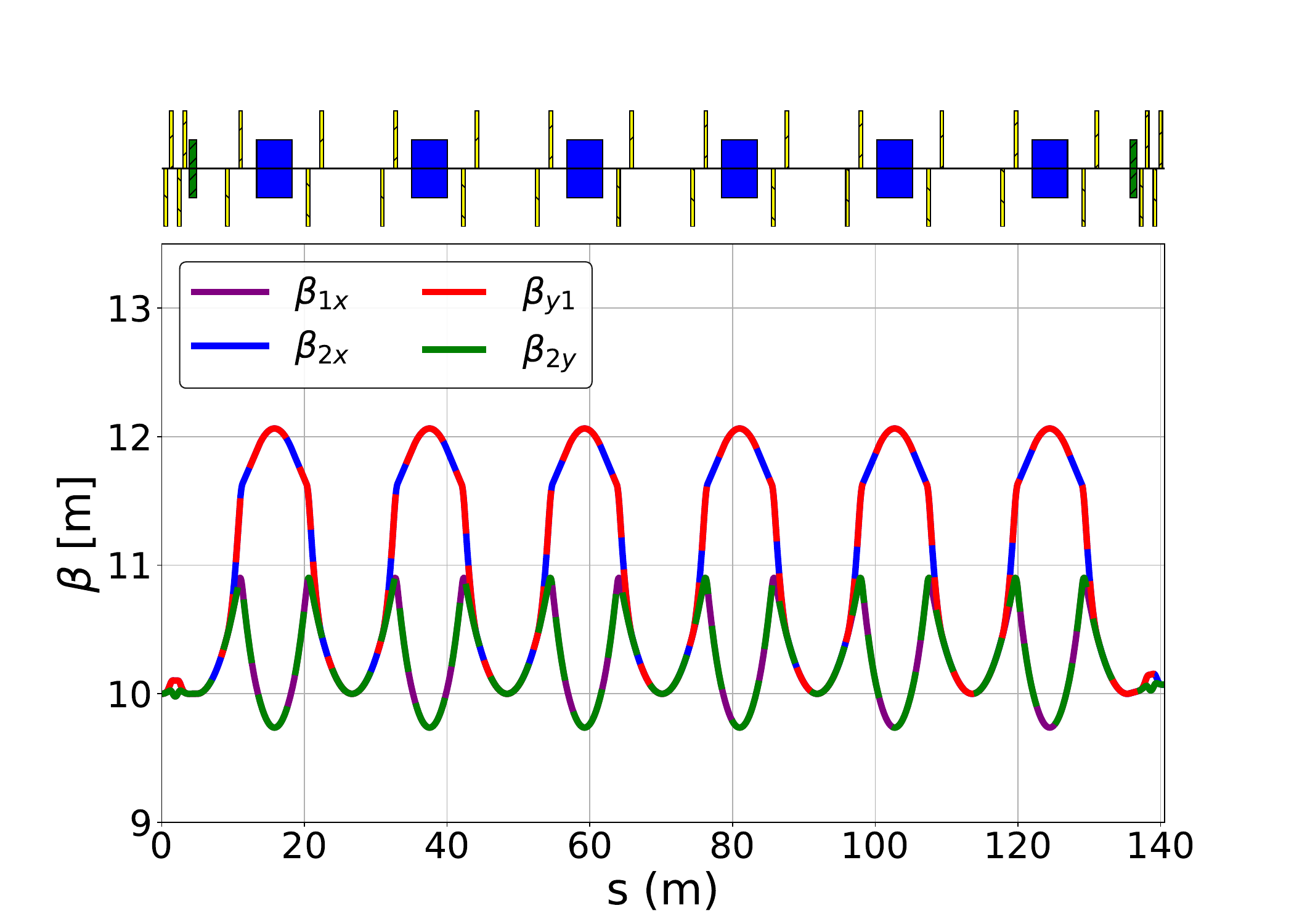}
	\includegraphics[width=0.49\linewidth]{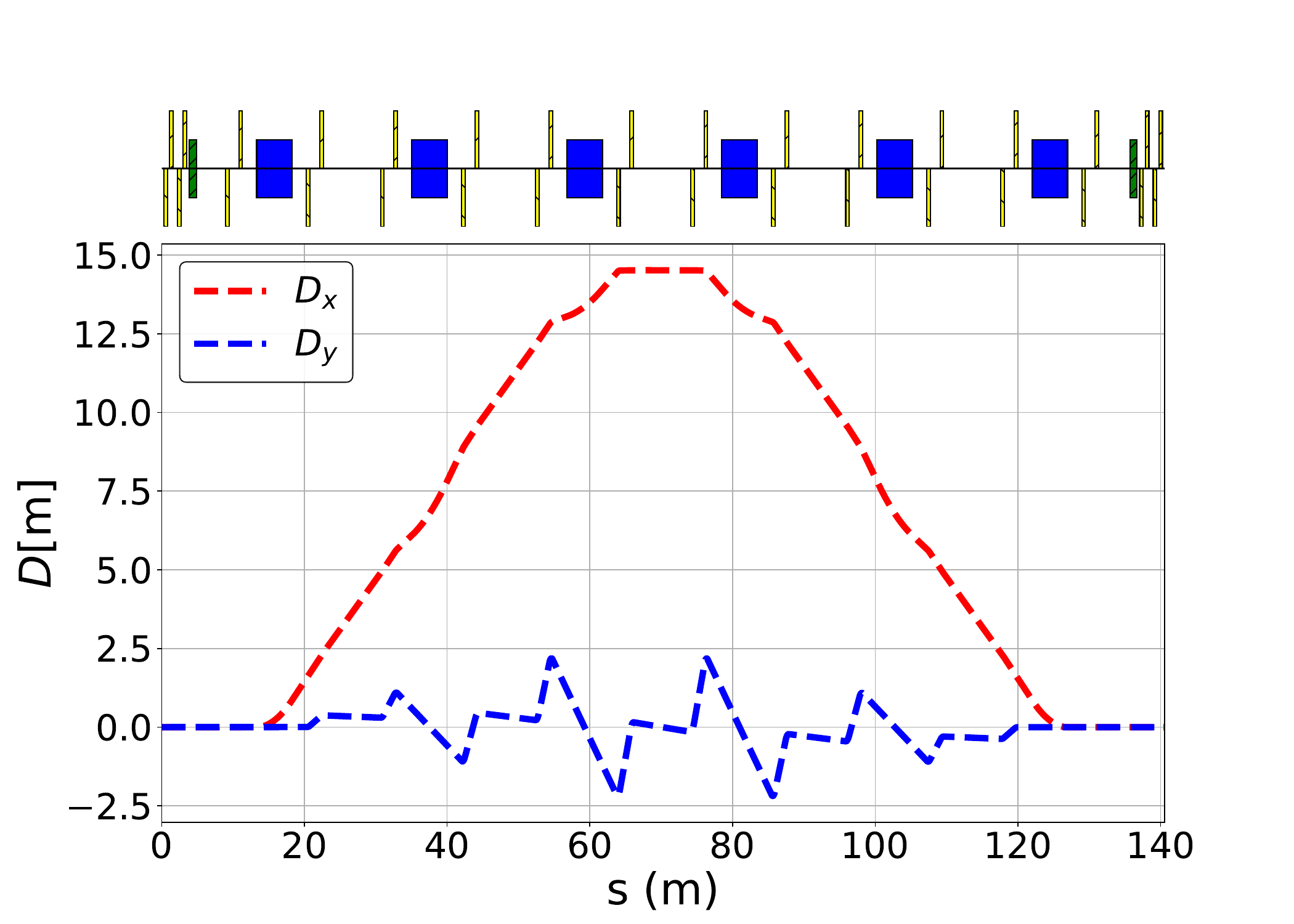}
	\includegraphics[width=0.49\linewidth]{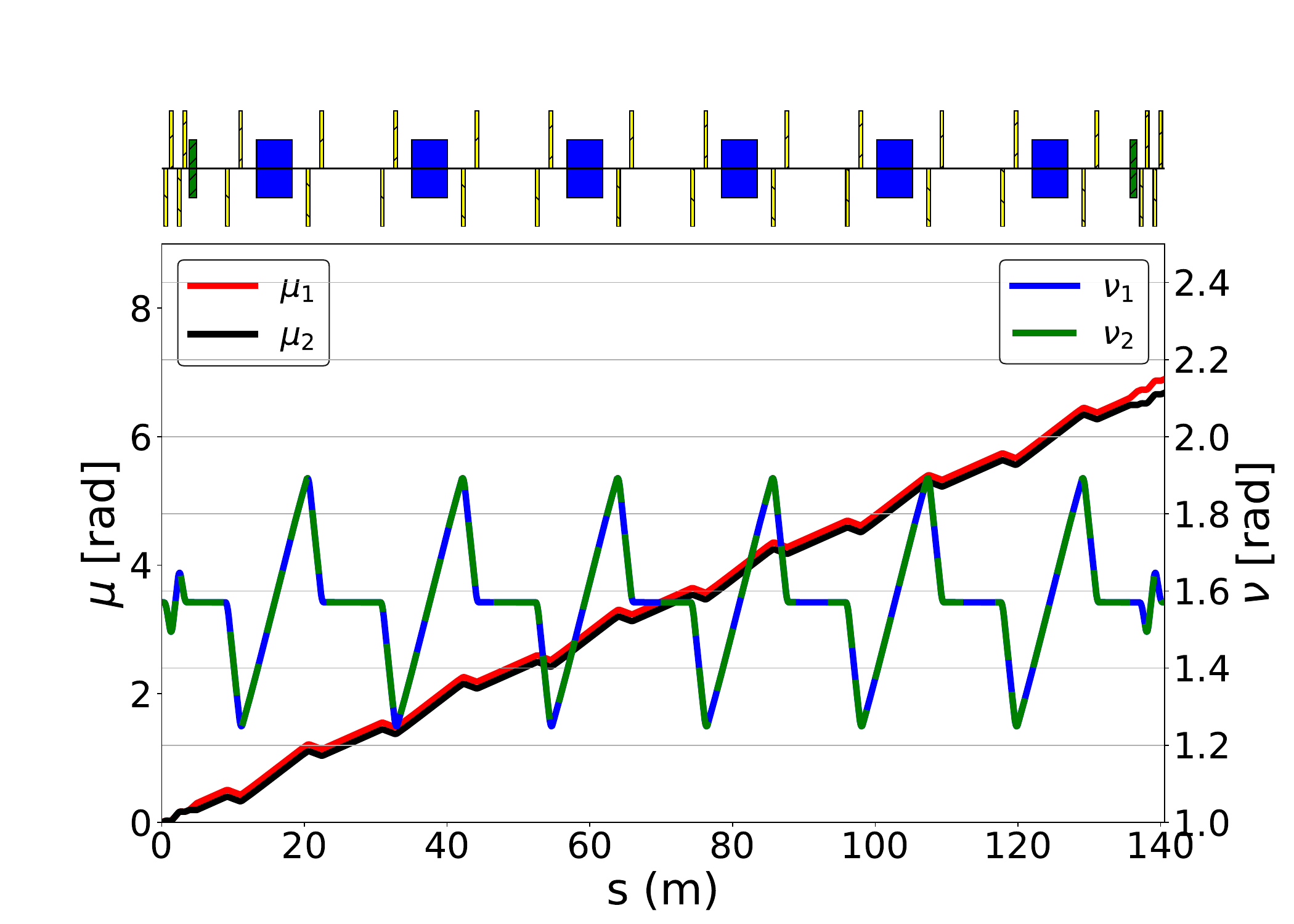}
	\includegraphics[width=0.49\linewidth]{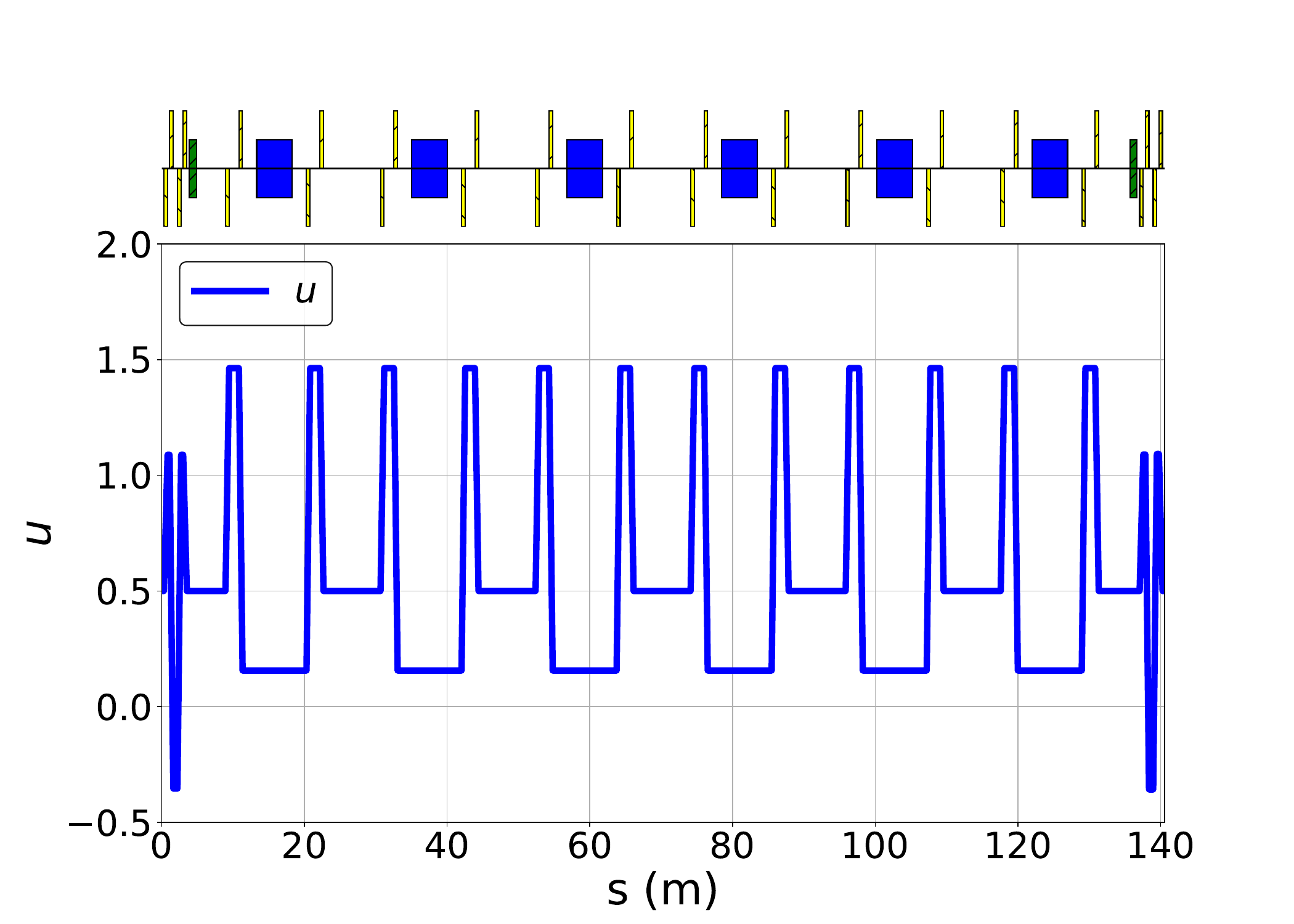}
	\caption{Periodic skew-quadrupole achromat cells showing straight sections and periodic bends with fixed phase advance. Coupled optics functions plot (top-left), dispersions in both planes (top-right), phase advances $\mu_{1,2}$ and phases of couplings $\nu_{1,2}$ (bottom-left), and coupling strength parameter $u$ (bottom-right).}
	\label{fig:achromatdesignwithphases}
\end{figure}

Two basic ring cells were designed and investigated. 
In the first design, we have used the achromatic condition achieved by fixing the phase advance for periodic bend cells to $\pi/3$. Setting the phase advance to $\pi/3$ creates a full achromat by suppressing the dispersion in both planes. The use of a solenoid in a straight section splits the tunes and breaks the degeneracy of the two planes. It also helps to avoid coupling resonances. The Achromat case is shown in Fig.~\ref{fig:achromatdesignwithphases}. With a dipole bending angle of $20^{\circ}$, the design includes three straight sections formed with a super periodicity of $3$. RF cavities can be added in the straight sections for acceleration, and a large solenoid can be used in one of the straight sections for beam cooling if needed. 

\begin{figure}[tbp]
	\centering
	\includegraphics[width=0.49\linewidth]{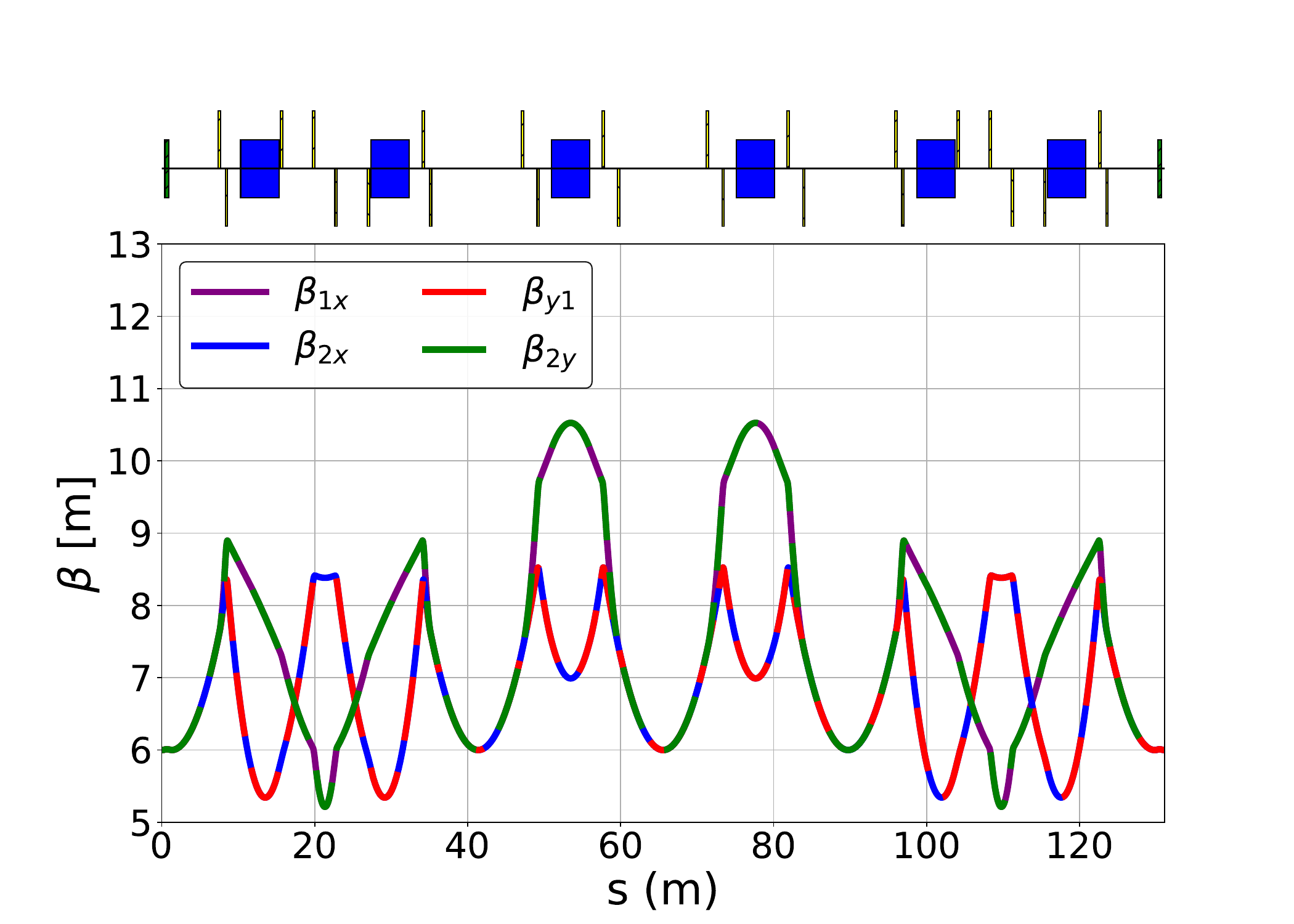}
	\includegraphics[width=0.49\linewidth]{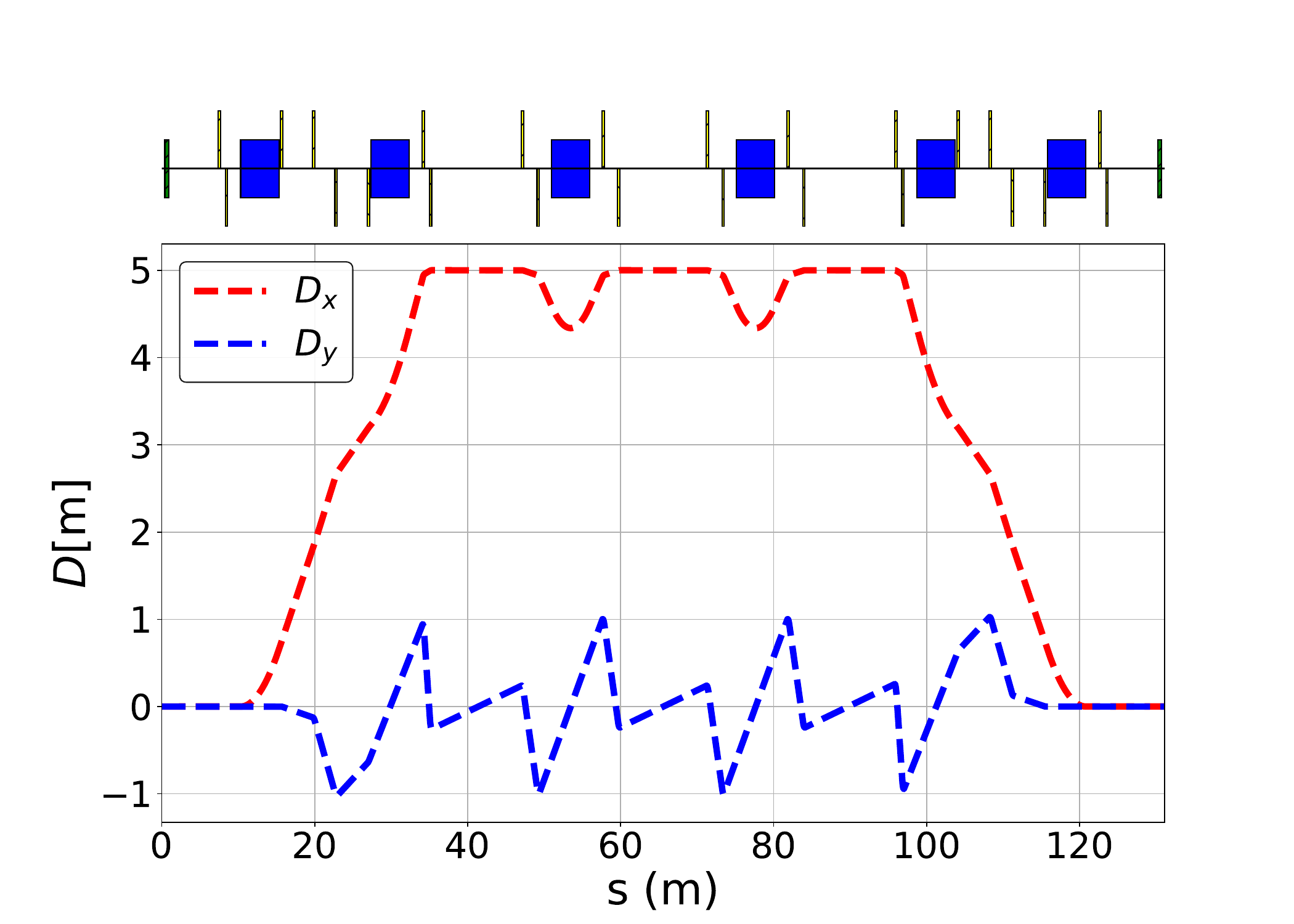}
	\includegraphics[width=0.49\linewidth]{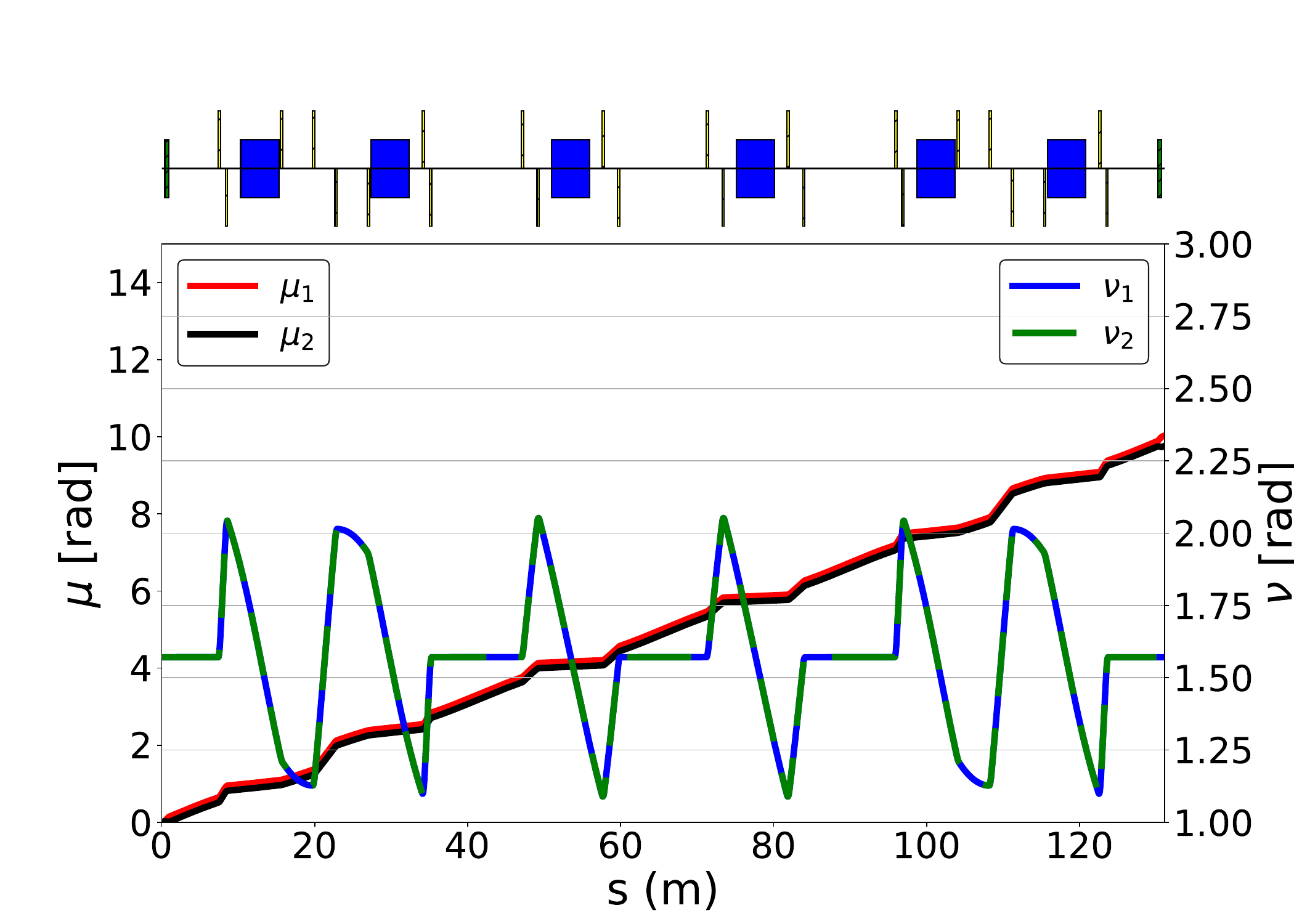}
	\includegraphics[width=0.49\linewidth]{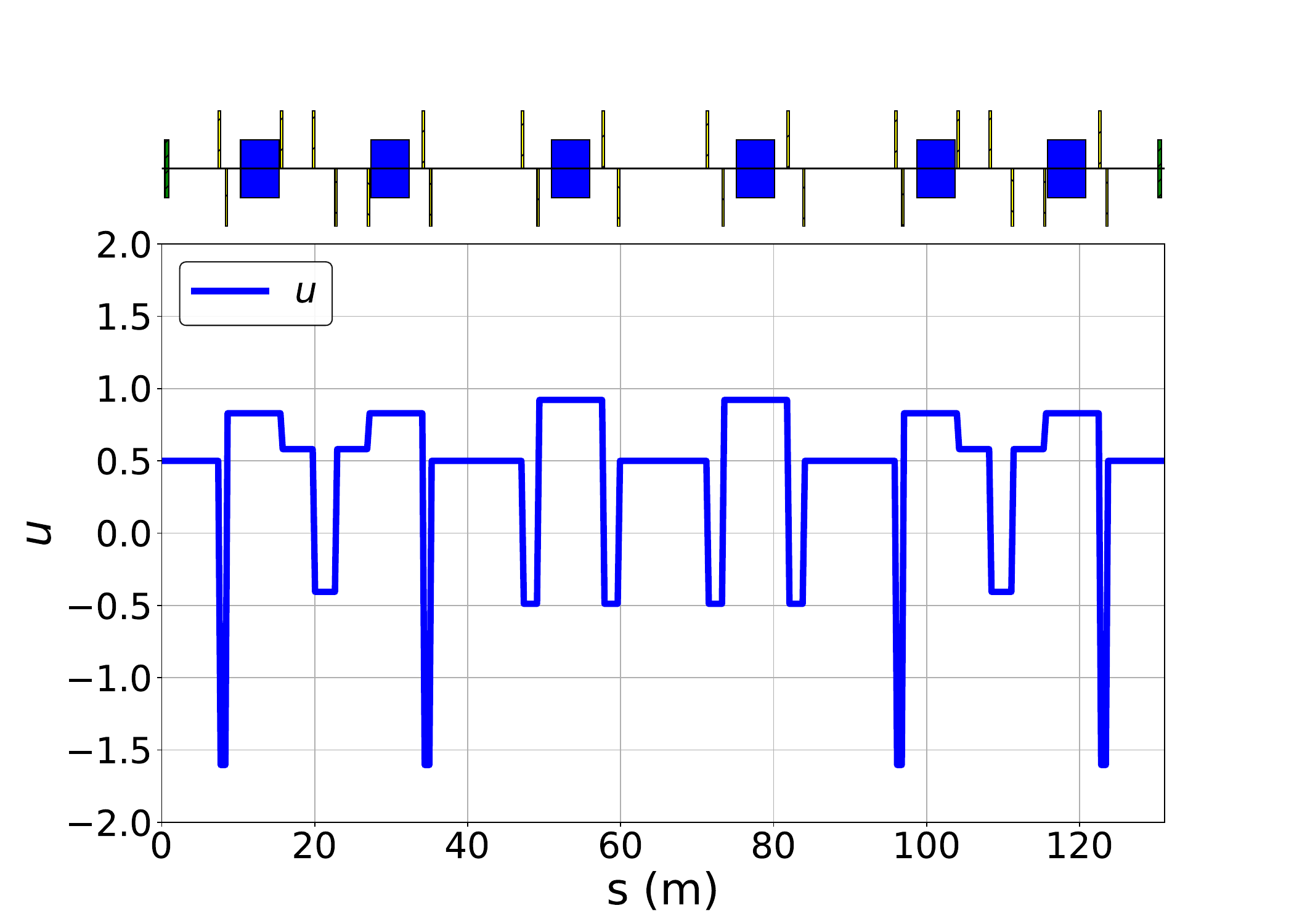}
	\caption{Periodic skew-quadrupole non-achromatic cells showing straight sections and periodic bends with matched optics functions. Coupled optics functions plot (top-left), dispersions in both planes (top-right), phase advances $\mu_{1,2}$ and phases of couplings $\nu_{1,2}$ (bottom-left), and coupling strength parameter $u$ (bottom-right).}
	\label{fig:nonachromatdesignskew}
\end{figure}

In the second design, we used the optics transformation method to find a periodic dispersion function in addition to periodic optics functions, where a fixed phase advance is not assumed. The periodic functions for this case are: $\beta_{1x},\beta_{2x},\beta_{1y},\beta_{2y}=6.0$ meters. The periodic dispersions are $D_{x}=5.0$\,m, $D_{y}=0.0$\,m, $D_{x}'=0.0$, and $D_{y}'=0.00$. The dipole bending angle is $\theta = 30^{\circ}$. The dispersion suppression and creation sections are matched separately and involve only skew quadrupoles. Figure~\ref{fig:nonachromatdesignskew} shows the mentioned design without including the full periodic bend cells.    

\begin{figure}[tbp]
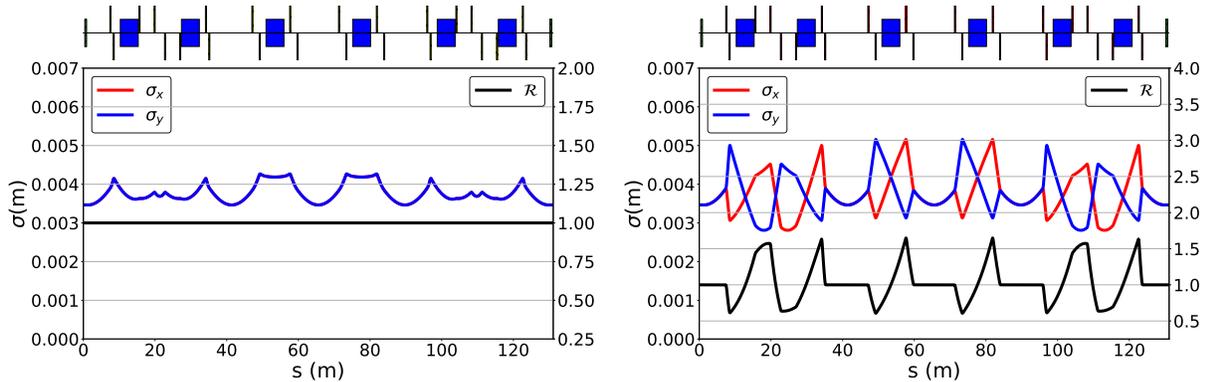

	\centering
	\includegraphics[width=0.49\linewidth]{Assets/comparingrmses/skewquadrupolecaseratios.png}
	\includegraphics[width=0.49\linewidth]{Assets/comparingrmses/normalquadrupolecaseratios.png}
	\caption{Comparison of rms beam sizes and their ratios along a skew-quadrupole super-period (left) and a normal-quadrupole super-period (right). The results are from beam tracking simulations using ImpactX.}
	\label{fig:comparingrmses}
\end{figure}

The general behavior of the $\beta$ functions in the two cases with normal quadrupoles and skew quadrupoles, as seen in Fig.~\ref{fig:dbas}, is similar for $u=1/2$. Using an uncoupled lattice with a strongly correlated beam does not change the strength of coupling, as shown in Eq.~\eqref{Eq:Fromcoupledtocoupleduncmatrix}. The transformation process becomes equivalent for both modes $1$ and $2$, suggesting $\beta_{1x}=\beta_{2x}$ and $\beta_{1y}=\beta_{2y}$. However, for skew quadrupoles we have $\beta_{1x}=\beta_{2y}$ and $\beta_{2x}=\beta_{1y}$, which corresponds to equal off-mode optics functions (mode $1$ projection on $y$ and mode $2$ projection on $x$). This results in interesting beam size correlations in the case of equal mode emittances, $\epsilon_{1}=\epsilon_{2}=\epsilon_{0}$, and uncoupled transport with normal quadrupoles: $\sigma_{x}^{2}=2\epsilon_{0}\beta_{1}$, $\sigma_{y}^{2}=2\epsilon_{0}\beta_{2}$, which yields a beam size ratio $\sigma_{x}/\sigma_{y} = \sqrt{\beta_{1}/\beta_{2}}$. Interestingly, for the skew quadrupole case, the beam sizes will be $\sigma_{x}^{2}=\epsilon_{0}(\beta_{1}+\beta_{2})$ and $\sigma_{y}^{2}=\epsilon_{0}(\beta_{2} + \beta_{1})$, which yields $\sigma_{x}/\sigma_{y}=1$. The difference between the two cases is that the uncoupled lattice transformation yields no rotation in the $(x,y)$ phase space. However, the coupled lattice transformation will produce an $(x,y)$ space rotation. This comparison is shown in Fig.~\ref{fig:comparingrmses} using ImpactX simulations~\cite{huebl2022next}, where the rms beam size ratio changes with normal quadrupoles but remains constant in a skew-quadrupole-based lattice. 

\subsection{A Fully Coupled Periodic Ring}
\begin{figure}[tbp]
	\centering
	\includegraphics[width=0.49\linewidth]{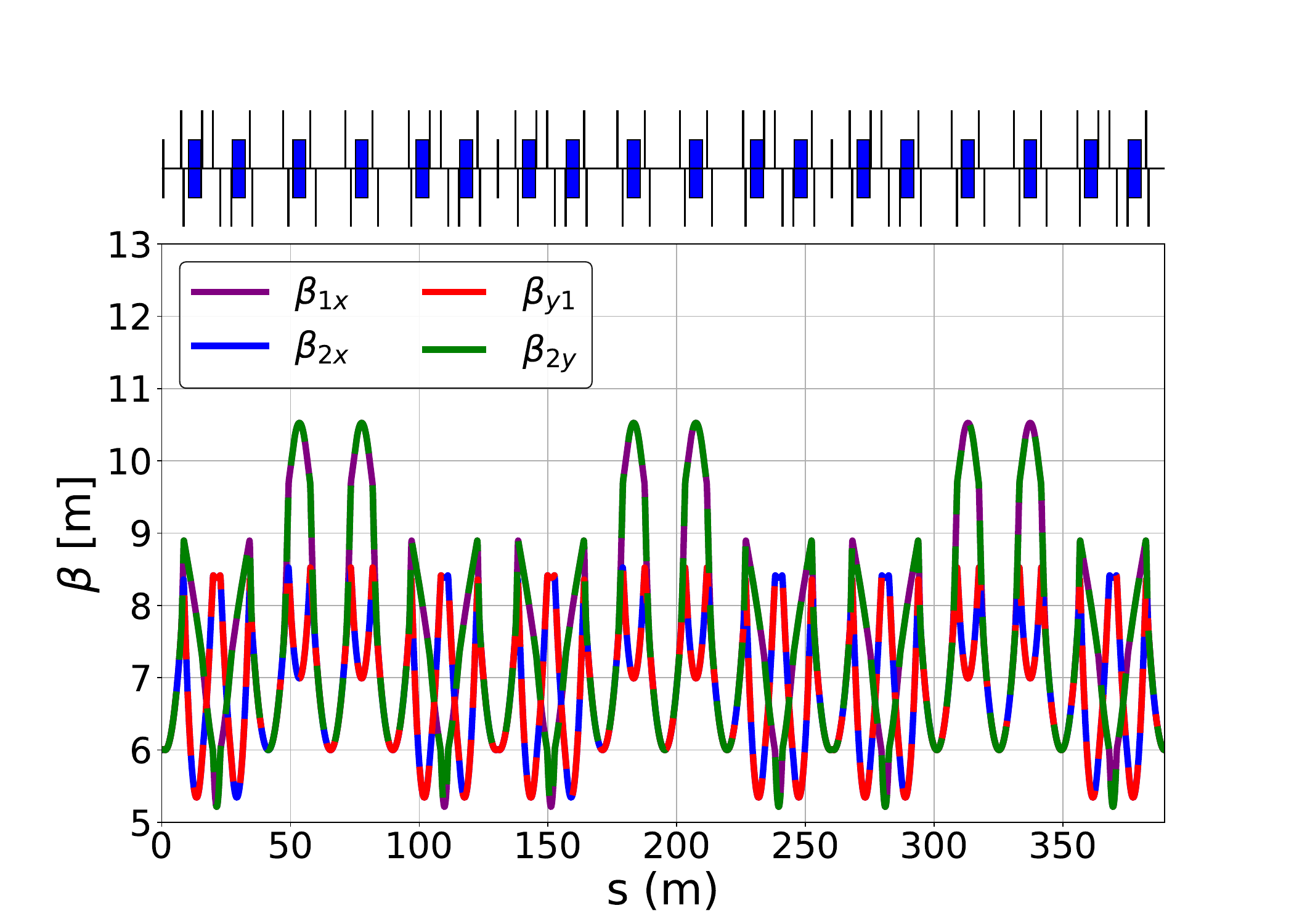}
	\includegraphics[width=0.49\linewidth]{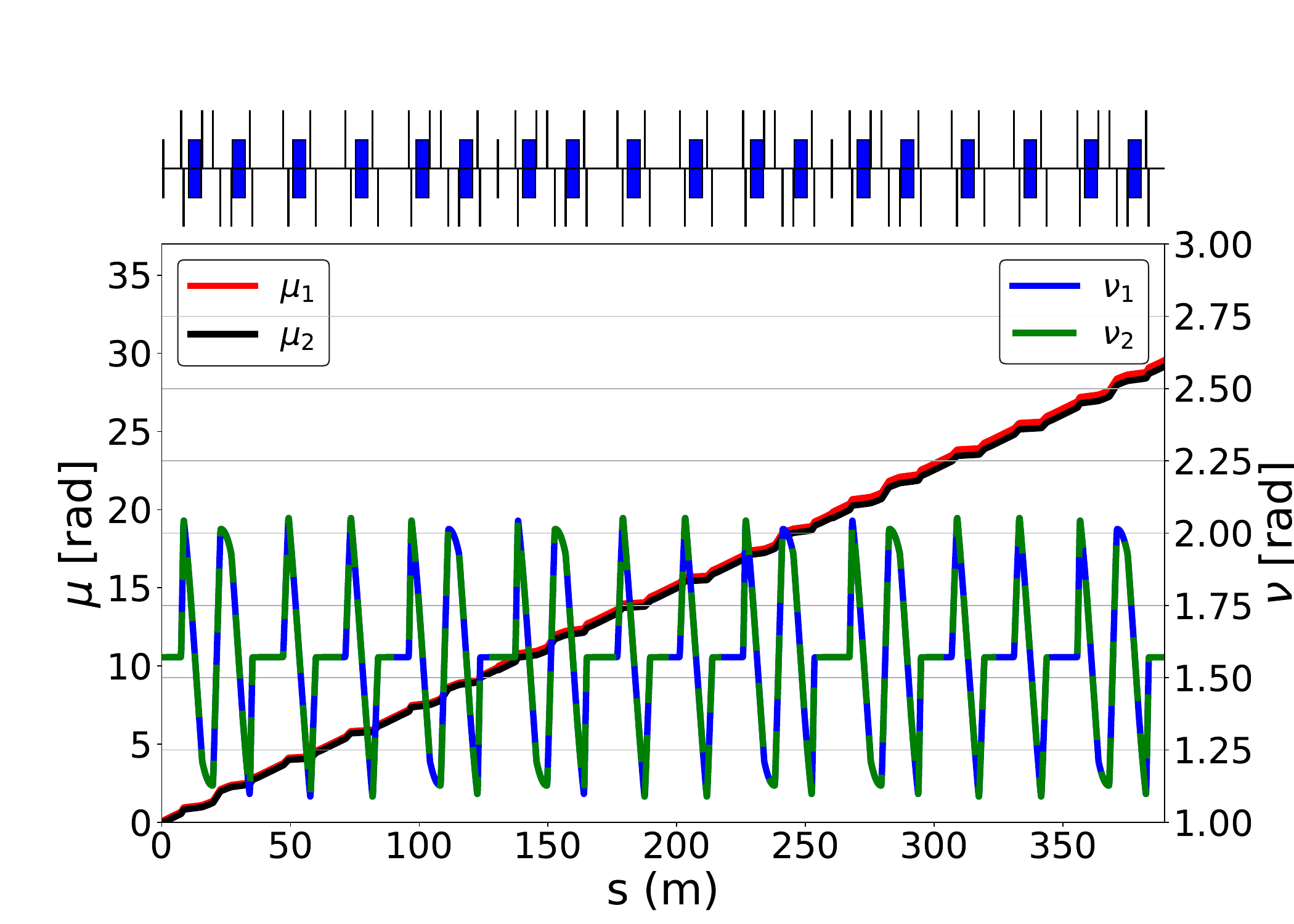}
	\includegraphics[width=0.49\linewidth]{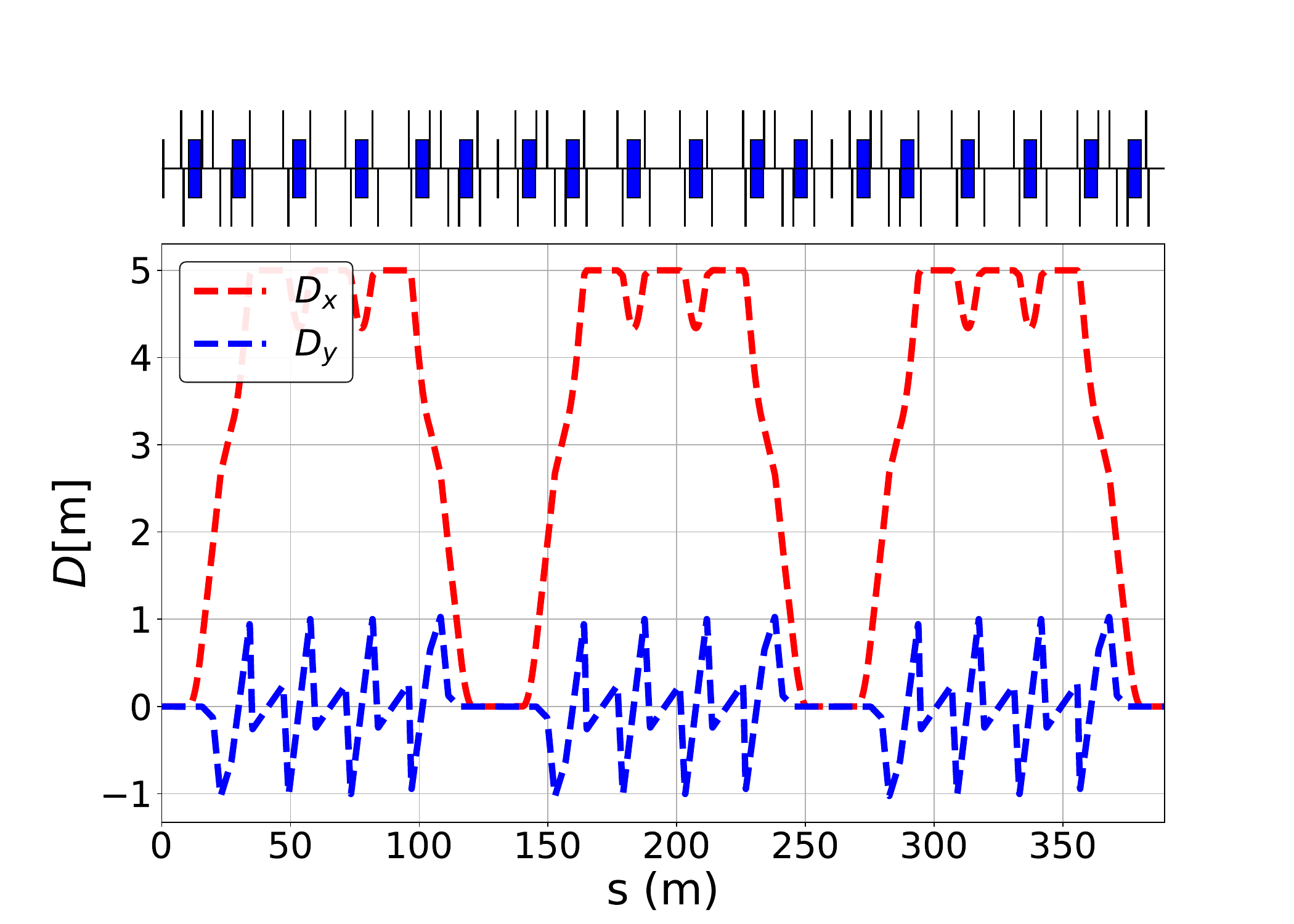}
	\includegraphics[width=0.49\linewidth]{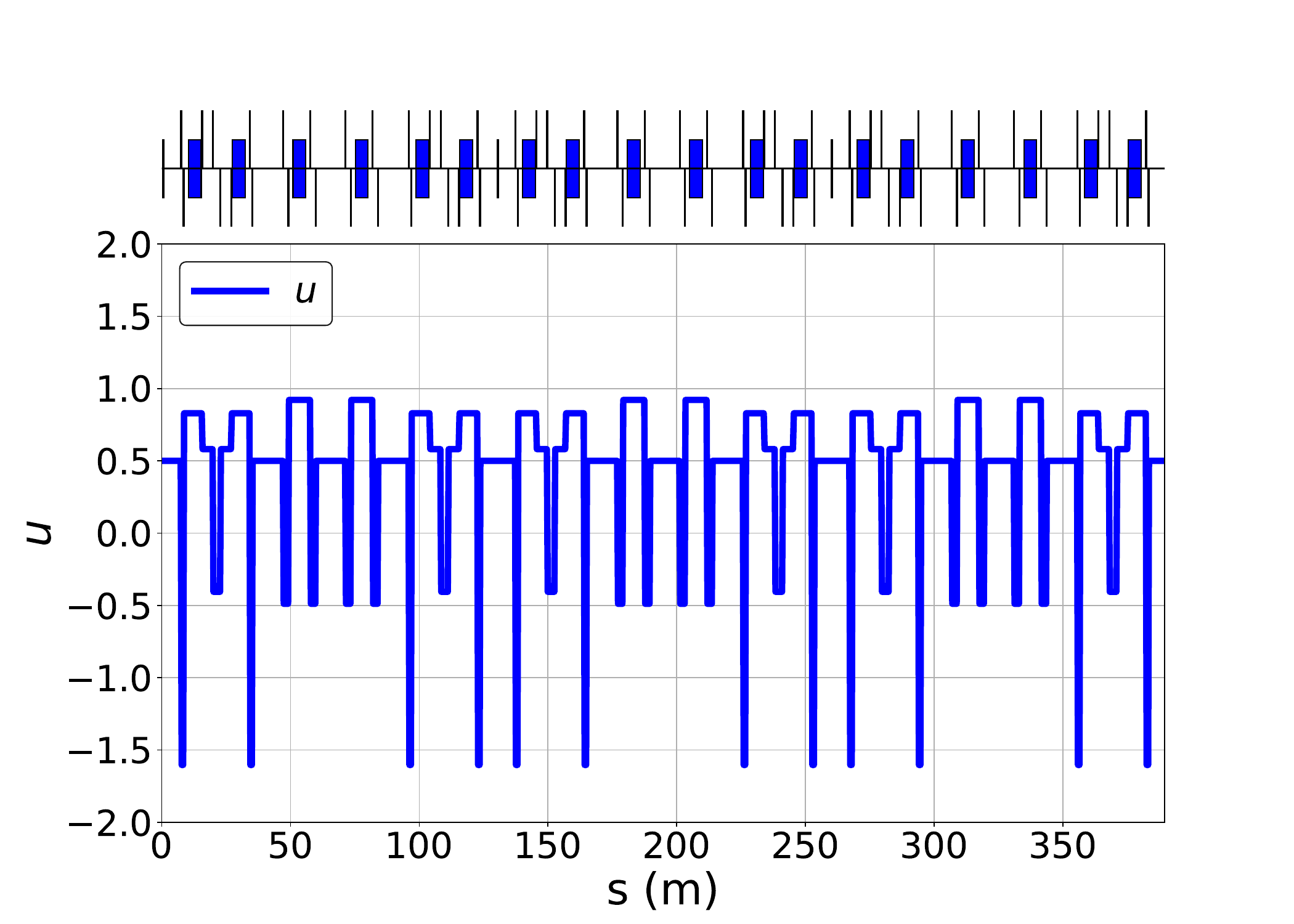}
	\caption{Fully coupled ring with skew quadrupoles and solenoids. Coupled beta functions (top-left), phases of coupling (top-right), dispersion functions (bottom-left), and strength of coupling (bottom-right).}
	\label{fig:fullcoupledring}
\end{figure}

The ring constructed in this section involves skew quadrupoles, solenoids, and combined function dipoles, resulting in the fully periodic optics shown in Fig.~\ref{fig:fullcoupledring}. Solenoids are used to produce and control tune splitting, and could be used for cooling sections as well. Due to the presence of skew quadrupoles in dispersive sections, dispersion exists and needs to be suppressed in both planes. With the SAG design approach mentioned earlier, the coupling phases are also periodically transported. The initial coupled $\beta$ functions are $\beta_{i}=6.0$ meters in all planes, and all initial $\alpha$ functions are zeros. The initial coupling phases are $\nu_{1,2} = \pi/2$. 

\section{Solutions for Beam Characterization and Diagnostics}
While we can design and match the optics in coupled dynamics, it is important to consider beam characterization and diagnostics methods for coupled beams and lattices; for example, to distinguish between eigenmode emittances for fully coupled lattices or special configurations such as Derbenev's Adapter. 
It is challenging to come up with a general diagnostics method for extracting eigenmode emittances from observables such as beam sizes. From our studies, we have identified the coupling strength, $u$, as an important parameter; see Section~\ref{subsec:specialsoln}. Through matching procedures, we have found that $u=1/2$, corresponding to fully coupled configurations, imposes certain constraints on coupled functions. These simplifications lead to possible methods of extracting the eigenmode emittances from beam observables. Eigenmode emittances can be calculated generally using the full knowledge of the beam matrix as shown in Appendix~\ref{appendix:B}, which yields:
\begin{equation}
	\begin{split}
		\epsilon_{1}&= \frac{1}{2}\sqrt{-Tr[(\Sigma S)^{2}] + \sqrt{Tr[(\Sigma S)^{2}]^{2} - 16\epsilon_{4D}^{2}}}, \\
		\epsilon_{2}&= \frac{1}{2}\sqrt{-Tr[(\Sigma S)^{2}] - \sqrt{Tr[(\Sigma S)^{2}]^{2} - 16\epsilon_{4D}^{2}}}.
	\end{split}
\end{equation}
Here $\Sigma$ is the $4D$ beam matrix, $S$ the symplectic unit matrix, $\epsilon_{4D}$ the 4D emittance, and $\epsilon_{4D}=\epsilon_{1}\epsilon_{2}$. However, this requires knowledge of the $4D$ emittance and the full $4D$ beam matrix. To find these parameters, we can exploit the symmetries around $u=1/2$ coupled beams. These provide an interesting method leading to the extraction of eigenmode emittances through observables in the accelerator system. It is important to point out one significant difference between conventional uncoupled dynamics and coupled dynamics. That is, uncoupled dynamics do not project a tilted ellipse on the $(x,y)$ plane, while coupled optics can project tilted ellipses, since it is based on projections from eigenplanes. These projections are controlled by the phases of coupling, $\nu_{1,2}$. The parametrization of the beam matrix is presented in~\cite{lebedev2010betatron}, it is also given in Appendix~\ref{appendix:B}. 

\subsection{Coupled Beam, Uncoupled Lattice ($u=1/2$)}
\begin{figure}[tbp]
	\centering
	\includegraphics[width=0.49\linewidth]{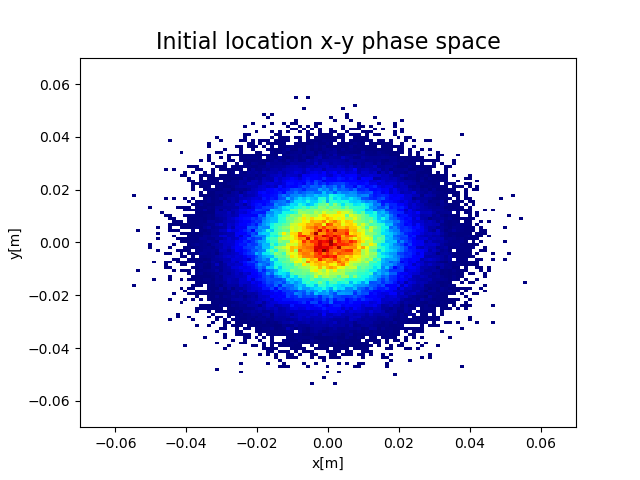}
	\includegraphics[width=0.49\linewidth]{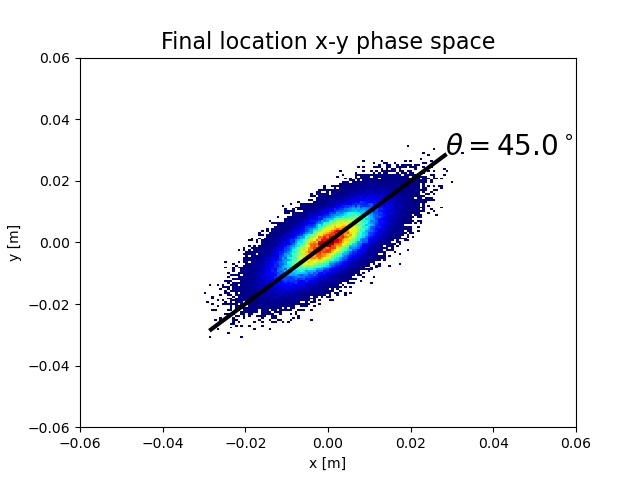}
	\caption{$(x,y)$ distributions at initial location (left) and final location (right).}
	\label{fig:xydistributionsdiagnostics}
\end{figure}

Applying the simplifications that arise from $u=1/2$ case, for a downstream transfer matrix that is uncoupled: $\beta_{1x}=\beta_{2x}=\beta_{1}$, $\beta_{1y}=\beta_{2y}=\beta_{2}$. For simplicity, the measurement can be taken at the beam waist location where $\alpha_{1x}=\alpha_{2x}=\alpha_{1}=0$ and $\alpha_{1y}=\alpha_{2y}=\alpha_{2}=0$. The initial location measurement is taken when the beam's transverse cross-section is aligned with $x-y$ plane with phases of coupling $\nu_{1,2}=\pi/2$. If the downstream lattice does not include any coupling elements, then the coupling strength parameter is always conserved, $u=1/2$. This leads to the important rms quantities:
\begin{equation}
	\begin{split}
		\sigma_{x}^{2} &= \beta_{1}(\epsilon_{1} + \epsilon_{2}), \qquad \sigma_{y}^{2}= \beta_{2}(\epsilon_{1} + \epsilon_{2}), \\
		\sigma_{xy} &= \sqrt{\beta_{1}\beta_{2}}(\epsilon_{1}\cos\nu_{1} + \epsilon_{2}\cos\nu_{2}).
	\end{split}
\end{equation}
Computing the projected emittance, $\epsilon_{x}$ as in uncoupled dynamics, yields:
\begin{equation}
	\begin{split}
		\epsilon_{x} &= \frac{1}{2}(\epsilon_{1} + \epsilon_{2}), \quad \epsilon_{y}=\frac{1}{2}(\epsilon_{1} + \epsilon_{2}).
	\end{split}
\end{equation}
This suggests that the sum of eigenmode emittances can be extracted from projected emittances. In order to find individual emittances, a difference of emittances is needed. With the appropriate choice of optics, the beam matrix at the initial location can be simplified:
\begin{equation}
	\Sigma_{i} = \begin{pmatrix}
		\sigma_{x}^{2} & 0 & 0 & \sigma_{xy'} \\
		0 & \sigma_{x'}^{2} & \sigma_{yx'} & 0 \\
		0 & \sigma_{yx'} & \sigma_{y}^{2} & 0 \\
		\sigma_{xy'} & 0 & 0 & \sigma_{y'}^{2}
	\end{pmatrix}.
\end{equation}
With a dedicated region in an accelerator system with known elements, the change in beam sizes can be expressed in terms of the transfer map $\mathcal{M}=\mathrm{diag}(M,N)$:
\begin{equation}
	\begin{split}
		\sigma_{xf}^{2} &= M_{11}^{2}\sigma_{xi}^{2} + M_{12}^{2}\sigma_{x'i}^{2}, \\
		\sigma_{yf}^{2} &= N_{11}^{2}\sigma_{yi}^{2} + N_{12}^{2}\sigma_{y'i}^{2}, \\
		\sigma_{xyf} &= M_{11}N_{12}\sigma_{xy'i} + M_{12}N_{11}\sigma_{yx'i}.
	\end{split}
\end{equation}
Interestingly, at the initial observation location with $\alpha$'s zero and aligned ellipse $\nu_{1,2}=\pi/2$, the second moments $\sigma_{xy'}$ and $\sigma_{yx'}$ are
\begin{equation}
	\begin{split}
		\sigma_{xy'}&=\frac{1}{2}\sqrt{\frac{\beta_{1}}{\beta_{2}}}(\epsilon_{1}-\epsilon_{2}), \quad \sigma_{yx'}=\frac{1}{2}\sqrt{\frac{\beta_{2}}{\beta_{1}}}(\epsilon_{1}-\epsilon_{2}).
	\end{split}
    \label{eq:second_moments}
\end{equation}
According to Eq.~\eqref{eq:second_moments}, $\sigma_{xyf}$, the final correlation between $x$ and $y$ coordinates, is
\begin{equation}
	\begin{split}
		\sigma_{xyf}&= \frac{1}{2}(\epsilon_{1}-\epsilon_{2})\left(M_{11}N_{12}\sqrt{\frac{\beta_{1}}{\beta_{2}}} - M_{12}N_{11}\sqrt{\frac{\beta_{2}}{\beta_{1}}}\right), \\ 
		&= \frac{1}{2}(\epsilon_1{-\epsilon_{2}})\left(M_{11}N_{12}\frac{\sigma_{xi}}{\sigma_{yi}} - M_{12}N_{11}\frac{\sigma_{yi}}{\sigma_{xi}}\right).
	\end{split}
\end{equation}
The correlation between $x$ and $y$ coordinates can be expressed in terms of the rotation angle:
\begin{equation}
	2\sigma_{xyf} = (\sigma_{xf}^{2}-\sigma_{yf}^{2})\tan2\theta_{f},
\end{equation}
where $\theta_{f}$ is the rotation angle of the $(x,y)$ space ellipse. This leads to
\begin{equation}
	\begin{split}
		\epsilon_{1}-\epsilon_{2} &= \frac{(\sigma_{xf}^{2}-\sigma_{yf}^{2})\tan2\theta_{f}}{M_{11}N_{12}\frac{\sigma_{xi}}{\sigma_{yi}} - N_{11}M_{12}\frac{\sigma_{yi}}{\sigma_{xi}}}.
	\end{split}
	\label{Eq:differenceofemittancescase1}
\end{equation}
Now, the individual eigenmode emittances can be expressed in terms of observables between the two points in the accelerator:
\begin{equation}
    \begin{split}
        2\epsilon_{1} &= 2\sqrt{\frac{\sigma_{xi}^{2}\sigma_{xf}^{2}}{M_{12}^{2}}-\frac{M_{11}^{2}}{M_{12}^{2}}\sigma_{xi}^{4}} + \frac{(\sigma_{xf}^{2}-\sigma_{yf}^{2})\tan2\theta_{f}}{M_{11}N_{12}\frac{\sigma_{xi}}{\sigma_{yi}}-M_{12}N_{11}\frac{\sigma_{yi}}{\sigma_{xi}}}, \\
        2\epsilon_{2}&= \frac{(\sigma_{xf}^{2}-\sigma_{yf}^{2})\tan2\theta_{f}}{M_{11}N_{12}\frac{\sigma_{xi}}{\sigma_{yi}}-M_{12}N_{11}\frac{\sigma_{yi}}{\sigma_{xi}}} - 2\sqrt{\frac{\sigma_{xi}^{2}\sigma_{xf}^{2}}{M_{12}^{2}}-\frac{M_{11}^{2}}{M_{12}^{2}}\sigma_{xi}^{4}},
    \end{split}
\end{equation}
where subscripts $i,f$ designate the initial and final locations. Therefore, measurements of the beam size and rotation angle in the $(x,y)$ space with knowledge of the linear transport matrix lead to the extraction of the eigenmode emittances. This is shown in Fig.~\ref{fig:xydistributionsdiagnostics}, where the rotation angle at the final location can be easily extracted.

\subsection{Extraction of Coupling Strength}
\begin{figure}[tbp]
    \centering
    \includegraphics[width=0.49\linewidth]{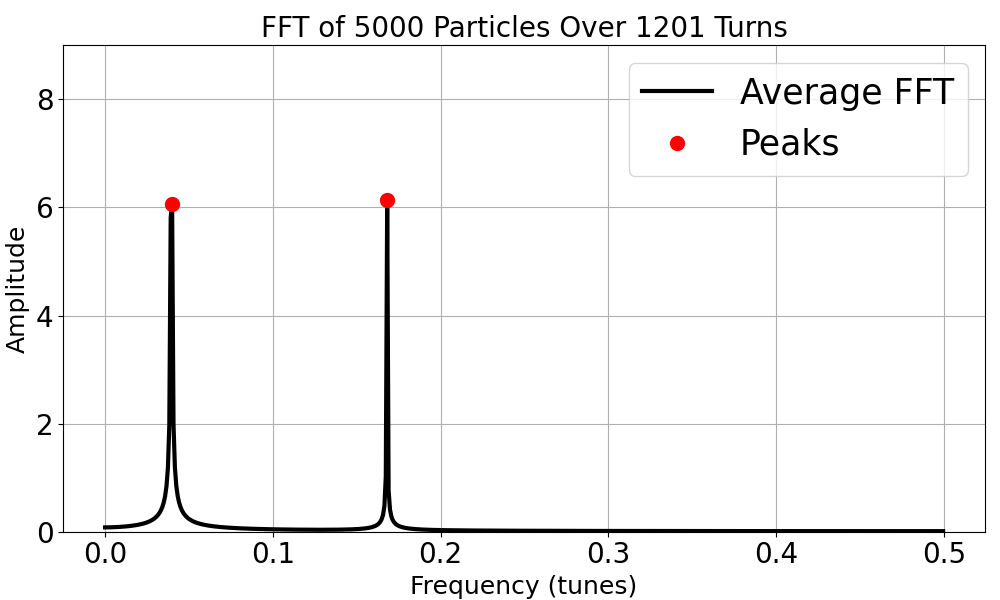}
    \includegraphics[width=0.49\linewidth]{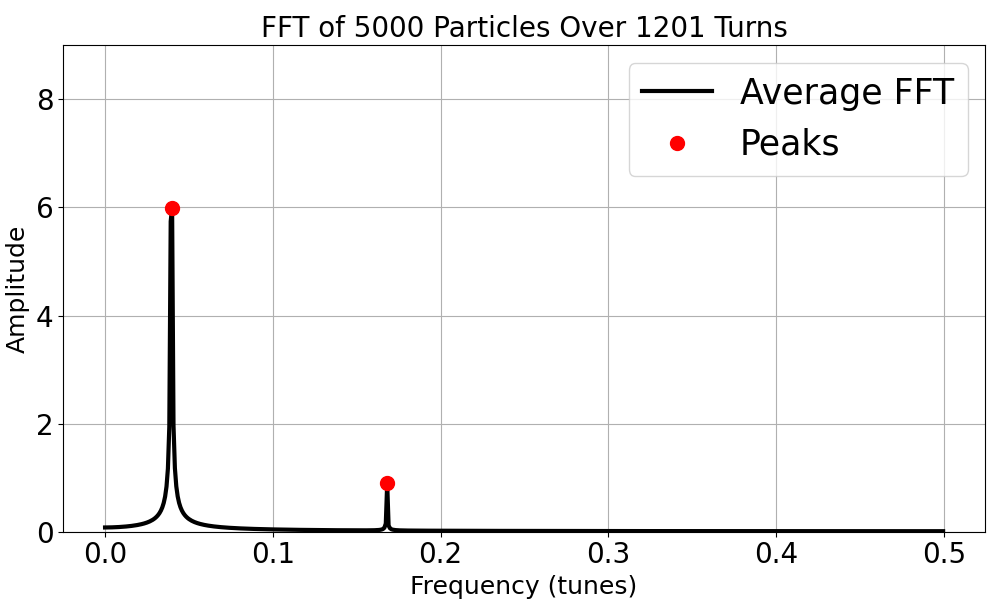}
    \caption{FFT analysis of $u=1/2$ with $\nu_{1,2}=\pi/2$ beam in the $x$ plane. Equal eigenmode emittances (left), and factor of $100$ difference between eigenmode emittances (right). Tracked with ImpactX code in a simple ring design.}
    \label{fig:fftanal}
\end{figure}

The coupling strength can be determined by calculating the phase advance between two points in the lattice. Equation~\eqref{Eq:phaseadvances} gives a direct relation between the matrix elements, phase advances, and the coupling strength. If the downstream cell is comprised of uncoupled elements and the phase advance is extracted between two points, the coupling strength can be determined as
\begin{equation}
    u = 1 - \tan\Delta\mu_{1}\left(\frac{M_{11}}{M_{12}}\beta_{1x} - \alpha_{1x}\right).
\end{equation} 
The coupling strength stays constant if the transfer matrix is uncoupled. 
The projections from mode $1$ and mode $2$ onto $x$ and $y$ planes will yield two Fourier peaks as a result of two modes of oscillations projecting onto real space as illustrated in Fig.~\ref{fig:fftanal}. The two peaks correspond to the tunes of modes $1$ and $2$, where the relative amplitudes of the peaks reflect the eigenmodes emittances ratio. Although Fig.~\ref{fig:fftanal} was produced for $u=1/2$, the behavior is general. The Fourier transformation of the phase space coordinate yields:
\begin{equation}
\begin{split}
    X(f) &= \frac{\sqrt{\epsilon_{1}\beta_{1x}}}{2}\delta(f-Q_{1}) + \frac{\sqrt{\epsilon_{2}\beta_{2x}}}{2}e^{i\nu_{2}}\delta(f-Q_{2}),\\
    Y(f) &= \frac{\sqrt{\epsilon_{1}\beta_{1y}}}{2}e^{i\nu_{1}}\delta(f-Q_{1}) + \frac{\sqrt{\epsilon_{2}\beta_{2y}}}{2}\delta(f-Q_{2}),
    \end{split}
\end{equation}
where the phases of coupling, $\nu_{1,2}$, contribute to the phase of the FFT, and $Q_{1,2}$ are the mode tunes. For $u=1/2$ lattices with phases of coupling $\pi/2$, the mode 2 tune is observable in FFT analysis of the $x$ plane and similarly in FFT of the $y$ plane.

\section{Linear Difference Resonance}\label{sec:lineardifres}

\begin{figure}[!htbp]
    \centering
    \includegraphics[width=0.42\linewidth]{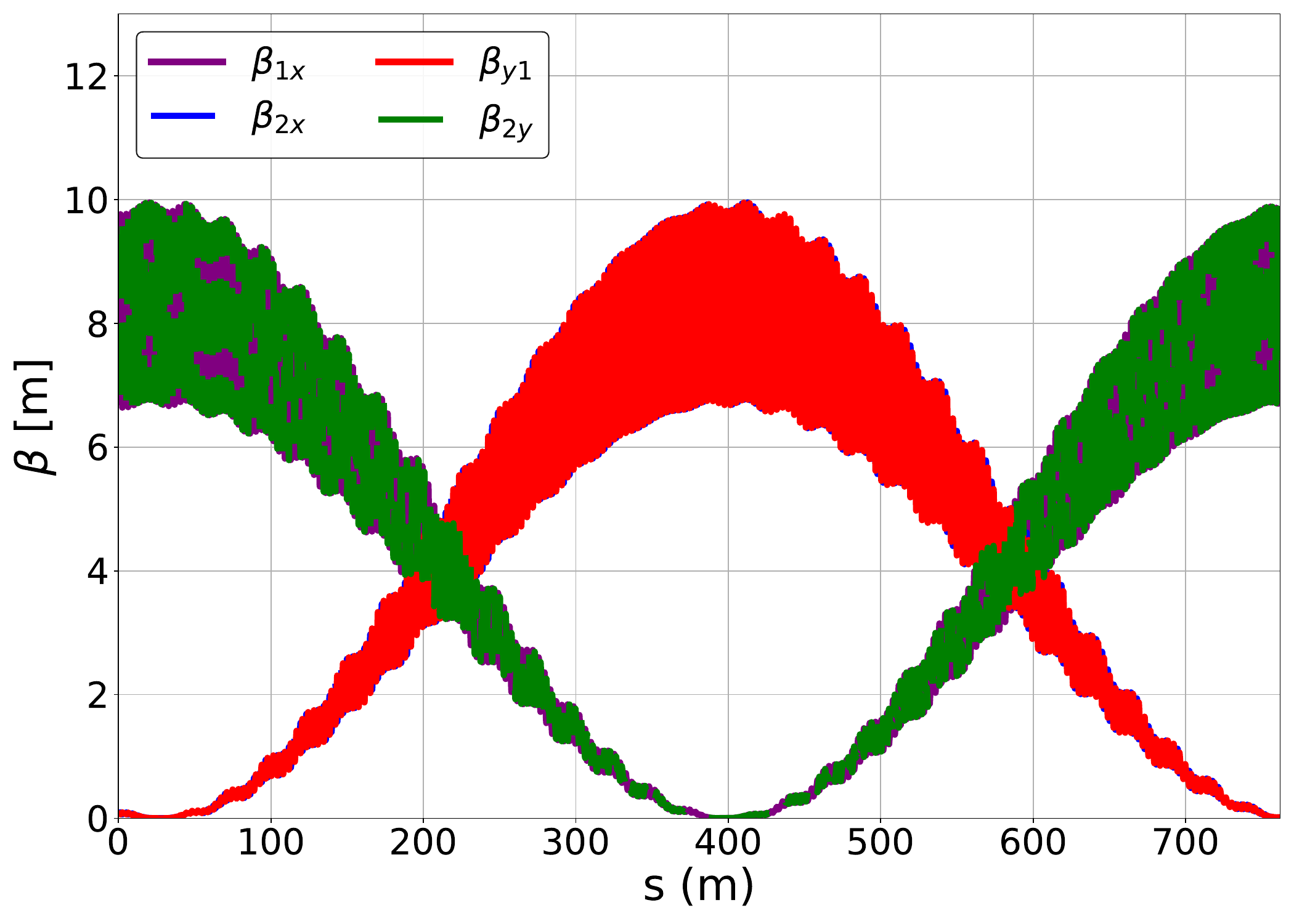}
    \includegraphics[width=0.42\linewidth]{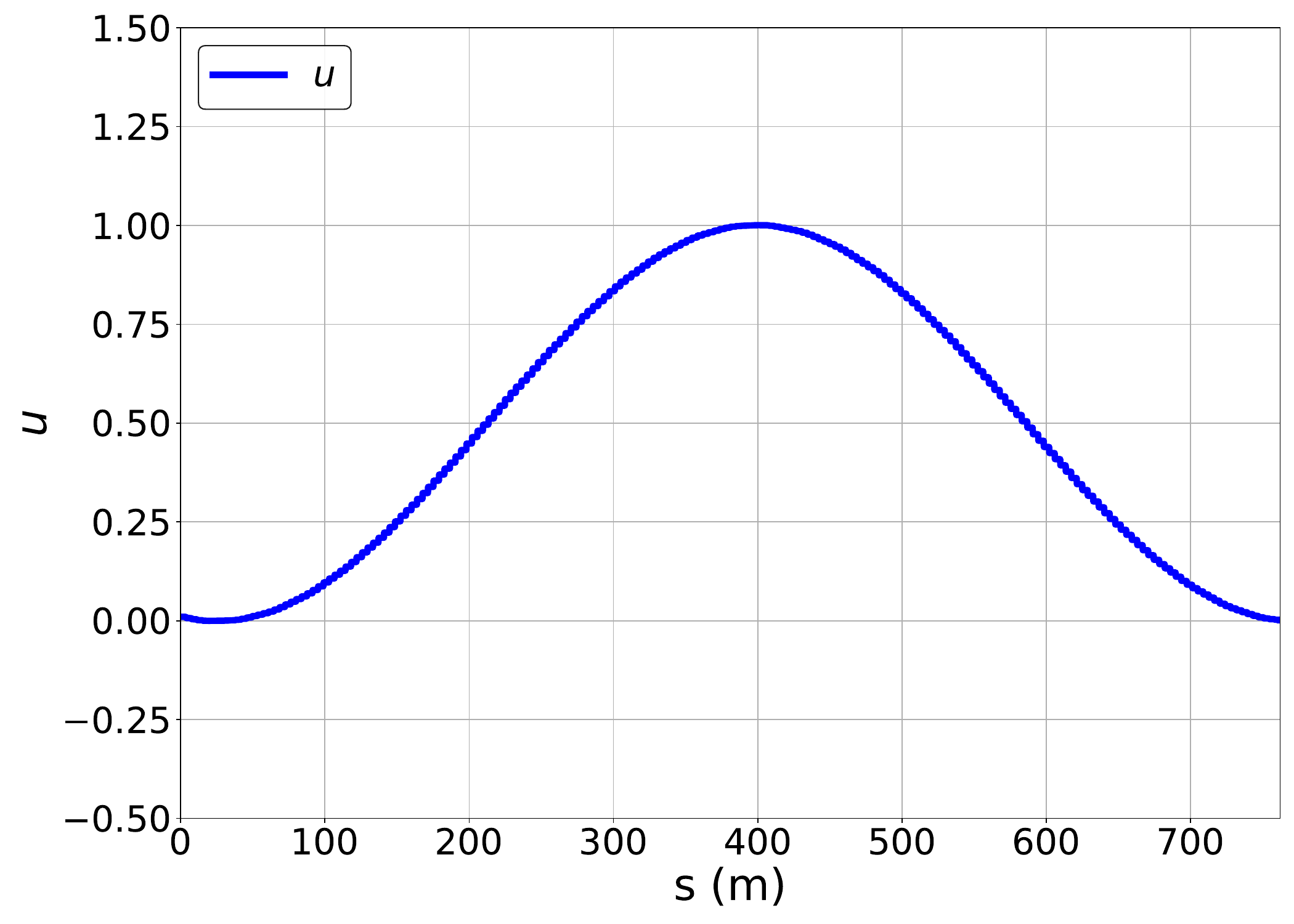}
    \includegraphics[width=0.42\linewidth]{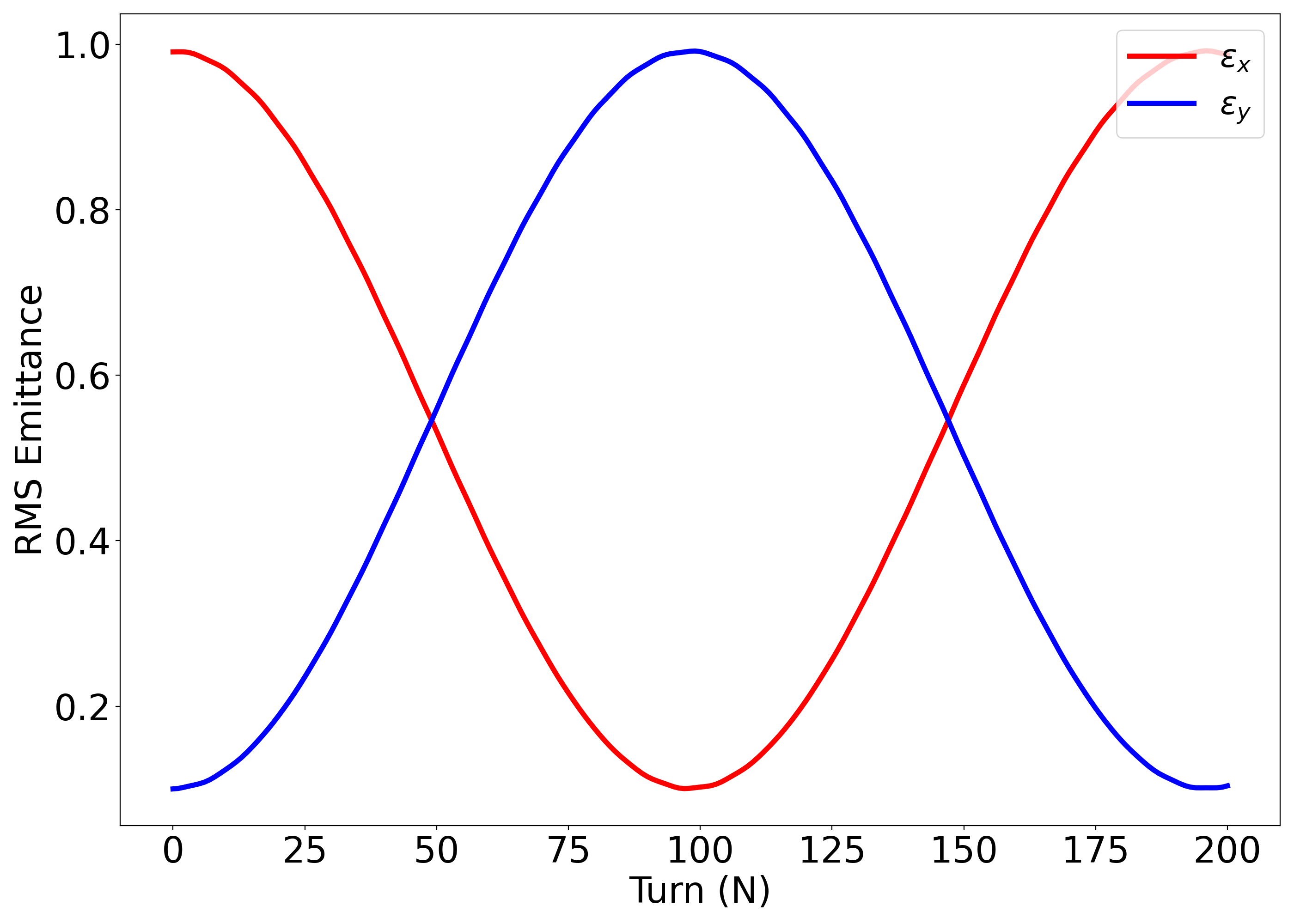}
    \includegraphics[width=0.42\linewidth]{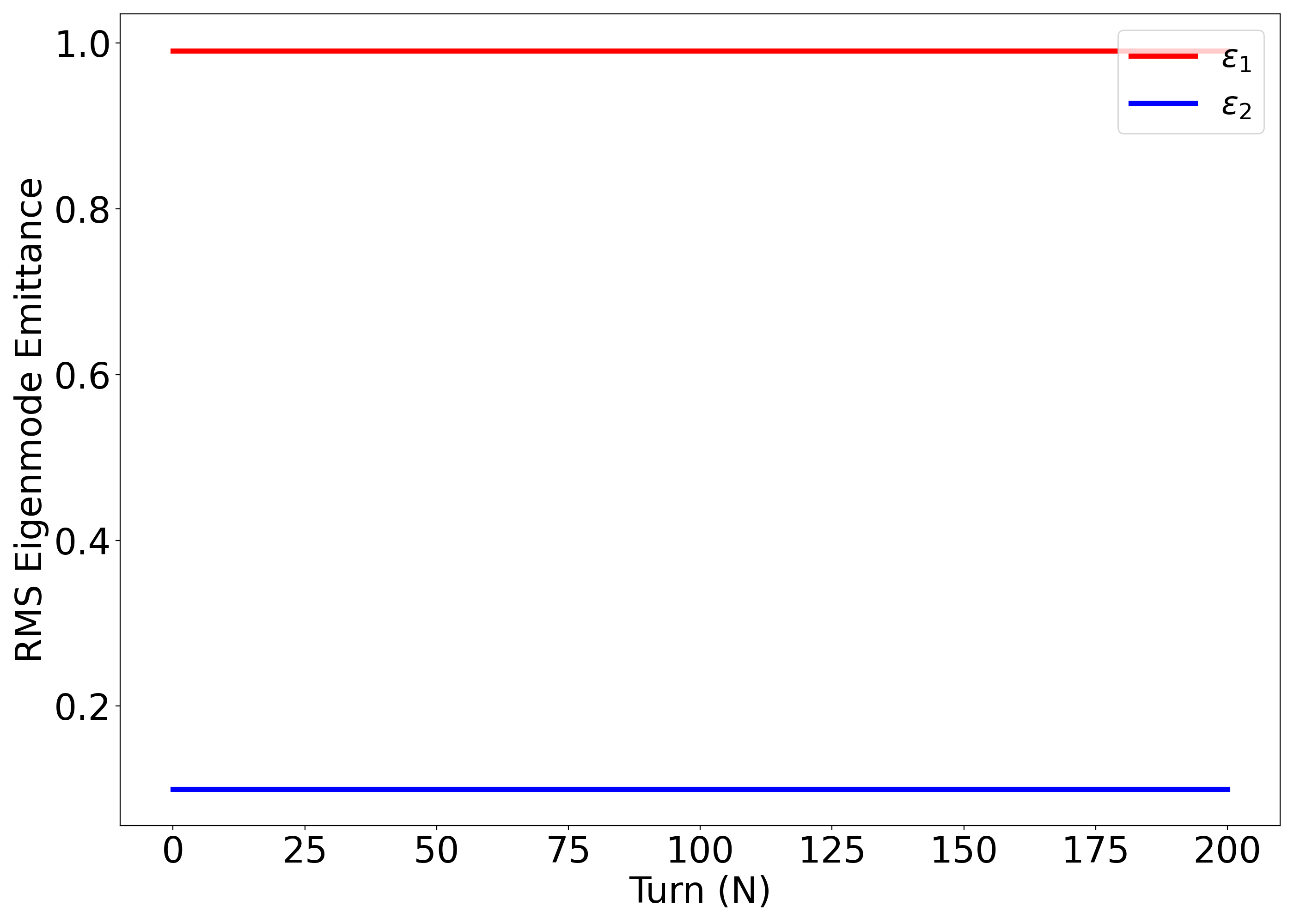}
    \caption{Tracking through a periodic lattice with skew perturbation. Coupled beta functions(top-left), coupling strength parameter(top-right),RMS projected emittances(bottom-left), RMS eigenmode emittances(bottom-right)}
    \label{fig:differenceresplot}
\end{figure}

The use of skew quadrupoles adds two more resonances to the linear resonance conditions, namely sum and difference resonances. The difference resonance relates to the condition when the tunes satisfy: $Q_{x}-Q_{y} = n$ for $n$ an integer. The difference resonance is introduced through a resonance proximity parameter and skew quadrupole coefficient in the literature, $\Delta = Q_{x}-Q_{y}-k$. The averaged slow coordinate rotation vectors yield the equation:
\begin{equation}
    \frac{d}{dn}\begin{pmatrix}
        a \\
        b
    \end{pmatrix} = -i\begin{pmatrix}
        \Delta & C \\
        C & -\Delta
    \end{pmatrix}\begin{pmatrix}
        a\\
        b
    \end{pmatrix},
\end{equation}
where $a,b$ are resonant frame vectors, $\Delta$ the resonance proximity parameter and $C$ the resonance driving term as shown in Appendix.~\ref{sec:linearskew}. The decoupling angle is computed with respect to the matrix shown above and yields:
\begin{equation}
    \tan2\alpha = \frac{\Delta}{C}.
\end{equation}
Since the angle decouples the normalized motion, it can be associated with the Edwards-Teng angle defined in Section~\ref{subsec:edwardstenglBrel}. The decoupling angle can be related to the coupling strength parameter, $u$, with the definition $\tan^{2}\phi = u/(1-u)$, which yields the solution for the strength parameter as:
\begin{equation}
    u = \frac{1}{2}(1-\frac{\Delta}{\sqrt{\Delta^{2}-C^{2}}}).
\end{equation}
As we see, $\Delta=0$ corresponds to on-resonance behavior, which yields $u=1/2$ as the coupling strength. This is when two emittances get ``shared'' which also corresponds to the fully coupled scenario in the coupled parametrization. The shared emittance is seen from the projection emittances from Eq.~\eqref{eq:projectedemittanceswithu}, as for $u=1/2$ case reduces to:
\begin{equation}
    \epsilon_{x,y} = \frac{1}{2}(\epsilon_{1} + \epsilon_{2}),
\end{equation}
where $\epsilon_{1,2}$ are the eigenmode emittances intrinsic to the beam and $\epsilon_{x,y}$ the projected emittances. When $ u=0$, the projected emittances are assigned the correct mode representations. The turn-by-turn emittance exchange rate is given in terms of resonance parameters in~\cite{franchi2007emittance}. However, since the coupling strength parameter depends on the resonance parameters, the exchange rate can be expressed in terms of the coupling strength parameter of the distribution, since the projected emittances can be written as:
\begin{equation}
    \begin{split}
        \epsilon_{x}(n) &= \epsilon_{1} -u_{b}(n)(\epsilon_{1}-\epsilon_{2}), \\
        \epsilon_{y}(n) &= \epsilon_{2} + u_{b}(n)(\epsilon_{1}-\epsilon_{2}),
    \end{split}
\end{equation}
where $u_{b}$ is the beams coupling strength and the rate of change of projected emittances with respect to turn number depends on the coupling strength parameter as:
\begin{equation}
    \begin{split}
        \frac{d\epsilon_{x}}{dn}&= -\frac{du_{b}}{dn}(\epsilon_{1}-\epsilon_{2}), \\
        \frac{d\epsilon_{y}}{dn}&= \frac{du_{b}}{dn}(\epsilon_{1}-\epsilon_{2}).
    \end{split}
\end{equation}
As the beam continuously goes through the skew perturbation, the coupling strength parameter of the beam changes. Therefore, we can write an ansatz for the evolution of the coupling strength parameter as:
\begin{equation}
    u_{b}(n) = \frac{1}{2}(1-\cos(2\phi_{0} + 2\pi n\chi)),
\end{equation}
where $\phi_{0}$ is initial phase relation where $\chi = \sqrt{\Delta^{2} + C^{2}}$ term. The propagation of the coupling strength parameter is then written as:
\begin{equation}
    u_{b}(n)=\frac{1}{2} + (u_{b0}-\frac{1}{2})\cos(2\pi n\chi) + \sqrt{u_{b0}(1-u_{b0})}\sin(2\pi n\chi)
\end{equation}
This results in the propagation of the projected emittances as:
\begin{equation}
	\begin{split}
		\epsilon_{x}(n) &= \frac{\epsilon_{1}}{2} + \frac{\epsilon_{2}}{2} - \left( (u_{b0}-\frac{1}{2})\cos(2\pi n \chi) + \sqrt{u_{b0}(1-u_{b0})}\sin(2\pi n \chi)\right )(\epsilon_{1}-\epsilon_{2}), \\
		\epsilon_{y}(n) &= \frac{\epsilon_{1}}{2} + \frac{\epsilon_{2}}{2} + \left( (u_{b0}-\frac{1}{2})\cos(2\pi n \chi) + \sqrt{u_{b0}(1-u_{b0})}\sin(2\pi n \chi)\right )(\epsilon_{1}-\epsilon_{2}).
	\end{split}
\end{equation}
Where $\epsilon_{1,2}$ are invariant eigenmode emittances, $u_{b0}$ is initial coupling strength of the beam. Parametrizing the linear difference resonance from the perspective of the coupling strength parameter suggests that it is the coupling strength parameter of the beam that changes, rather than the emittances, which yields the exchange principle. An example of this dynamic behavior is observed in the generating function as it tracks through a linear periodic system with equal tunes $Q_{x}=Q_{y}$, featuring a skew perturbation at the system's end. The results are shown in Fig.~\ref{fig:differenceresplot}, where the coupled optics functions illustrate the behavior of the difference resonance. The off-mode beta functions start from zero, indicating no coupling, and become equal at the $u=1/2$ location. The on-mode functions tend to zero as the coupling strength parameter reaches one and periodically oscillate in that pattern. The minimum emittance exchange is fully realized when the coupling strength parameter $u$ reaches one, and the RMS emittance in $x$ swaps with the RMS emittance in $y$. As we see from the figure, the eigenmode emittances are invariant in this process due to the linearity of the transformation. The RMS projected emittances and eigenmode emittances are tracked and computed using ImpactX simulation~\cite{huebl2022next}. 
\section{Conclusion}
In conclusion, we have presented an explicit formalism that governs the transformation of coupled beam optics functions in a similar way to the transformation of uncoupled beam Twiss functions. The coupled optics are derived from both Mais-Ripken parametrization and Lebedev-Bogacz eigenvector parametrization. We have shown the connection and equivalence between the Mais-Ripken and Lebedev-Bogacz parameters. Using the transfer matrix method with the generating vectors as the basis, it is straightforward to create an optimization routine to find the optimal periodic functions. Using a method for matching and optimizing coupled functions, we have shown that a fully coupled lattice can be created. 
The capability of producing matched coupled lattices will help explore new options for future accelerator and collider designs that could benefit from potential advantages in the unexplored area of coupled dynamics, such as the application of circular-mode beams and what they might offer. We have also given a picture of linear difference resonance that arises from skew perturbation to a system and related the emittance exchange principle to the coupled optics functions with visualizations.
We have also presented a method for extracting eigenmode emittances from observables that could be measured in fully coupled lattices. Having a transfer matrix method for coupled optics functions enables the integration of this method into well-known design codes. It could also enable future AI/ML modeling for coupled lattice design, which would eventually lead to interesting new designs that we have not yet thought of. 

\section{Acknowledgments}
This work was supported by the U.S. Department of Energy, under Contract No. DE-AC02-06CH11357.

\appendix

\section{Phase Advances From Coupled Transfer Matrix}
The enerating vectors found in Eq.~\eqref{Eq:genvecscoupledLB} provide straightforward definitions for optics functions and phase advances as in Eq.~\eqref{Eq:genvectorscoupledfunctions}. We can rewrite the generating vectors as:
\begin{equation}
	\begin{split}
		\vec{z}_{1} &=\begin{pmatrix}
		    x_{1}\\
            x'_{1}\\
            y_{1}\\
            y'_{1}
		\end{pmatrix} =\begin{pmatrix}
		\sqrt{\beta_{1x}}\cos\mu_{1} \\
		\frac{-\alpha_{1x}\cos\mu_{1}-(1-u)\sin\mu_{1}}{\sqrt{\beta_{1x}}} \\
		\sqrt{\beta_{1y}}\cos(\mu_{1}-\nu_{1}) \\
		\frac{-\alpha_{1y}\cos(\mu_{1}-\nu_{1})-u\sin(\mu_{1}-\nu_{1})}{\sqrt{\beta_{1y}}}
	\end{pmatrix}, \qquad \vec{z}_{2}= \begin{pmatrix}
	    x_{2}\\
        x'_{2}\\
        y_{2}\\
        y'_{2}
	\end{pmatrix}=\begin{pmatrix}
		\sqrt{\beta_{1x}}\sin\mu_{x} \\
		\frac{(1-u)\cos\mu_{1} - \alpha_{1x}\sin\mu_{1}}{\sqrt{\beta_{1x}}} \\
		\sqrt{\beta_{1y}}\sin(\mu_{1}-\nu_{1})\\
		\frac{u\cos(\mu_{1}-\nu_{1})-\alpha_{1y}\sin(\mu_{1}-\nu_{1})}{\sqrt{\beta_{1y}}}
	\end{pmatrix}, \\
		\vec{z}_{3}&= \begin{pmatrix}
		    x_{3}\\
            x'_{3}\\
            y_{3}\\
            y'_{3}
		\end{pmatrix}=\begin{pmatrix}
		\sqrt{\beta_{2x}}\cos(\mu_{2}-\nu_{2}) \\
		\frac{-\alpha_{2x}\cos(\mu_{2}-\nu_{2})-u\sin(\mu_{2}-\nu_{2})}{\sqrt{\beta_{2x}}} \\
		\sqrt{\beta_{2y}}\cos\mu_{2} \\
		\frac{-\alpha_{2y}\cos\mu_{2}-(1-u)\sin\mu_{2}}{\sqrt{\beta_{2y}}}
	\end{pmatrix}, \qquad \vec{z}_{4}= \begin{pmatrix}
	    x_{4}\\
        x'_{4}\\
        y_{4}\\
        y'_{4}
	\end{pmatrix}=\begin{pmatrix}
		\sqrt{\beta_{2x}}\sin(\mu_{2}-\nu_{2}) \\
		\frac{u\cos(\mu_{2}-\nu_{2})-\alpha_{2x}\sin(\mu_{2}-\nu_{2})}{\sqrt{\beta_{2x}}} \\
		\sqrt{\beta_{2y}}\sin\mu_{2} \\
		\frac{(1-u)\cos\mu_{2}-\alpha_{2y}\sin\mu_{2}}{\sqrt{\beta_{2y}}}
	\end{pmatrix}.
	\end{split}
\end{equation}
The phase advances can be extracted from the generating vectors as:
\begin{equation}
	\begin{split}
		\mu_{1} &= \arctan(x_{2}/x_{1}), \qquad \mu_{2}= \arctan(y_{4}/y_{3}), \\
		\mu_{1}-\nu_{1} &= \arctan(y_{2}/y_{1}), \qquad \mu_{2}-\nu_{2}= \arctan(x_{4}/x_{3}). 
	\end{split}
\end{equation}
Thus, the propagation of the phase advances depends on the propagation of the generating vectors. The change in phase advance is $\Delta\mu_{1}=\mu_{1f}- \mu_{1i}$, and can be written as:
\begin{equation}
	\begin{split}
		\tan(\Delta\mu_{1}) &= \frac{\tan\mu_{f}-\tan\mu_{i}}{1 + \tan\mu_{f}\tan\mu_{i}} = \frac{\frac{x_{2f}}{x_{1f}} - \frac{x_{2i}}{x_{1i}}}{1 + \frac{x_{2f}}{x_{1f}}\frac{x_{2i}}{x_{1i}}} 
		= \frac{x_{2f}x_{1i} - x_{1f}x_{2i}}{x_{1f}x_{1i} + x_{2f}x_{2i}}.
	\end{split}
\end{equation}
Since the generating vector components are known , the final state of the generating vectors can be obtained from the transfer matrix, which will yield:
\begin{equation}
\tan(\Delta\mu_{1}) = \frac{M_{12}(1-u) - m_{11}\sqrt{\beta_{1x}\beta_{1y}}\sin\nu_{1} +m_{12}\sqrt{\frac{\beta_{1x}}{\beta_{1y}}}\left (u\cos\nu_{1} + \alpha_{1y}\sin\nu_{1} \right )}{M_{11}\beta_{1x}-M_{12}\alpha_{1x} + m_{11}\xi_{1xy} + m_{12}\xi_{1xy'}}.
\end{equation}
The same procedure can be carried out for other phase parameters. 

\section{Coupled Beam Matrix and Optics Functions}\label{appendix:B}
Using the normalization conditions of eigenvectors, $\vec{v}_{i}^{\dagger}S\vec{v}_{i}=-2i$, the strength of coupling can be deduced as:
\begin{equation}
    \begin{split}
        \kappa_{x}&= \sqrt{\frac{\beta_{2x}}{\beta_{1x}}}, \quad \kappa_{y}=\sqrt{\frac{\beta_{1y}}{\beta_{2y}}}, \\
        A_{x}&= \alpha_{1x}\kappa_{x} - \alpha_{2x}\kappa_{x}^{-1}, \\
        A_{y}&=\alpha_{2y}\kappa_{y}-\alpha_{1y}\kappa_{y}^{-1}, \\
        u&= \frac{-\kappa_{x}^{2}\kappa_{y}^{2}\pm\sqrt{\kappa_{x}^{2}\kappa_{y}^{2}(1+\frac{A_{x}^{2}-A_{y}^{2}}{\kappa_{x}^{2}-\kappa_{y}^{2}}(1-\kappa_{x}^{2}\kappa_{y}^{2}))}}{1-\kappa_{x}^{2}\kappa_{y}^{2}}.
    \end{split}
\end{equation}
The normalization condition also allows one to extract single particle geometric amplitudes. Multiplying the phase space vector on the left hand side with $\vec{v}_{1}^{\dagger}S$ leads to deducing the single particle amplitude:
\begin{equation}
    \begin{split}
        2J_{1}&= \vec{v}_{1}^{\dagger}S\vec{z}\vec{z}^{T}S^{T}\vec{v}_{1} \\
             & = (\frac{(1-u)^{2}+\alpha_{1x}^{2}}{\beta_{1x}})x^{2} + 2\alpha_{1x}xx' + \beta_{1x}x'^{2} + (\frac{u^{2}+\alpha_{1y}^{2}}{\beta_{1y}})y^{2} + 2\alpha_{1y}yy' + \beta_{1y}y'^{2} \\
             & + \frac{2xy}{\sqrt{\beta_{1x}\beta_{1y}}}((u(1-u)+\alpha_{1x}\alpha_{1y})\cos\nu_{1} + (\alpha_{1y}-u(\alpha_{1x}+\alpha_{1y}))\sin\nu_{1}) \\
             & + 2yx'\sqrt{\frac{\beta_{1x}}{\beta_{1y}}}(\alpha_{1y}\cos\nu_{1}-u\sin\nu_{1}) + 2xy'\sqrt{\frac{\beta_{1y}}{\beta_{1x}}}(\alpha_{1x}\cos\nu_{1}+(1-u)\sin\nu_{1}) \\
             & +2x'y'\sqrt{\beta_{1x}\beta_{1y}} \cos\nu_{1}.
    \end{split}
\end{equation}
The first two entries in the equation above show two Courant-Snyder like ellipses on $x-x'$ and $y-y'$ planes. The rest of the terms show the correlations between different phase planes that depend on the phases of coupling.

The beam matrix is parametrized using the Lebedev-Bogacz parametrization
in~\cite{lebedev2010betatron}:
\begin{equation}
	\begin{split}
		\sigma_{xx} &= \epsilon_{1}\beta_{1x} + \epsilon_{2}\beta_{2x}, \\
		\sigma_{xx'} &= -\alpha_{1x}\epsilon_{1} - \alpha_{2x}\epsilon_{2},\\
		\sigma_{x'x'} &= \epsilon_{1}\frac{(1-u)^{2} + \alpha_{1x}^{2}}{\beta_{1x}} + \epsilon_{2}\frac{u^{2} + \alpha_{2x}^{2}}{\beta_{2x}}, \\
		\sigma_{yy}&= \epsilon_{1}\beta_{1y} + \epsilon_{2}\beta_{2y}, \\
		\sigma_{yy'} &= -\alpha_{1y}\epsilon_{1} - \alpha_{2y}\epsilon_{2}, \\
		\sigma_{y'y'} &= \epsilon_{1}\frac{u^{2}+\alpha_{1y}^{2}}{\beta_{1y}} + \epsilon_{2}\frac{(1-u)^{2} + \alpha_{2y}^{2}}{\beta_{2y}}, \\
		\sigma_{xy}&= \epsilon_{1}\sqrt{\beta_{1x}\beta_{1y}}\cos\nu_{1} + \epsilon_{2}\sqrt{\beta_{2x}\beta_{2y}}\cos\nu_{2}, \\
		\sigma_{xy'} &= \epsilon_{1}\sqrt{\frac{\beta_{1x}}{\beta_{1y}}}(u\sin\nu_{1} - \alpha_{1y}\cos\nu_{1}) - \epsilon_{2}\sqrt{\frac{\beta_{2x}}{\beta_{2y}}}((1-u)\sin\nu_{2}+\alpha_{2y}\cos\nu_{2}), \\
		\sigma_{yx'} &= -\epsilon_{1}\sqrt{\frac{\beta_{1y}}{\beta_{1x}}}((1-u)\sin\nu_{1}+\alpha_{1x}) + \epsilon_{2}\sqrt{\frac{\beta_{2y}}{\beta_{2x}}}(u\sin\nu_{2} - \alpha_{2x}\cos\nu_{2}), \\
		\sigma_{x'y'} &= \epsilon_{1}\frac{1}{\sqrt{\beta_{1x}\beta_{2x}}}((\alpha_{1y}(1-u)-\alpha_{1x}u)\sin\nu_{1} + (u(1-u) + \alpha_{1x}\alpha_{2y})\cos\nu_{1}) \\
		&+ \epsilon_{2}\frac{1}{\sqrt{\beta_{2x}\beta_{2y}}}\left( 
				(\alpha_{2x}(1-u) - \alpha_{2y}u)\sin\nu_{2} + (u(1-u) + \alpha_{2x}\alpha_{2y})\cos\nu_{2}
		\right).
	\end{split}
\end{equation}
The $(x,y)$ plane distribution for a general tilted ellipse is
\begin{equation}
\begin{split}
    &\frac{x^{2}}{\sigma_{x}^{2}} - \frac{2\tilde{\alpha}}{\sigma_{x}\sigma_{y}}xy + \frac{y^{2}}{\sigma_{y}^{2}} = 1-\tilde{\alpha}^{2}, \\
    & \tilde{\alpha} = \frac{\sigma_{xy}}{\sigma_{x}\sigma_{y}}.
\end{split}
\end{equation}
In order to have no tilt in $x-y$ plane, $\sigma_{xy}=0$. In coupled optics, no tilt suggests the phases of coupling are $\pi/2$.
\begin{equation}
    \frac{x^{2}}{\sigma_{x}^{2}} + \frac{y^{2}}{\sigma_{y}^{2}} = 1.
\end{equation}

Under symplectic transformations the beam matrix transforms linearly, $\Sigma_{f}=\mathcal{M}^{T}\Sigma_{i}\mathcal{M}$. The 4D emittance is computed through the determinant of the beam matrix:
\begin{equation}
    det(\Sigma_{f}) = det(\mathcal{M})det(\Sigma_{i})det(\mathcal{M}) = \epsilon_{4D}^{2}=(\epsilon_{1}\epsilon_{2})^{2}.
\end{equation}
Since the product of emittances is known from the determinant of the beam matrix, a sum of emittances is needed so computing individual eigenmode emittances is possible. The diagonalized beam matrix will yield RMS eigenmode emittances as:
\begin{equation}
    \tilde{\Sigma} = T^{-1}\Sigma T = \mathrm{diag}(\epsilon_{1},\epsilon_{1},\epsilon_{2},\epsilon_{2}),
\end{equation}
where $T$ is a symplectic matrix that diagonalizes the beam matrix. Computing trace of the diagonalized beam matrix and symplectic unit matrix will yield the sum of RMS eigenmode emittances as:
\begin{equation}
    Tr[(\tilde{\Sigma}S)^{2}]=-2(\epsilon_{1}^{2}+\epsilon_{2}^{2}).
\end{equation}
Using the cyclic property of the trace and identities of the symplectic unit matrix and the diagonalizing matrix $T$, we can prove that the two traces are identical in value: $Tr[(\tilde{\Sigma}S)^{2}] = Tr[(\Sigma S)^{2}]=-2(\epsilon_{1}^{2} + \epsilon_{2}^{2})$. Knowing the sum of the two RMS eigenmode emittances and the product of them allows us to express each individual eigenmode emittances from the beam matrix, as for instance for mode 1:
\begin{equation}
    \begin{split}
        \epsilon_{2}^{2} &= det(\Sigma)/\epsilon_{1}^{2}, \\
        Tr[(\Sigma S)^{2}]&= -2(\epsilon_{1}^{2} + \frac{det(\Sigma)}{\epsilon_{1}^{2}}), \\
    \end{split}
\end{equation}
\begin{equation}
    \begin{split}
        2\epsilon_{1}^{4} + \epsilon_{1}^{2}Tr[(\Sigma S)^{2}] + 2det(\Sigma)=0,
    \end{split}
\end{equation}
the solution will yield:
\begin{equation}
    \epsilon_{1} = \frac{1}{2}\sqrt{-Tr[(\Sigma S)^{2}] + \sqrt{Tr^{2}[(\Sigma S)^{2}] -16det(\Sigma)}}.
\end{equation}

\section{Symmetric Alternating Gradient Design Condition}\label{Appendix:C}
\begin{figure}[tbp]
    \centering
    \includegraphics[width=0.8\linewidth]{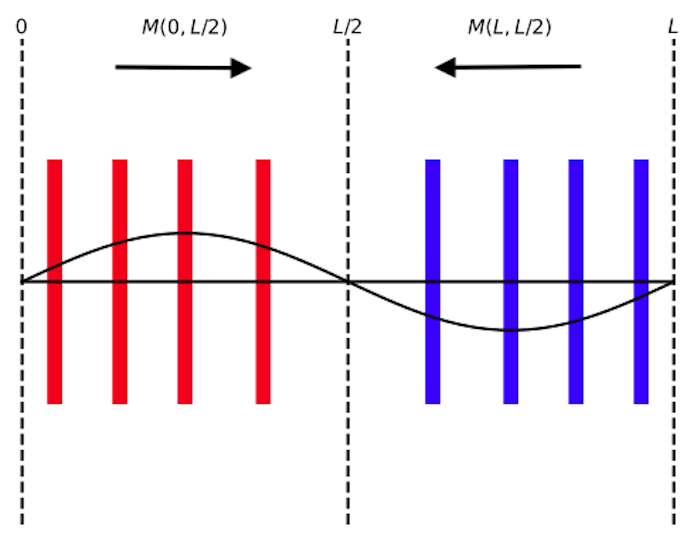}
    \caption{Symmetric alternating gradient design principle.}
    \label{fig:mirrorsymmetry}
\end{figure}

    Symmetric alternating gradient~(SAG) design is a midpoint plane symmetry that ensures equal phase advance transformation for quadrupole-based periodic cells. The design principle is shown in Fig.~\ref{fig:mirrorsymmetry}. The magnetic elements to the left of the middle of the cell are identical in strength but opposite in polarity, $K=-K$, with respect to the elements to the right. The transfer matrices are expressed as
\begin{equation}
    \begin{split}
        \mathcal{M}(0,L/2)= \begin{pmatrix}
            M & 0 \\
            0 & N
        \end{pmatrix}, \qquad \mathcal{M}(L,L/2)= \begin{pmatrix}
            N & 0 \\
            0 & M
        \end{pmatrix}.
    \end{split}
    \label{Eq:midplanesymmetryeq}
\end{equation}
A symmetric design ensures every optics function is periodic through the transformation.

\section{Linear Difference Resonance Driven By Skew Quadrupole Perturbation}\label{sec:linearskew}
The linear difference resonance is observed when uncoupled Hamiltonian is perturbed by skew quadrupole potential, $V_{skew}=gxy$. The perturbed Hamiltonian is written as:
\begin{equation}
    H = Q_{x}J_{x} + Q_{y}J_{y} + V_{skew}(\psi_{x},\psi_{y},J_{x},J_{y}).
\end{equation}
Where, for the skew potential Courant-Snyder coordinate parametrization is used:
\begin{equation}
    H_{skew} = g\sqrt{J_{x}J_{y}}\cos(\Psi_{x}-\Psi_{y}) + g\sqrt{J_{x}J_{y}}\cos(\Psi_{x}+\Psi_{y}),
\end{equation}
which yields the two conditions for sum and difference resonances. Introducing a change of variable, $G=\frac{1}{2}(J_{x}+J_{y})$ and $I=\frac{1}{2}(J_{x}-J_{y})$ transforms the Hamiltonian:
\begin{equation}
    H = \delta Q_{x} + \Delta I + \tilde{g}\sqrt{G^{2}-I^{2}}\cos\xi,
\end{equation}
where $\delta = Q_{x}+Q_{y}$ and $\Delta = Q_{x}-Q_{y}-m$. The angle $\xi$ depends on the difference of betatron phases, since there is no dependence on the sum of betatron phases, the Hamiltonians equations yield an invariant, $\dot{G}=0$, which is the sum of actions of $x$ and $y$ planes, the rate of change of actions depends on $\dot{I}$ which gives opposite signs: $\dot{J}_{x}=-\dot{J}_{y}$ which yields in the emittance exchange formalism. The equations of motion yields the Hill's equations. The solution for the skew potential is carried out in normal form coordinates:
\begin{equation}
    h_{x}=\frac{1}{\sqrt{2}}(X+iP_{x})= \sqrt{J_{x}}e^{-i\Psi_{x}}, \qquad h_{y}=\frac{1}{\sqrt{2}}(Y+iP_{y})=\sqrt{J_{y}}e^{-i\Psi_{y}}.
\end{equation}
The skew perturbation is written as: $\Delta p_{x}=gy$ and $\Delta p_{y}=gx$. Expressing the coordinates in terms of normalized coordinates and expanding to normal-form coordinates yields:
\begin{equation}
\begin{split}
    \Delta h_{x} &= \frac{i}{2}g\sqrt{\beta_{x}\beta_{y}}(h_{x} + h_{x}^{*}), \\
    \Delta h_{y} &= \frac{i}{2}g\sqrt{\beta_{x}\beta_{y}}(h_{y} + h_{y}^{*}).
    \end{split}
\end{equation}
One turn map rotates the normalized vectors, hence; the normal-form vectors are written as: $h_{x}(n)=h_{x}e^{-i2\pi n Q_{x}}$. Moving to resonant co-rotating frame by introducing slow variables:
\begin{equation}
    a_{n} =h_{x}e^{i2\pi Q_{x}n}, \quad b_{n} = h_{y}^{*}e^{-i2\pi(Q_{y}+k)n},
\end{equation}
with detuning resonance proximity parameter as: $\Delta = Q_{x}-Q_{y}-k$, the averaging yields:
\begin{equation}
    \frac{d}{dn}\begin{pmatrix}
        a\\
        b
    \end{pmatrix} = -i \begin{pmatrix}
        \Delta & C \\
        C & -\Delta
    \end{pmatrix}\begin{pmatrix}
        a\\
        b
    \end{pmatrix}
\end{equation}
Which yields the decoupling angle as:
\begin{equation}
    \tan2\alpha = \frac{\Delta}{C},
\end{equation}
Where, $C$ is defined as the driving term:
\begin{equation}
    C = \frac{1}{2\pi}\oint g(s)\sqrt{\beta_{x}\beta_{y}}e^{i(\Psi_{x}-\Psi_{y} - k\theta)},
\end{equation}
for $\theta = 2\pi s/C$. 
\clearpage

\bibliographystyle{unsrt}      
\bibliography{main}

\end{document}